\documentclass[11pt,a4paper]{article}
\DeclareMathAlphabet{\scr}{U}{rsfs}{m}{n}
\pdfoutput=1
\usepackage{latexsym}
\usepackage{epsfig}
\usepackage[mathscr]{eucal}
\usepackage{amsfonts}
\usepackage{amscd}
\usepackage{cite}
\usepackage{array}   
\usepackage{amssymb}
\usepackage{colordvi}
\usepackage[centertags]{amsmath}
\usepackage{enumerate}
\usepackage{graphicx}
\usepackage{booktabs}
\usepackage{theorem}
\usepackage{multirow}
\usepackage[footnotesize]{caption}
\usepackage{bbm}
\usepackage[labelfont=bf,textfont={sl,bf}]{subfig}
 \usepackage{slashed}
\usepackage[obeyFinal]{todonotes}
\usepackage{color}
\usepackage{ulem}

\usepackage{mathtools}

\DeclarePairedDelimiter\abs{\lvert}{\rvert}%
\newcommand{\newc}{\newcommand}
\newc{\be}{\begin{equation}}
\newc{\ee}{\end{equation}}
\newc{\bea}{\begin{eqnarray}}
\newc{\eea}{\end{eqnarray}}
\newc{\ol}{\overline}
\newc{\wt}{\widetilde}
\newc{\bs}{\boldsymbol}
\newc{\m}{\mathcal}
\newc{\lan}{\langle}
\newc{\ra}{\rangle}
\newc{\pa}{\partial}

\newcommand{\sbeta}{{s_{\beta}}}
\newcommand{\cbeta}{{c_{\beta}}}
\newcommand{\tbeta}{{t_{\beta}}}

\newcommand{\non}{\nonumber} 
\newcommand{\crn}{\nonumber \\}
\newcommand{\beq}{\begin{eqnarray}} 
\newcommand{\eeq}{\end{eqnarray}} 
\newcommand{\bpmatrix}{\begin{pmatrix}}
\newcommand{\epmatrix}{\end{pmatrix}}
\newcommand{\ba}{\begin{array}}
\newcommand{\ea}{\end{array}}
\newcommand{\braket}[1]{\left(#1\right)}
\newcommand{\sbraket}[1]{\left[#1\right]}

\newcommand{\fr}{\frac}

\newcommand{\diag}{\text{diag}}
\newcommand{\al}{\alpha}

\newcommand{\calM}{{\cal M}}

\newcommand{\calL}{{\cal L}}
\newcommand{\calO}{{\cal O}}

\renewcommand{\ol}{\text{1l}}

\newcommand{\mueff}{\mu_{\text{eff}}}
\newcommand{\figref}[1]{Fig.~\ref{#1}}
\renewcommand{\eqref}[1]{Eq.~(\ref{#1})}
\newcommand{\bib}[1]{Ref.~\cite{#1}}
\newcommand{\tab}[1]{Table~\ref{#1}}
\newcommand{\sect}[1]{Section~\ref{#1}}
\newcommand{\ssect}[1]{Subsection~\ref{#1}}

\newcommand{\appen}[1]{Appendix~\ref{#1}}

\newcommand{\DRb}{\overline{\text{DR}}}

\newcommand{\MSb}{\overline{\text{MS}}}
\newcommand{\Retilde}{\widetilde{\text{Re}}}
\renewcommand{\Re}{\text{Re}\,}
\renewcommand{\Im}{\text{Im}\,}

\newcommand{\ie}{{\it i.e.\;}}

\newcommand{\bc}{\begin{center}}
\newcommand{\ec}{\end{center}}

\newcommand{\gev}{~\text{GeV}}
\newcommand{\mev}{~\text{MeV}}

\newcommand{\ti}{\tilde}

\newcommand{\Ga}{\Gamma}

\newcommand{\ghqqs}{g_{h_i q\bar q}^{S}}
\newcommand{\ghqqa}{g_{h_i q\bar q}^{P}}

\newcommand{\ga}{\gamma}
\newcommand{\calR}{{\cal R}}

\newcommand{\deQED}{\Delta_{\text{QED}}}
\newcommand{\de}{\delta}
\newcommand{\De}{\Delta}
\newcommand{\la}{\lambda}

\newcommand{\tree}{\text{tree}}

\newcommand{\mi}{m_{\ti\chi_{i}}}
\newcommand{\mj}{m_{\ti\chi_{j}}}

\newcommand{\ZH}{{\bf Z}^{H}}

\newcommand{\expeta}{e^{i\varphi_u }}
\newcommand{\expetam}{e^{-i\varphi_u }}
\newcommand{\expetas}{e^{i\varphi_s }}

\newcommand{\slashp}{\slash{\!\!\! p} }

\newcommand{\gsim}{\raisebox{-0.13cm}{~\shortstack{$>$ \\[-0.07cm]
      $\sim$}}~}
\newcommand{\lsim}{\raisebox{-0.13cm}{~\shortstack{$<$ \\[-0.07cm]
      $\sim$}}~}
\newcommand{\s}{\newline \vspace*{-3.5mm}}

\begin{document}
\title{
\vspace*{-3cm}
\phantom{h} \hfill\mbox{\small KA-TP-15-2019}
\\[-1.1cm]
\phantom{h} \hfill\mbox{\small IFIRSE-TH-2019-3}
\\[1cm]
\textbf{One-Loop Corrections to the Two-Body Decays of the Neutral
  Higgs Bosons in the Complex NMSSM\\[4mm]}}

\date{}
\author{Julien Baglio$^{1}$\footnote{E-mail: \texttt{julien.baglio@uni-tuebingen.de}}\;, 
Thi Nhung Dao$^{2}$\footnote{E-mail: \texttt{dtnhung@ifirse.icise.vn}}\;,
Margarete M\"{u}hlleitner$^{3}$\footnote{E-mail: \texttt{margarete.muehlleitner@kit.edu}}
\\[9mm]
{\small\it
$^1$Institute for Theoretical Physics, Eberhard Karls University T\"ubingen, } \\
{\small\it Auf der Morgenstelle 14, D-72076 T\"ubingen, Germany.}\\[3mm]
{\small\it
$^2$Institute For Interdisciplinary Research in Science and Education, ICISE,}\\
{\small\it Khu Vuc 2, Ghenh Rang, Quy Nhon, Binh Dinh, Vietnam.}\\[3mm]
{\small\it $^3$Institute for Theoretical Physics, Karlsruhe Institute of Technology,} \\
{\small\it Wolfgang-Gaede-Str. 1, 76131 Karlsruhe, Germany.}\\[3mm]
}
\maketitle

\begin{abstract}
\noindent
Since no direct signs of new physics have been observed so far
indirect searches in the Higgs sector have become increasingly 
important. With the discovered Higgs boson behaving very 
Standard Model (SM)-like,
however, indirect new physics manifestations are in general 
expected to be small. On the theory side, this makes 
precision predictions for the Higgs parameters and observables
indispensable. In this paper, we provide in the framework of the
CP-violating Next-to-Minimal Supersymmetric extension of the SM
(NMSSM) the complete
next-to-leading order (SUSY-)electroweak corrections to the neutral
Higgs boson decays that are on-shell and non-loop
induced. Together with the also provided SUSY-QCD corrections to
colored final states, they are implemented in the Fortran code {\tt
  NMSSMCALC} which already includes the state-of-the art QCD
corrections. The new code is called {\tt NMSSMCALCEW}. This way we
provide the NMSSM Higgs boson decays and 
branching ratios at presently highest possible precision and thereby
contribute to the endeavor of searching for New Physics at present
and future colliders. 
\end{abstract}
\thispagestyle{empty}
\vfill
\newpage

\section{Introduction}
The discovery of a scalar particle by the LHC experiments
ATLAS~\cite{Aad:2012tfa} and CMS~\cite{Chatrchyan:2012xdj} and the
subsequent investigation of its properties revealed a Higgs boson that
behaves very Standard Model (SM)-like. Also years after its discovery
there are no evidences for new physics from direct searches. In this
situation the precise investigation of the Higgs sector plays an
important role. Indirect effects of physics beyond the SM (BSM) might
show up in the properties of the discovered Higgs boson. With a mass
of 125.09~GeV~\cite{Aad:2015zhl} it does not exclude the possibility
for the Higgs boson of a supersymmetric (SUSY) extension of the SM,
like the minimal (MSSM) or the next-to-minimal (NMSSM)
ones. Supersymmetry certainly belongs to the best motivated and most 
intensively studied BSM extensions, and the NMSSM, with a Higgs sector
consisting of seven Higgs bosons arising after electroweak symmetry
breaking (EWSB) from the two doublet and singlet fields of
the Higgs sector, provides a rich
phenomenology~\cite{Maniatis:2009re,Ellwanger:2009dp}. The
experimental limits strongly restrict possible new physics effects in
the Higgs sector and call for precision in the theory predictions for
the Higgs boson observables. This is also necessary in order to be
able to distinguish between new physics extensions in case of
discovery. \s

In this paper, we concentrate on the NMSSM Higgs boson decays. 
While the (SUSY-)QCD corrections can be taken over from the MSSM case
with the appropriate modifications and a minimum of effort, this is
not the case for the electroweak (EW) corrections. In the recent years
there has been some progress on this subject. In the CP-conserving
NMSSM, members of our group computed the next-to-leading order (NLO) SUSY-EW 
and SUSY-QCD corrections to the decays of CP-odd NMSSM Higgs bosons
into stop pairs and found that the both the EW and the SUSY-QCD
corrections are significant and can be of opposite
sign~\cite{Baglio:2015noa}. The authors of~\cite{Belanger:2016tqb,Belanger:2017rgu}
provided in the framework of the CP-conserving NMSSM its full one-loop
renormalization and the two-body Higgs decays at one-loop order
in the on-shell (OS) renormalization scheme.  A
generic calculation of the two-body partial decays widths at full
one-loop level was provided
in~\cite{Goodsell:2017pdq} in the $\DRb$ scheme. In~\cite{Domingo:2018uim} the full
one-loop corrections for the 
neutral CP-violating NMSSM Higgs bosons were calculated to their
decays into fermions and gauge bosons and
combined with the leading QCD corrections. For the Higgs-to-Higgs
decays, we provided in previous papers the complete
one-loop~\cite{Nhung:2013lpa} and the order ${\cal O} 
  (\alpha_t \alpha_s)$ two-loop~\cite{Muhlleitner:2015dua} corrections
  in the CP-conserving and CP-violating NMSSM, respectively.
\s

In this work, we compute, in the framework of the CP-violating NMSSM,
the complete next-to-leading order (SUSY-)electroweak 
corrections to the neutral NMSSM Higgs boson decays into all
tree-level induced SM final states, {\it i.e.} into fermion and
massive gauge boson pairs, but 
also into non-SM pairs, namely gauge and Higgs boson final states, 
chargino and neutralino pairs, and into squarks. Where applicable we
combine our corrections with the already available (SUSY-)QCD
corrections. We furthermore include the complete
  one-loop corrections to the decays into Higgs boson pairs, {\it cf.}~Refs.~\cite{Nhung:2013lpa,Muhlleitner:2015dua}. 
For the loop-induced decays into gluon and photon pairs
as well as a photon and a $Z$ boson no corrections are provided as
they would be of two-loop order. For the first time, we present the
one-loop corrections to the 
electroweakino, stop and sbottom masses in the context of the CP-violating
NMSSM, by applying both OS and $\DRb$ schemes.
We have implemented our corrections in our original code {\tt
  NMSSMCALC}~\cite{Baglio:2013iia}, 
which calculates, based on a mixed OS-$\DRb$ scheme, the NMSSM Higgs
mass corrections and decays in both 
the CP-conserving and CP-violating case. This way we provide the NMSSM 
Higgs boson decays and branching ratios at presently highest possible
precision including the state-of-the-art (SUSY-)QCD and the computed
(SUSY-)EW corrections. In the EW higher-order corrections we not only include the NLO
vertex corrections but also take into account the proper on-shell
conditions of the external decaying Higgs bosons up to two-loop order
${\cal O}(\alpha_t \alpha_s+\alpha_t^2)$. This is the
order up to which the mass corrections 
for the NMSSM Higgs bosons both in the CP-conserving~\cite{Ender:2011qh} and
CP-violating case~\cite{Graf:2012hh,Muhlleitner:2014vsa,Dao:2019qaz} have been
implemented in {\tt NMSSMCALC}. The new program is
  called {\tt NMSSMCALCEW} can be obtained at the
url: https://www.itp.kit.edu/$\sim$maggie/NMSSMCALCEW/. 
Here also a detailed description of the program and its structure are
given, instructions on how to compile and run it as well as
information on modifications, which is constantly updated. 
A brief description of the code is given in Appendix~\ref{app:nmssmcalcew}. \s

The paper is organized as follows. In Section~\ref{sec:nmssmtree} we
introduce the NMSSM sectors at tree level, that are
  relevant for our computation, and set our
notation before we move on to the NMSSM at one-loop level 
in Section~\ref{sec:oneloop}. We here describe the renormalization of
the Higgs, chargino/neutralino, and squark sectors as well as the loop
corrections to the Higgs boson masses and mixings, to the neutralino
and chargino masses, and finally to the squark masses and their
mixings.
Section \ref{sec:oneloopdecays} is devoted to the detailed
presentation of our calculation of the one-loop corrections to the
neutral non-loop induced Higgs boson decays into on-shell final
states, namely into fermion pairs, massive gauge boson pairs, final
states with one gauge and one Higgs boson, neutralino and chargino
pairs, and squark final states. In Section~\ref{sec:numerical} we
present the numerical analysis of the one-loop corrections to the
Higgs boson branching ratios into SM and SUSY final states, where
we discuss in particular the size of the newly implemented
corrections to both the branching ratios and to the
  electroweakino and third generation 
squark masses. Our conclusions are given in
Section~\ref{sec:conclusions}. Explicit expressions of the
counterterm couplings for the decays of  neutral Higgs bosons into a
squark pair are displayed in Appendix~\ref{appendix1}. 

\section{The NMSSM at Tree Level \label{sec:nmssmtree}}
We are working in the complex NMSSM with a preserved $\mathbb{Z}_3$
symmetry. The Lagrangian of the NMSSM can be divided into the Lagrangian
of the MSSM and the additional part coming from the NMSSM. For
convenience of the reader and to set our notation, we give here the
parts of the Lagrangian that are relevant for our calculations. For
the Higgs sector we need the NMSSM Higgs potential. It is derived
from the NMSSM superpotential $W_{\text{NMSSM}}$, the corresponding
soft  SUSY-breaking terms, and the $D$-term contributions. With the Higgs
doublet superfields $\hat{H}_u$ and $\hat{H}_d$ coupling to the up-
and down-type quark superfields, respectively, and the singlet
superfield $\hat{S}$, we have for the NMSSM superpotential
\begin{align} 
W_{\text{NMSSM}} = W_{\text{MSSM}} - \epsilon_{ab} \lambda \hat{S} \hat{H}_d^a
\hat{H}_u^b + \frac{1}{3} \kappa \hat{S}^3~,
\end{align}
where $a,b=1,2$ are the indices of the fundamental $SU(2)_L$
representation and $\epsilon_{ab}$ is the totally antisymmetric tensor
with $\epsilon_{12}=\epsilon^{12}=1$ . The MSSM part reads
\begin{align}
W_{\text{MSSM}} = - \epsilon_{ab} \bigl( y_u \hat{H}_u^a \hat{Q}^b \hat{U}^c -
y_d \hat{H}_d^a \hat{Q}^b \hat{D}^c - y_e \hat{H}_d^a \hat{L}^b
\hat{E}^c\bigr)
\; ,
\end{align}
in terms of the left-handed quark and lepton superfield doublets
$\hat{Q}$ and $\hat{L}$ and the right-handed up-type, down-type, and
electron-type superfield singlets $\hat{U}$, $\hat{D}$, and $\hat{E}$,
respectively. Charge conjugation is denoted by the superscript $c$,
and color and generation indices have been omitted. The NMSSM
superpotential contains the coupling $\kappa$ of the self-interaction
of the new singlet superfield and the coupling $\lambda$ for the
$\hat{S}$ interaction with the two Higgs doublet
superfields. Both couplings are complex. The quark and
lepton Yukawa couplings $y_d,$ $y_u$, and $y_e$ are in general
complex. However, in case of no generation mixing, as assumed in this paper,
the phases of the Yukawa couplings can be absorbed through a
redefinition of the quark fields, so that the phases can be
chosen arbitrarily without changing the physical
meaning~\cite{Kobayashi:1973fv}. 
The soft SUSY-breaking NMSSM Lagrangian in terms of the component
fields $H_u, H_d$ and $S$ reads
\begin{align}
\mathcal L_{\text{NMSSM}}^{\text{soft}} = \mathcal
L_{\text{MSSM}}^{\text{soft}} - m_S^2 |S|^2 
   + 
 (\epsilon_{ab} A_{\lambda} \lambda S H_d^a H_u^b - \frac{1}{3}
   A_{\kappa}\kappa S^3 + h.c.%
) \; .
\end{align}
It contains two more complex parameters specific to the NMSSM, the
soft SUSY-breaking trilinear couplings $A_\lambda$ and 
$A_\kappa$. The soft SUSY-breaking MSSM contribution can be cast into
the form
\begin{align}\nonumber
\mathcal L_{\text{MSSM}}^{\text{soft}} &= - m_{H_d}^2 |H_d|^2 - m_{H_u}^2 
|H_u|^2  
- m_{\tilde{Q}}^2 |\tilde{Q}|^2  - m_{\tilde{u}_R}^2 |\tilde{u}_R|^2  - m_{\tilde{d}_R}^2 |\tilde{d}_R|^2 
- m_{\tilde{L}}^2 |\tilde{L}|^2 - m_{\tilde{e}_R}^2 |\tilde{e}_R|^2 \\& \quad \nonumber
+ \epsilon_{ab} ( y_u A_u H_u^a \tilde{Q}^b \tilde{u}^*_R 
                -  y_d A_d H_d^a \tilde{Q}^b \tilde{d}^*_R 
                -  y_e A_e H_d^a \tilde{Q}^b \tilde{e}^*_R + h.c.)\\& \quad
- \frac{1}{2} (M_1  \tilde{B} \tilde{B} +   M_2 \tilde{W}_i \tilde{W}_i + M_3
\tilde{G} \tilde{G} + h.c) \; ,
\end{align}
where the SM-type and SUSY fields corresponding to a superfield
(denoted with a hat) are represented by a letter without and with a
tilde, respectively.  The indices $\tilde{Q}$
  ($\tilde{L}$) of the soft SUSY-breaking masses denote, exemplary
for the first generation, the left-handed quark (lepton) doublet
component fields of the corresponding quark and lepton superfields,
and $\tilde{u}_R,\tilde{d}_R,\tilde{e}_R$ the right-handed component
fields for the up-type and down-type quarks, and charged leptons, respectively,.
The trilinear couplings $A_u$, $A_d$ and $A_e$ of the up-type and
down-type quarks and charged leptons are in general complex,
whereas the soft SUSY-breaking mass terms $m_x^2$
($x=S,H_u,H_d,\tilde{Q},\tilde{u}_R,\tilde{d}_R,\tilde{L},\tilde{e}_R$)
are real. The soft SUSY-breaking mass 
parameters of the gauginos, $M_1$, $M_2$, $M_3$, for the bino, the winos, and
the gluinos, $\tilde{B}$, $\tilde{W}_i$ ($i=1,2,3$), and $\tilde{G}$,
corresponding to the weak hypercharge $U(1)$, the weak isospin $SU(2)$,
and the color $SU(3)$ symmetry, are in general complex. The
$R$-symmetry  can  be  exploited  to  choose  either $M_1$ or $M_2$ to  be real. 
In this paper we keep both $M_1$ and $M_2$ complex. \s 

Expanding the scalar Higgs fields about their vacuum expectation
values (VEVs) $v_u$, $v_d$, and $v_s$, two further phases, $\varphi_u$ and
$\varphi_s$, are introduced which describe the phase differences between the VEVs,
 \begin{align}
H_d = \begin{pmatrix}\frac{1}{\sqrt{2}}(v_d + h_d + i a_d) \\ 
                      h_d^- \end{pmatrix}~, \;\;\; 
H_u = e^{i \varphi_u} \begin{pmatrix} h_u^+ \\
            \frac{1}{\sqrt{2}}(v_u + h_u + i a_u) \end{pmatrix}~, \;\;\; 
S = \frac{e^{i \varphi_s}}{\sqrt{2}} (v_s + h_s + i a_s)~.
\label{Higgsdecomp}
\end{align}
For vanishing phases, the fields $h_i$ and  $a_i$ with 
$i = d, u, s$  correspond to the CP-even and CP-odd part of the
neutral entries of $H_d$, $H_u$ and $S$. The charged
components are denoted 
by $h_{d,u}^\pm$. In this paper, we set the phases of
the Yukawa couplings to zero. We furthermore re-phase the left- and
right-handed up-quark fields as $u_L \to e^{-i \varphi_u} u_L$ and $u_R \to e^{i\varphi_u}
u_R$, so that the quark and lepton mass terms yield real masses. \s

After electroweak symmetry breaking (EWSB) the six Higgs interaction states
mix and in the basis $\phi=(h_d, h_u, h_s, a_d, a_u, a_s)$ the mass
term is given by 
\be 
{\cal L}^m_{\text{neutral}}= \fr 12 \phi^T M_{\phi\phi}\phi \;.
\ee
The mass matrix $M_{\phi\phi}$ is obtained from the second derivative
of the Higgs potential with respect to the Higgs fields in the vacuum. 
The explicit expression of the mass matrix $M_{\phi\phi}$ can be found
in \bib{Graf:2012hh}.  The transformation into mass eigenstates
at tree level can be performed with orthogonal matrices $\mathcal R,
\mathcal R^{G}$,
\bea
{\text{diag}}(m_{h_1}^2, m_{h_2}^2,m_{h_3}^2,m_{h_4}^2,
m_{h_5}^2,0)&=& \mathcal R\mathcal R^{G} M \braket{\mathcal R\mathcal
  R^{G}}^T,\\ 
(h_1,h_2,h_3,h_4,h_5,G)^T& = & \mathcal R\mathcal R^{G} (h_d, h_u,
h_s, a_d, a_u, a_s)^T \crn
&=& \calR  (h_d, h_u,
h_s, a, a_s, G)^T \;, \label{eq:rotgaugemasstree}
\eea
where the matrix $\mathcal R^{G}$ is used first to single out the
Goldstone boson $G$. The tree-level Higgs mass eigenstates are denoted
by the small letter $h$, and their masses are ordered as $m_{h_1} \leq
... \leq m_{h_5}$. \s

The mass term for the charged components of the Higgs doublets in 
\be  
{\cal L}^m_{\text{charged}} = \bpmatrix h_d^+, h_u^+ \epmatrix
M_{h^+h^+} \bpmatrix h_d^-\\ h_u^- \epmatrix,
\ee
is given by 
\begin{align} 
M_{h^+h^+}= \fr12 \bpmatrix \tbeta & 1 \\
                                1 & 1/\tbeta \epmatrix \bigg[& M_W^2 s_{2\beta} +
\fr{\abs{\lambda}v_s}{\cos( \varphi_\lambda+\varphi_u+ \varphi_s) }
\braket{\sqrt{2} \,\Re A_\lambda+\abs{\kappa}v_s \cos(\varphi_\kappa+3\varphi_s) } \crn  
& -\fr{2 |\lambda|^2 M_W^2 s_{\theta_W}^2}{e^2}s_{2\beta} \bigg] \;, 
\end{align}
where $M_W$ is the mass of the $W$ boson, $\theta_W$ the
electroweak mixing angle, $e$ the electric charge and
$\varphi_\lambda,$ $\varphi_\kappa$ the complex phases of $\lambda$ and
$\kappa$, respectively. The angle $\beta$ is defined as
\beq
\tan \beta = v_u/v_d \;.
\eeq
Here and in the following we use the short hand notation $c_x = \cos
x$, $s_x = \sin x$ and $t_x = \tan x$. Diagonalizing 
this mass matrix by a rotation matrix with the angle $\beta_c$, for
which at Born level $\beta_c=\beta$, one
obtains the charged Higgs boson mass as
\be 
M_{H^\pm}^2 = M_W^2 +\fr{\abs{\lambda}v_s}{s_{2\beta}\cos(
  \varphi_\lambda+\varphi_u+ \varphi_s) } \braket{\sqrt{2} \Re
  A_\lambda+\abs{\kappa}v_s \cos( \varphi_\kappa+ 3\varphi_s)
}-\fr{2 |\lambda|^2 M_W^2 s_{\theta_W}^2}{e^2} .
\ee  
The charged Goldstone boson $G^\pm$, on the other hand, is massless. \s

The fermionic superpartners of the neutral Higgs bosons,
$\tilde{H}_d^0$, $\tilde{H}_u^0$, $\tilde{S}$, and of the neutral gauge
bosons, $\tilde{B}$, $\tilde{W}_3$, mix, and in the Weyl spinor basis
 $\psi^0 =  (\tilde{B},\tilde{W}_3, \tilde{H}^0_d,\tilde{H}^0_u,  \tilde{S})^T$ 
the neutralino mass matrix $M_N$ is given by
\begin{align}
&M_N = \nonumber \\& \begin{pmatrix} 
M_1               & 0        & - c_\beta M_Z s_{\theta_W} &   
                                 M_Z s_\beta s_{\theta_W} e^{-i \varphi_u}
               & 0\\
0                 & M_2      &    c_\beta M_W    & - M_W s_\beta e^{-i \varphi_u} 
               & 0\\
- c_\beta M_Z s_{\theta_W} & c_\beta M_W & 0                &
                 - \lambda \frac{v_s}{\sqrt{2}} e^{i \varphi_s}   & 
                - \frac{\sqrt{2} M_W s_\beta s_{\theta_W} \lambda e^{i\varphi_u}}{e}
  \\
 M_Z s_\beta s_{\theta_W} e^{-i \varphi_u} &  - M_W s_\beta e^{-i \varphi_u} & 
                            - \lambda \frac{v_s}{\sqrt{2}}e^{i \varphi_s} & 0 &
                          -  \frac{\sqrt{2} M_W c_\beta s_{\theta_W} \lambda}{e} 
   \\
0   & 0     & -  \frac{\sqrt{2}M_W s_\beta s_{\theta_W}\lambda e^{i \varphi_u}}{e}  & 
                            - \frac{\sqrt{2} M_W c_\beta s_{\theta_W}\lambda}{e} &
                             \sqrt{2} \kappa v_s e^{i \varphi_s} 
\end{pmatrix} \label{eq:neuMass}       
\end{align}
after EWSB, where  $M_Z$ is the $Z$ boson mass.   
The symmetric neutralino mass matrix can be
diagonalized by a $5 \times 5$ matrix $N$, yielding   
$\text{diag}(m_{\tilde{\chi}^0_1},
m_{\tilde{\chi}^0_2},m_{\tilde{\chi}^0_3},m_{\tilde{\chi}^0_4},
m_{\tilde{\chi}^0_5}) = N^* M_N N^\dagger$,  
where the absolute mass values are ordered as $|m_{\tilde{\chi}^0_1}|\leq ... 
\leq |m_{\tilde{\chi}^0_5}|$. The neutralino mass eigenstates $\tilde{\chi}^0_i$, expressed 
as Majorana spinors, can then be obtained by 
\begin{align}\label{eq:neuspinor}
\tilde{\chi}_i^0 = \left( \begin{array}{c} \chi_i^0\smallskip 
                                            \\ \overline{\chi_i^0} \end{array}
  \right) \quad\ \text{with}\quad\ \chi^0_i = N_{ij} \psi^0_j,\quad\  i,\,j 
  = 1,\dots,5~,
\end{align}
where, in terms of the Pauli matrix $\sigma_2$,
\beq
\overline{\chi_i^0} = i \sigma_2 \chi_i^{0*} \;.
\eeq
The fermionic superpartners of the charged Higgs and gauge bosons are
given in terms of the Weyl spinors $\tilde{H}_d^\pm$,
$\tilde{H}_u^\pm$, $\tilde{W}^-$ and $\tilde{W}^+$. With
\beq
\psi^-_R = \begin{pmatrix} \tilde{W}^- \\ \tilde{H}_d^- \end{pmatrix}
\quad \mbox{and} \quad
\psi^+_L = \begin{pmatrix} \tilde{W}^+ \\ \tilde{H}_u^+ \end{pmatrix}
\eeq
the mass term for these spinors is of the form 
\be 
{\cal L}= (\psi^-_R)^T M_C \psi^+_L + h.c.  \;,
\ee
where
\begin{align}
M_C = \begin{pmatrix} M_2 & \sqrt{2}  s_\beta M_W e^{-i \varphi_u}\\
   \sqrt{2} c_\beta M_W & \lambda \frac{v_s}{\sqrt{2}} e^{i \varphi_s}
    \end{pmatrix}~. \label{eq:chaMass} 
\end{align}
The chargino mass matrix $M_C$ can be diagonalized with the help of two unitary 
$2 \times 2$ matrices, $U$ and $V$, yielding
\beq
\text{diag}(m_{\tilde{\chi}^\pm_1}, m_{\tilde{\chi}^\pm_2}) = U^* M_C
V^\dagger \;,
\eeq 
with $m_{\tilde{\chi}^\pm_1} \leq  m_{\tilde{\chi}^\pm_2}$. The
left-handed and the right-handed part of the mass eigenstates are 
\begin{align}
\tilde{\chi}^+_L = V \psi^+_L \quad \mbox{and} \quad \tilde{\chi}^-_R = U \psi^-_R~,
\end{align}
respectively, with the mass eigenstates ($i=1,2$)
\beq
\tilde{\chi}^+_i = \left( \begin{array}{c} 
\tilde{\chi}^+_{L_i}\smallskip \\ \overline{\tilde{\chi}^-_{R_i}} \end{array} \right)
\eeq
written as Dirac spinors. In summary, the bilinear terms in the chargino and
neutralino mass eigenstates are given by  
\begin{align}
{\cal L}=& \overline{\tilde{\chi}^+_i} \slashp P_L\tilde{\chi}^+_i+\overline{\tilde{\chi}^+_i} \slashp P_R\tilde{\chi}^+_i - \overline{\tilde{\chi}^+_i}  \sbraket{U^* M_C
V^\dagger }_{ij}P_L \tilde{\chi}^+_j - \overline{\tilde{\chi}^+_i}  \sbraket{V M_C^\dagger
U^T }_{ij}P_R \tilde{\chi}^+_j\crn
& +
\overline{\tilde{\chi}^0_k} \slashp P_L\tilde{\chi}^0_k+\overline{\tilde{\chi}^0_k} \slashp P_R\tilde{\chi}^0_k - \overline{\tilde{\chi}^0_k}  \sbraket{N^* M_N
N^\dagger }_{kl}P_L \tilde{\chi}^0_l -
  \overline{\tilde{\chi}^0_k}  \sbraket{N M_N^\dagger N^T
  }_{kl}P_R \tilde{\chi}^0_l \;,
\end{align}  
where the left- and right-handed projectors are defined as $P_{L/R} =
\braket{1\mp \ga_5}/2$ and $i,j=1,2$ and $k,l=1,\ldots,5$. \s

The scalar partners of the left- and right-handed quarks are denoted by
$\ti q_L$ and $\ti q_R$, respectively. The mixing matrix for the top squark
is given  by 
\be 
M_{\ti t} = \bpmatrix m_{\ti Q_3}^2 + m_t^2 + M_Z^2c_{ 2\beta}(\fr12 -
\fr23 s^2_{\theta_W}) & 
m_t\braket{A_t^* \expetam - \mueff/\tbeta} \\
m_t\braket{A_t \expeta - \mueff^*/\tbeta} & m_t^2 + m_{\ti t_R}^2
+\fr23 M_Z^2 c_{ 2\beta} s^2_{\theta_W}   \epmatrix,
\label{eq:mstopmat}
\ee 
while the bottom squark matrix reads 
\be 
M_{\ti b} = \bpmatrix m_{\ti Q_3}^2 + m_b^2 + M_Z^2c_{ 2\beta}(-\fr12
+ \fr13 s^2_{\theta_W}) & 
m_b\braket{A_b^*  - \expeta\mueff \tbeta} \\
m_b\braket{A_b  -\expetam \mueff^* \tbeta} & m_b^2 + m_{\ti b_R}^2
-\fr13 M_Z^2 c_{ 2\beta}s^2_{\theta_W}   \epmatrix,
\label{eq:msbotmat}
\ee 
where
\be 
\mueff = \fr{\lambda v_s \expetas}{\sqrt{2}}.
\ee
The mass eigenstates are obtained by diagonalizing these squark
matrices with the unitary transformations
\be   
\diag(m_{\ti q_1}^2,m_{\ti q_2}^2 ) = U^{\ti q}  M_{\ti q}  U^{\ti
  q\dagger}, \quad \bpmatrix 
\ti q_1\\ \ti q_2\epmatrix =  U^{\ti q} \bpmatrix
\ti q_L\\ \ti q_R\epmatrix, \quad q=t,b,
\ee
with the usual convention $m_{\ti q_1}\le m_{\ti q_2}$.

\section{The NMSSM at One-Loop Level \label{sec:oneloop}}
\subsection{Renormalization}
\subsubsection{The Higgs Sector}
For the Higgs sector  we follow the mixed on-shell OS-$\DRb$
renormalization scheme  
described and applied in
  Refs.~\cite{Ender:2011qh,Graf:2012hh,Muhlleitner:2014vsa,Dao:2019qaz}. We
  do not repeat all details here but quote the most important formulae. There are
eighteen parameters entering the Higgs sector at tree level,  
\begin{align}
& m_{H_d}^2, m_{H_u}^2, m_S^2,M_W^2, M_Z^2, e, 
  \tan\beta, v_s, \varphi_s, \varphi_u, 
 \abs\lambda, \varphi_\lambda, \abs\kappa,
\varphi_\kappa, \Re A_{\lambda},\Im A_\lambda, \Re A_{\kappa},
\Im A_\kappa~. \label{eq:orgparset}
\end{align}
Note that for the sake of convenience we decompose the complex
trilinear couplings $A_\lambda$ and $A_\kappa$ into a real and
imaginary part in contrast to \bib{Graf:2012hh} where the 
absolute values and complex phases were used. It was found in
\bib{Graf:2012hh} that the four complex phases $\varphi_s,$ 
$\varphi_u,$ $\varphi_\lambda$ and $\varphi_\kappa$ do not need to be
renormalized. We verified this statement and will discard them in our
renormalization procedure. \s

In our introduction of the NMSSM Higgs sector in
  Sec.~\ref{sec:nmssmtree} we have already replaced the $U(1)$ and
  $SU(2)$ gauge couplings $g'$ and $g$ and the VEV $v$ by the physical
  observables $M_W$, $M_Z$ and $e$. It is convenient to further
  convert, where possible, the input parameters of the set
  \eqref{eq:orgparset} to parameters that can be interpreted more
easily in terms of physical quantities. Thus we trade
the three soft SUSY-breaking mass parameters
  $m_{H_d}^2, m_{H_u}^2, m_S^2$ as well as $\Im
  A_\lambda$ and $\Im A_\kappa$ for the tadpole parameters $t_\phi$
  ($\phi=h_d, h_u, h_s, a_d, a_s$). These coefficients of the terms
of the Higgs potential $V_{\text{Higgs}}$ are linear in the 
Higgs boson fields and have to vanish, in order to ensure the minimum at
non-vanishing VEVs $v_u,$ $v_d$, $v_s$,
\beq
t_\phi \equiv \left.\frac{\partial V_{\text{Higgs}}}{\partial \phi}
\right|_{\text{Min.}} \stackrel{!}{=} 0 \;.
\eeq
It is debatable whether the tadpole parameters can be called physical
quantities, but certainly their introduction is motivated by physical
interpretation. In the same way, in a slight abuse of the language, we
will call the renormalization conditions for the tadpole parameters
on-shell. With the new set of input parameters, we allow for two
possible renormalization schemes in our Higgs mass calculation. The
difference between the two schemes relates to the treatment of the
charged Higgs mass. In the first scheme the charged Higgs mass is an OS
input parameter, 
\begin{align}
\underbrace{t_{h_d}, t_{h_u}, t_{h_s}, t_{a_d}, t_{a_s}, M_{H^\pm}^2,
  M_W^2, M_Z^2, e}_{\text{on-shell}}, \underbrace{ \tan \beta,   v_s, \abs\lambda, 
\abs\kappa, \Re A_{\kappa}}_{\overline{\text{DR}}} \;,
\label{eq:defparset1}
\end{align}
while in the second scheme $\Re A_{\lambda}$ is an input
  parameter renormalized as a $\DRb$ parameter and
the charged Higgs mass is a derived quantity, 
\begin{align}
\underbrace{t_{h_d}, t_{h_u}, t_{h_s}, t_{a_d}, t_{a_s}, M_W^2, M_Z^2, e}_{\text{on-shell}}, 
\underbrace{ \tan \beta,   v_s,  |\lambda|,
   |\kappa|,  \Re A_{\lambda}, \Re A_{\kappa} }_{\overline{\text{DR}}}~.
\label{eq:defparset2}
\end{align}
For the definition of the one-loop OS counterterms we refer the
reader to \bib{Graf:2012hh}. The counterterms of the $\DRb$ parameters do
not contribute to the final physical results but they are
kept to ensure UV
finiteness. We define these counterterms solely in the Higgs sector by
requiring that all renormalized self-energies of the Higgs bosons be
finite. This is different with respect to the definition in \bib{Graf:2012hh}
where the chargino and neutralino sectors were used. The numerical
results between the two definitions are identical, however. We renormalize the
Higgs fields in the $\DRb$ scheme as described in \bib{Graf:2012hh} 
at one-loop level, and in Refs.~\cite{Muhlleitner:2014vsa,Dao:2019qaz}
at two-loop order ${\cal O}(\alpha_t \alpha_s)$ and ${\cal O}(\alpha_t^2)$, respectively. 

\subsubsection{The Chargino and Neutralino Sector}
\label{sssect:NCrenor}
The chargino and neutralino sectors are described by fourteen real
parameters: $M_W, M_Z,$ $\tan \beta,   v_s, \varphi_s,
\varphi_u,\abs\lambda, $ $\varphi_\lambda, \abs\kappa, \varphi_\kappa,
|M_1|,\varphi_{M_1}, |M_2|,\varphi_{M_2}$.  Since the first ten of
these already appear in the Higgs sector, there remains to define the
renormalization conditions for the four parameters $|
M_1|,\varphi_{M_1}, |M_2|,\varphi_{M_2}$.\footnote{In the case of
  the MSSM one needs to renormalize three complex parameters $M_1,M_2,
  \mu$.} There are no physical renormalization conditions to
fix the counterterms of the phases $\varphi_{M_1}, \varphi_{M_2}$. It has
been found in \bib{Graf:2012hh} that the complex phases of
$M_1$ and $M_2$ do not need to be renormalized. We verified this statement in
our computation.\footnote{The same holds true in the complex
  MSSM~\cite{Bharucha:2012nx}. The three complex phases of $M_1$,
  $M_2$ and $\mu$ do not need to be renormalized in order to render
  all Green functions finite.} 
In addition, we have to renormalize the chargino and neutralino fields
in order to obtain finite self-energies. In the literature there exist two
descriptions for the introduction of wave function renormalization
(WFR) constants. In the Espriu-Manzano-Talavera (EMT) description  
two independent renormalization constants were introduced for incoming
and outgoing fermions~\cite{Espriu:2002xv,Fowler:2009ay,Bharucha:2012nx}. Thanks to more
degrees of freedom one can keep contributions arising from absorptive
parts of the loop integrals and eliminate completely the mixing self-energies 
thereby fulfilling the standard OS conditions. However, the hermicity
of the renormalized Lagrangian is not satisfied anymore.  In the
Denner description~\cite{Denner:1991kt}, one WFR constant was used
instead. It preserves the hermicity constraint, but the absorptive
part of the loop integral must be eliminated. We want to investigate
the effect of the absorptive part and therefore apply both
descriptions. In the following we will derive the formulae in the EMT
method. From these formulae, one can easily obtain the ones in the Denner description. 
The bare parameters and fields are replaced by the
renormalized ones and the corresponding counterterms as
\bea 
M_1&\to& M_1 + \de \abs{M_1} e^{i\varphi_{M_1}},\\
M_2&\to& M_2 + \de \abs{M_2} e^{i\varphi_{M_2}},\\
P_L\tilde{\chi}^+_i&\to & \braket{ 1+ \fr12 \de
  Z_L^{\tilde{\chi}^+}}_{ij} P_L
\tilde{\chi}^+_j,\quad 
\overline{\tilde{\chi}^+_i}P_L\to
\overline{\tilde{\chi}^+_j}\braket{1+ \fr12 \de \bar
  Z_R^{\ti\chi^+}}_{ji} P_L,\\ 
P_R\tilde{\chi}^+_i&\to & \braket{ 1+ \fr12 \de
  Z_R^{\tilde{\chi}^+}}_{ij}P_R\tilde{\chi}^+_j, \quad 
\overline{\tilde{\chi}^+_i}P_R\to
\overline{\tilde{\chi}^+_j}\braket{1+ \fr12 \de \bar
  Z_L^{\ti\chi^+}}_{ji} P_R, \\ 
P_L\tilde{\chi}^0_k&\to & \braket{1+ \fr12 \de Z^{\tilde{\chi}^0}_L}_{kl}P_L\tilde{\chi}^0_l,\quad
\overline{\tilde{\chi}^0_{k}}P_L\to 
\overline{\tilde{\chi}^0_{l}}\braket{1+ \fr12 \de \bar
  Z_R^{\ti\chi^0}}_{lk} P_L, \\ 
P_R\tilde{\chi}^0_k&\to & \braket{1+ \fr12 \de Z^{\tilde{\chi}^0}_R}_{kl}P_R\tilde{\chi}^0_l,\quad
\overline{\tilde{\chi}^0_k}P_R\to
\overline{\tilde{\chi}^0_l}\braket{1+ \fr12 \de \bar
  Z_L^{\ti\chi^0}}_{lk} P_R \;, 
\eea
where $i,j=1,2$ and $k,l=1,\ldots,5$. Since the neutralinos are
Majorana fermions we have 
\beq
\de Z^{\ti\chi^0}_R= \braket{\de \bar Z^{\ti\chi^0}_L}^{T}\; \quad
\mbox{and} \quad \de Z^{\ti\chi^0}_L= \braket{\de \bar Z^{\ti\chi^0}_R}^{T} \;.
\eeq 
Note that we do not need to renormalize the rotation matrices $U,$ $V$
and $N$ because their counterterms are redundant. They always appear in
combination with WFR constants. One can therefore always redefine the WFR
constants to absorb the counterterms of the rotation
  matrices, as shown in \bib{Fritzsche:2002bi}. In general, the
renormalized self-energies $\hat{\Sigma}$ of the fermions can be cast
into the following form, {\it cf.}~\bib{Denner:1991kt}, 
\bea 
\hat{\Sigma}_{ij}(p)= \slashp \hat{\Sigma}^L_{ij} (p^2) { P}_L + 
\slashp \hat{\Sigma}^R_{ij} (p^2) { P}_R + 
 \hat{\Sigma}^{Ls}_{ij} (p^2) { P}_L + \hat{\Sigma}^{Rs}_{ij} (p^2) {
   P}_R \; , \label{eq:sigmadecomposition}
\eea
with 
\bea
\hat{\Sigma}^L_{ij} (p^2)&=& \Sigma^L_{ij} (p^2) + \fr 12\braket{ \de
  Z_L + \de \bar Z_L}_{ij}, \label{eq:sigmaL}\\ 
\hat{\Sigma}^R_{ij} (p^2)&=& \Sigma^R_{ij} (p^2) + \fr 12\braket{ \de
  Z_R + \de \bar Z_R}_{ij},\label{eq:sigmaR}\\ 
\hat{\Sigma}^{Ls}_{ij} (p^2)&=& \Sigma^{Ls}_{ij} (p^2) - \braket{ \fr
  12 (\de \bar Z_{R})_{ij} m_{\tilde{\chi}_j}  +\fr12
  (\de Z_L)_{ij}m_{\tilde{\chi}_i} +(\de
  M^\tree_{\tilde{\chi}})_{ij}},\label{eq:sigmaLs}\\ 
\hat{\Sigma}^{Rs}_{ij} (p^2)&=& \Sigma^{Rs}_{ij} (p^2) - \braket{ \fr 12 (\de \bar Z_L)_{ij} m_{\tilde{\chi}_j}  +\fr12
  (\de Z_R)_{ij}m_{\tilde{\chi}_i} +\braket{\de
    M^\tree_{\tilde{\chi}}}^\dagger_{ij}},\label{eq:sigmaRs} 
\eea
where 
$\Sigma_{ij}(p)$ denotes the unrenormalized self-energy of the
transition $\ti\chi^+_i\to \ti\chi^+_j, ~ i,j=1,2$, for the charginos
and  $\ti\chi^0_i\to \tilde{\chi}^0_j, ~i,j=1,\cdots,5$, for the
neutralinos. For the charginos, the tree-level mass matrix
$M^\tree_{\tilde{\chi}}$ and its counterterm $\delta M^\tree_{\tilde{\chi}}$ are given by
\be 
M^\tree_{\tilde{\chi}}=  U^* M_C
V^\dagger, \quad \de M^\tree_{\tilde{\chi}}=  U^* \de M_C
V^\dagger 
\ee
and for the neutralinos by 
  \be  
M^\tree_{\tilde{\chi}}=N^* M_N N^\dagger, \quad 
\de M^\tree_{\tilde{\chi}}=N^* \de M_N N^\dagger \;. 
\ee
In the following, we will discuss the OS conditions for the general
fermion fields $\ti\chi_i$ having the tree-level masses
$m_{\ti\chi_i}$.  The renormalized fermion propagator matrix is
given by 
\be 
\hat{S}(p)= - \hat{\Ga}(p)^{-1} \;,
\ee 
where the renormalized one-particle irreducible (1PI) two-point
functions $\hat{\Gamma}$ are related to the
renormalized self-energies as 
\be 
\hat\Ga_{ij}(p)= i\de_{ij}\braket{\slashp -m_{\ti\chi_i}} + i
\hat\Sigma_{ij}(p) \;.
\ee
The propagator matrix has complex poles at ${\cal M}^2_{\ti\chi_i}=
M_{\ti\chi_i}^2 -i M_{\ti\chi_i} \Ga_{\ti\chi_i}$, where
$\Ga_{\ti\chi_i}$ denotes the decay widths. 
In the OS scheme we require that 
the tree-level masses are equal to the physical masses, the mixing
terms are vanishing at the poles and that the residues of the
propagators are unity\footnote{In the Denner description, $\hat\Sigma_{ii}$ in  \eqref{eq:OS1}, $\hat\Sigma_{ij}$ in
  \eqref{eq:OSMixingkill} and $\hat\Gamma_{ii}$ in \eqref{eq:OS3} are replaced by
 $\Retilde\hat\Sigma_{ii}$, $\Retilde\hat\Sigma_{ij}$ and $\Retilde\hat\Gamma_{ii}$, respectively. $\Retilde$ means that one takes only the
  real part of the loop integrals but leaves the couplings
  unaffected.},
\begin{align}
& \Re\hat\Sigma_{ii}(p) \ti\chi_i\bigg|_{p^2=m_{\ti\chi_i}^2}=0 \;,& \label{eq:OS1} \\
 &\hat\Sigma_{ij}(p) \ti\chi_{j}\bigg|_{p^2=m_{\ti\chi_j}^2}=0 \; ,
&\overline{\ti\chi_i} \hat\Sigma_{ij}(p)\bigg|_{p^2=m_{\ti\chi_i}^2}=0
  \; , \quad i\neq j\label{eq:OSMixingkill}\\
&\lim_{p^2\to m_{\ti\chi_i } }\fr{1}{\slashp - m_{\ti\chi_i }
  }\Re\hat\Gamma_{ii}\ti\chi_{i}= i\ti \chi_i \; ,
& \lim_{p^2\to m_{\ti\chi_i}}
  \overline{\ti\chi_{i}}\Re\hat\Gamma_{ii} \fr{1}{\slashp -
  m_{\ti\chi_i}}= i\overline{\ti\chi_{i}} \; .\label{eq:OS3}
\end{align}
In addition, we require the chiral structure to vanish in the OS
limit\footnote{If we use the Denner OS conditions in
  \bib{Denner:1991kt} then these relations are automatically
  satisfied for the real parts only.},
\be 
\hat\Sigma^L_{ii}(m_{\ti\chi_i}^2) =
\hat\Sigma^R_{ii}(m_{\ti\chi_i}^2) \;,\quad  
\hat\Sigma^{Ls}_{ii}(m_{\ti\chi_i}^2)=
\hat\Sigma^{Rs}_{ii}(m_{\ti\chi_i}^2) \;.
\ee 
Applying the decomposition in \eqref{eq:sigmadecomposition} and
the tree-level relations
\be 
(\slashp -m_{\ti\chi_i} )\ti\chi_i=0 \;, \quad
\overline{\ti\chi_i}(\slashp + m_{\ti\chi_i} )=0 \;,
\ee
one obtains the mass counterterms $\delta m_{\tilde{\chi}_i} =
\mbox{Re} \, (\delta M_{\tilde{\chi}}^{\text{tree}})_{ii}$, with
\bea 
\hspace*{-0.4cm}
\Re (\delta M^\tree_{\tilde{\chi}})_{ii}  &=
\fr12\braket{m_{\ti\chi_i} \Re\Sigma^L_{ii} 
  (m_{\ti\chi_i}^2)+m_{\ti\chi_i} \Re\Sigma^R_{ii} (m_{\ti\chi_i}^2) +
  \Re\Sigma^{Ls}_{ii} (m_{\ti\chi_i}^2)+ \Re\Sigma^{Rs}_{ii}
  (m_{\ti\chi_i}^2) }, \label{eq:massCT1}
\eea
the off-diagonal wave function renormalization constants,
$\delta Z_{L/R,ij}, \delta \bar{Z}_{L/R,ij}$,\footnote{In the Denner description
  $\Sigma_{ij}^{L/R/Ls/Rs}$ are replaced by
  $\Retilde\Sigma_{ij}^{L/R/Ls/Rs}$.}$^,$\footnote{Note that the $\delta
  \bar{Z}$ are obtained from $\delta Z$ by interchanging the
  $m_{\tilde{\chi}}^2$ of the electroweakinos, but not the $m_{\tilde{\chi}}$.} 
\begin{align}
\delta Z_{L,ij} &= \frac{2}{\mi^2-\mj^2} \bigg[\mj^2
 \Sigma_{ij}^L (m_{\tilde{\chi}_j}^2)+\mi \mj \Sigma_{ij}^R (m_{\tilde{\chi}_j^2})
+\mi \Sigma^{Ls}_{ij} (m_{\tilde{\chi}_j}^2)\crn&+\mj
\Sigma^{Rs}_{ij} (m_{\tilde{\chi}_j}^2)  -\mj (\delta
M^{\text{tree}}_{\tilde{\chi}})^\dagger_{ij}
-\mi (\delta M^{\text{tree}}_{\tilde{\chi}})_{ij}
   \bigg]\,, \\
\delta Z_{R,ij}&=\frac{2}{\mi^2-\mj^2} \bigg[\mj^2 \Sigma^R_{ij}
                 (m_{\tilde{\chi}_j}^2) +\mi \mj \Sigma^L_{ij} (m_{\tilde{\chi}_j}^2) +\mi
   \Sigma^{Rs}_{ij} (m_{\tilde{\chi}_j}^2) \crn &
+\mj  \Sigma^{Ls}_{ij} (m_{\tilde{\chi}_j}^2)  - \mi (\delta
M^{\text{tree}}_{\tilde{\chi}})^\dagger_{ij}
  -\mj (\delta M^{\text{tree}}_{\tilde{\chi}})_{ij}  \bigg]\,, \\
\delta \bar{Z}_{L,ij}&=\delta Z_{L,ij}(\mj^2 \leftrightarrow \mi^2)\,,
  \\
\delta \bar{Z}_{R,ij}&=\delta Z_{R,ij}(\mj^2 \leftrightarrow \mi^2)\,,
\end{align}
and the diagonal wave function renormalization constants\footnote{In
  the Denner description $a=b=0$ and the $\Sigma^{L/R/Ls/Rs}_{ii}$
  are replaced by $\Retilde\Sigma^{L/R/Ls/Rs}_{ii}$.} 
\begin{align}
\delta Z_{L,ii}&= - \Sigma^{L}_{ii}(m_{\ti\chi_{i}}^2) - \mi
                 \frac{\partial}{\partial p^2} \left.\left[ \mi
                 \Sigma^{L}_{ii}(p^2) + \mi \Sigma^{R}_{ii}(p^2)
                 + \Sigma^{Ls}_{ii}(p^2)+ \Sigma^{Rs}_{ii}(p^2) \right]\right|_{p^2=\mi^2}+a\\
\delta Z_{R,ii}&= - \Sigma^{R}_{ii}(m_{\ti\chi_{i}}^2) - \mi \frac{\partial}{\partial p^2}
                  \left.\left[ \mi \Sigma^{L}_{ii}(p^2)+ \mi
                  \Sigma^{R}_{ii}(p^2)+ \Sigma^{Ls}_{ii}(p^2)+
                  \Sigma^{Rs}_{ii}(p^2) \right]\right|_{p^2=\mi^2}+  b +a\\
\delta \bar{Z}_{L,ii} &=\delta Z_{L,ii}-2a \\
\delta \bar{Z}_{R,ii}&= \delta Z_{R,ii} - 2b -2a \,,
\end{align}
where
\bea  
b&=&  - \fr{1}{\mi} \big[  \Sigma^{Ls}_{ii}(m_{\ti\chi_{i}}^2)  
-\Sigma^{Rs}_{ii}(m_{\ti\chi_{i}}^2) - (\de
M^\tree_{\tilde{\chi}})_{ii}+ \braket{\de 
  M^\tree_{\tilde{\chi}}}^*_{ii}\big]\;\\
a&=& -\fr{b}{2} \;.
\eea
Our results coincide with those
given in \bib{Bharucha:2012nx}. The above wave
function renormalization constants have been chosen such that for
all fermions the mixing terms are cancelled and the correct propagators are
produced at the tree-level mass values. 
Note that in case we do not have enough parameters to
renormalize all fermions on-shell, only some of these fermions satisfy
OS conditions, {\it i.e.}~their tree-level masses are equal to
the physical masses. The remaining fermions have loop-corrected
masses. This is the case for the electroweakino sector. Given the fact
that we have already fixed 
the renormalization scheme for the Higgs sector, only the two gaugino
masses $M_1$ and $M_2$ remain to be renormalized, while we have 7
masses (5 neutralino and 2 chargino masses), so that only two of them
can be set OS. The remaining 5 particles receive loop-corrections to
their masses. At loop level the mixing between
fermions is in general not vanishing any more, and the residues of the
propagators are not unity. These effects should be taken into account
if the loop corrections to the masses are large. This is not the case
for the renormalization schemes chosen here, therefore we 
neglect these effects. \s

In the chargino and neutralino sector hence only the 
two counterterms $\de \abs{M_1}$ and $\de 
\abs{M_2}$ remain to be determined. There are 21 different ways to
choose two out of the 7 masses for the OS conditions. We will consider here two 
different schemes. In the first scheme (OS1), we require the masses of
the wino-like chargino $\tilde{\chi}_i^+$ and the bino-like neutralino
$\tilde{\chi}_k^0$ to be OS. The  bino-like neutralino is sensitive to $M_1$
while the  wino-like chargino is sensitive to $M_2$. Note that we do
not choose the chargino and neutralino by referring to a fixed index order since
they may not be sensitive to $M_1$ or $M_2$. This can then lead to
numerical instability, as was found in the
MSSM~\cite{Fritzsche:2002bi,Guasch:2002ez,Baro:2009gn,Chatterjee:2011wc} 
and in \cite{Belanger:2016tqb,Belanger:2017rgu} for the NMSSM. 
We denote the tree-level masses for the neutralinos (charginos) by a
small letter $m_{\ti \chi^{0(+)}_i}$, and the loop corrected masses by
a capital letter $M_{\ti \chi^{0(+)}_i}$. In the OS scheme the tree-level masses are equal to
the loop-corrected ones. We define the counterterm mass matrices of the
chargino and neutralino sector in the interaction basis, $\de M_N$ and
$\de M_C$, through
\be 
M_N\to M_N +\de M_N \quad \mbox{and} \quad M_C\to M_C +\de M_C \;,
\ee
with the neutralino mass matrix given in \eqref{eq:neuMass} and the
chargino mass matrix in \eqref{eq:chaMass}. The counterterms  for
$M_1$ and $M_2$ are then given by
\begin{align}
\de \abs{M_2}  &=\fr{1}{\Re[U_{i1}^*V^*_{i1}
                 e^{i\phi_{M_2}}]}\bigg[\fr12 m_{\ti\chi_i^+}
                 \braket{\Re{\Sigma}_{ii}^{\ti\chi_i^+,
                 L}(m_{\ti\chi_i^+}^2)+ \Re{\Sigma}_{ii}^{\ti\chi_i^+,
                 R}(m_{\ti\chi_i^+}^2)}+ 
\fr 12 \Re{\Sigma}_{ii}^{\ti\chi_i^+, Ls}(m_{\ti\chi_i^+}^2)\crn
&+ \fr 12 \Re{\Sigma}_{ii}^{\ti\chi_i^+, Rs}(m_{\ti\chi_i^+}^2)-\Re
  \sbraket{U^*\de M_C V^\dagger}_{ii}\bigg|_{\de M_2=0} \bigg] \\ 
\de \abs{M_1} &=\fr{1}{\Re[N_{k1}^*N^*_{k1}
                e^{i\phi_{M_1}}]}\bigg[\fr12
                m_{\ti\chi_k^0}\braket{\Re{\Sigma}_{kk}^{\ti\chi_k^0,
                L}(m_{\ti\chi_k^0}^2)+ \Re{\Sigma}_{kk}^{\chi_k^0,
                R}(m_{\ti\chi_k^0}^2)}+  
\fr12 \Re{\Sigma}_{kk}^{\ti\chi_k^0, Ls}(m_{\ti\chi_k^0}^2)\crn
&+\fr12 \Re{\Sigma}_{kk}^{\ti\chi_k^0, Rs}(m_{\ti\chi_k^0}^2)  -
  \Re\sbraket{N^* \de M_N N^\dagger}_{kk}\bigg|_{\de
  M_1=0}\bigg] \;.
\end{align}

In the second scheme (OS2), we use the masses of the bino-like
neutralino, denoted by $\tilde{\chi}_k^0$, and of the wino-like neutralino,
denoted by $\tilde{\chi}_l^0$, as inputs. The renormalization conditions for
their OS masses are given by 
\bea 
 \hat{\Sigma}_{kk}^{\ti\chi_i^0}(p^2)\ti\chi_k^0
 \bigg|_{p^2=m_{\ti\chi^0_k}^2}&=&0\\ 
 \hat{\Sigma}_{ll}^{\ti\chi_i^0}(p^2)\ti\chi_l^0
 \bigg|_{p^2=m_{\ti\chi^0_l}^2}&=&0 \;. 
\eea 
This results in the two solutions for the counterterms $\de |M_1|$ and
$\de |M_2|$, 
\bea 
\de \abs{M_1}&=& \fr{a_1 \Re[N_{l2}^*N^*_{l2}e^{i\phi_{M_2}}] - a_2  \Re[N_{k2}^*N^*_{k2}e^{i\phi_{M_2}} ]}{
  \Re[N_{k1}^*N^*_{k1}e^{i\phi_{M_1}}] \Re[N_{l2}^*N^*_{l2}e^{i\phi_{M_2}}]-
\Re[N_{k2}^*N^*_{k2}e^{i\phi_{M_2}}] \Re[N_{l1}^*N^*_{l1}e^{i\phi_{M_1}}] }  \\
\de \abs{M_2}&=& -\fr{a_1 \Re[N_{l1}^*N^*_{l1} e^{i\phi_{M_1}}] - a_2  \Re[N_{k1}^*N^*_{k1}e^{i\phi_{M_1}} ] }
{
  \Re[N_{k1}^*N^*_{k1}e^{i\phi_{M_1}}] \Re[N_{l2}^*N^*_{l2}e^{i\phi_{M_2}}]-
\Re[N_{k2}^*N^*_{k2}e^{i\phi_{M_2}}] \Re[N_{l1}^*N^*_{l1}e^{i\phi_{M_1}}] } 
\eea
with 
\beq
a_1&=&\bigg[\fr12 \left(
m_{\ti\chi_k^0} \left(\Re{\Sigma}_{kk}^{\ti\chi_k^0,
    L}(m_{\ti\chi_k^0}^2)+ \Re{\Sigma}_{kk}^{\ti\chi_k^0,
    R}(m_{\ti\chi_k^0}^2) \right) +  
\Re{\Sigma}_{kk}^{\tilde{\chi}_k^0, Ls}(m_{\ti\chi_k^0}^2)+
\Re{\Sigma}_{kk}^{\tilde{\chi}_k^0, Rs}(m_{\ti\chi_k^0}^2)
\right)
\crn &&
- \Re\braket{N^* \de M_N N^\dagger}_{kk}\bigg|_{\de M_1=\de M_2=0}\bigg]  \\
a_2&=&\bigg[\fr12 \left( m_{\tilde{\chi}_l^0} \left( \Re{\Sigma}_{ll}^{\tilde{\chi}_l^0,
    L}(m_{\ti\chi_l^0}^2)+ \Re{\Sigma}_{ll}^{\tilde{\chi}_l^0,
    R}(m_{\ti\chi_l^0}^2) \right) +  
\Re{\Sigma}_{ll}^{\tilde{\chi}_l^0, Ls}(m_{\ti\chi_l^0}^2)+\Re
{\Sigma}_{ll}^{\tilde{\chi}_l^0, Rs}(m_{\ti\chi_l^0}^2) \right)
\crn &&- \Re\braket{N^* \de M_N 
  N^\dagger}_{ll}\bigg|_{\de M_1=\de M_2=0}\bigg] \;.
\eeq

For the field renormalization constants of the charginos and
neutralinos, we impose the OS conditions for the tree-level
masses. Besides the two OS
schemes, we will also adopt the $\DRb$ renormalization scheme for
$M_1$ and $M_2$, while for the field renormalization constants we use
the OS conditions.

\subsubsection{The Squark Sector}
\label{sect:Resquark}
We consider here only the third-generation squarks. 
The results for the first- and second-generation
  squarks are obtained analogously.
There are seven parameters to be renormalized in this sector,  
\be m_t,m_b,m_{\ti Q_3}^2, m_{\ti t_R}^2,m_{\ti b_R}^2,A_t,A_b, \ee
where $A_t,A_b$ are complex and the mass terms are real.
We denote their corresponding coun\-ter\-terms as
\be \de m_t,\de m_b, \de m_{\ti Q_3}^2, \de m_{\ti t_R}^2, \de m_{\ti b_R}^2, \de A_t, \de A_b, \ee
and  define the squark-mass counterterm matrices as
\be M_{\ti q}\to M_{\ti q} + \de M_{\ti q},\ee
with $M_{\ti q}$ given in \eqref{eq:mstopmat} for the stops and in \eqref{eq:msbotmat}
for the sbottoms. The renormalization of the
  remaining parameters appearing in the squark mass matrices has been
  specified in the renormalization of the Higgs sector, more specifically see
  Refs.~\cite{Ender:2011qh,Graf:2012hh,Muhlleitner:2014vsa,Dao:2019qaz}. \s

We have to renormalize the squark fields in order to make the squark
self-energies finite. Here we use both the EMT and the Denner
description. For the EMT description we have to introduce two separate
WFR constants, one for the particle and one for the anti-particle.
We introduce the squark WFR constants for the
  particles and anti-particles in the mass eigenstate basis as\footnote{In the Denner
description, we have $\de \bar{Z}_{\ti q}= \de Z_{\ti q}^\dagger$.}  
\be
\bpmatrix
\ti q_1\\ \ti q_2\epmatrix \to \braket{1 + \fr 12 \de Z_{\ti q}} \bpmatrix
\ti q_1\\ \ti q_2\epmatrix \,, \quad  \bpmatrix
\ti q_1\\ \ti q_2\epmatrix^\dagger \to  \bpmatrix
\ti q_1\\ \ti q_2\epmatrix^\dagger \braket{1 + \fr 12 \de \bar{Z}^{\ti
    q}}.
\ee  
The renormalized self-energies in the mass eigenstate basis are given by ($i,j=1,2$)
\be  
\hat\Sigma^{\ti q}_{ij}(p^2) =  \Sigma^{\ti q}_{ij}(p^2)
+ \fr 12 \braket{\de \bar{Z}^{\ti q}_{ij}+\de Z^{\ti q}_{ij} } p^2 
- \fr12 \braket{\de \bar{Z}^{\ti q}_{ij}
  m_{\ti q_j}^2 + m_{\ti q_i}^2\de Z^{\ti q}_{ij} }- \braket{
  U_{\ti q}\de  M_{\ti q} U_{\ti q }^\dagger} _{ij}
,\label{eq:selfenergysquark}
\ee 
where we denote by $\Sigma^{\ti q}_{ij}$ the unrenormalized
self-energies for the $\ti q_i^* \to  \ti q_j^*$
transition.\footnote{Note that in the real NMSSM the unrenormalized
  self-energies for the $\ti q_i^* \to  \ti q_j^*$ transition and for
  the  $\ti q_i \to  \ti q_j$ transition are identical. They are
  different, however, in the complex case.}  
In the following, we give the OS counterterms. The  $\DRb$ counterterms
are then easily obtained by taking only the divergent parts of the
corresponding OS counterterms. \s

Applying the decomposition 
of the fermionic self-energies as given in \eqref{eq:sigmadecomposition},
the mass counterterm in the OS scheme for the top and bottom quark,
respectively, is given by $(q=t,b)$\footnote{In the Denner
  description, $\Re$ is replaced by $\Retilde$.}  
\be 
\de m_q = \fr12\Re\left\{(\Sigma_q^{L}(m_q^2)
  +\Sigma_q^{R}(m_q^2))m_q + \Sigma_q^{Ls}(m_q^2) +\Sigma_q^{Rs} (m_q^2)
\right\}.
\ee 
The OS conditions for the scalar renormalized self-energies are
($i,j=1,2$)\footnote{In the Denner description, $\Re\hat\Sigma^{\ti
    q}_{ii}$ is replaced by $\Retilde\hat\Sigma^{\ti q}_{ii}$ in
  Eqs.~(\ref{eq:OS1b}) and (\ref{eq:OS3b}) while 
in \eqref{eq:OS2} $\Re\hat\Sigma^{\ti q}_{ij}$ is replaced by
$\Retilde\hat\Sigma^{\ti q}_{ij}$.} 
\begin{align}
& \Re\hat\Sigma^{\ti q}_{ii}(m_{\ti q_i}^2) =0
\label{eq:OS1b}\\
&\hat\Sigma^{\ti q}_{ij}(m_{\ti q_i}^2) =0 \;,& \hspace*{-2.5cm}
\hat\Sigma^{\ti q}_{ij}(m_{\ti q_j}^2) =0 \;,& \quad i\neq j 
\label{eq:OS2}\\
&\fr{\pa \Re\hat\Sigma^{\ti q}_{ii}(p^2)}{\pa
  p^2}\bigg{|}_{p^2 = m_{\ti q_i}^2} =0 \;. \label{eq:OS3b} 
 \end{align}
We apply the conditions in Eqs.~(\ref{eq:OS1b}) and (\ref{eq:OS2})  for
the top squark to determine $\de A_t$, $\de m_{\ti Q_3}^2$ and $\de
m_{\ti t_R}^2$,\footnote{In the Denner description,
  $\Re\Sigma^{\tilde{t}}_{ii}$ in Eqs.~(\ref{eq:deltam11}) and
  (\ref{eq:deltam22}) 
  is replaced by $\Retilde\Sigma^{\tilde{t}}_{ii}$. Note, however,
  that \eqref{eq:deltaY} is the same in both the EMT and the
  Denner description. We use $\Retilde$ in the
    definition of $\delta Y_{\tilde{q}}$ ($q=t,b$) so that $\delta A_q
    = (\delta A_q^*)^*$. The contribution from the imaginary part of the loop integrals is
    then moved into $\delta Z_{ij}^{\ti q}$ and
    $\delta \bar Z_{ij}^{\ti q}$.} 
\bea
 \de m_{\ti Q_3}^2&=& \abs{U^{\ti t}_{11}}^2 \de m_{\ti t_1}^2+\abs{U^{\ti t}_{12}}^2 \de m_{\ti t_2}^2 + U^{\ti t}_{21}U^{\ti t *}_{11} \de Y +  U^{\ti t}_{11}U^{\ti t *}_{21} \braket{\de Y}^*-2 m_t \de m_t \\
&&+  \fr23 \sin\beta\cos^3\beta M_Z^2(3-4\sin^2\theta_W) \;\de\tan\beta + \fr16\cos2\beta \;\de M_Z^2 -\fr23 \cos2\beta \; \de M_W^2\crn
 \de m_{\ti t_R}^2&=& \abs{U^{\ti t}_{12}}^2 \de m_{\ti t_1}^2+\abs{U^{\ti t}_{22}}^2 \de m_{\ti t_2}^2 + U^{\ti t}_{22}U^{\ti t*}_{12} \de Y +  U^{\ti t}_{12}U^{\ti t*}_{22} \braket{\de Y}^*
-2 m_t \de m_t \\
&&+  \fr8{3} \sin\beta\cos^3\beta M_Z^2\sin^2\theta_W \;\de\tan\beta - \fr23\cos2\beta \;\de M_Z^2 \crn
&&+\fr2{3} \cos2\beta \; \de M_W^2\crn
\de A_t &= &\fr{\expetam}{m_t}\bigg[ U^{\ti t}_{11}U^{\ti t*}_{12}\braket{\de m_{\ti t_1}^2 -
\de m_{\ti t_2}^2} + U^{\ti t}_{11}U^{\ti t*}_{22} (\de Y_{\ti t})^* +  U^{\ti t}_{21}U^{\ti t*}_{12} \de Y_{\ti t}  \crn
&-& \braket{A_t \expeta -\fr{\mueff^*}{\tan\beta}}\de m_t\bigg]
 - \fr{\expetam \mueff^* \de\tan\beta}{\tan^2\beta} +\fr{\expetam\de \mueff^*}{\tan\beta}\,,
\eea
where
\bea 
\de m_{\ti t_1}^2& =&\Re\Sigma^{\ti t}_{11}(m_{\ti t_1}^2) \label{eq:deltam11}\\
\de m_{\ti t_2}^2& =&\Re\Sigma^{\ti t}_{22}(m_{\ti t_2}^2)\label{eq:deltam22}\\
\de Y_{\ti t} &=& \sbraket{U^{\ti t} \de M_{\ti t} U^{\ti
    t\dagger}}_{12} = \sbraket{U^{\ti t} \de M_{\ti t} U^{\ti
    t\dagger}}_{21}^* = \fr{1}{2}\Retilde \braket{\Sigma^{\ti
    t}_{12}(m_{\ti t_1}^2)+ \Sigma^{\ti t}_{12}(m_{\ti t_2}^2)} \,.\label{eq:deltaY}
\eea

There remain two parameters from the bottom squark sector to be
determined, $A_b$, $m_{\ti b_R}^2$. We choose the OS scheme where the
bottom squark with the dominant contribution from the right-handed
sbottom, which we denote by $\ti b_{i_R}$, is OS and the mixing
between the two bottom squark states vanishes. The three counterterms 
 $\de m_{\ti b_R}^2,\Re \de A_b,\Im \de A_b$ are then obtained by solving 
three linear equations
\begin{align} \abs{U^{\ti b}_{i_R2}}^2 x + 2m_b\Re[U^{\ti b}_{i_R2}U^{\ti b*}_{i_R1}]y - 2m_b\Im[U^{\ti b}_{i_R2}U^{\ti b*}_{i_R1}]z
&= d_1 \\
\Re[U^{\ti b}_{12}U^{\ti b*}_{22} ] x + m_b\Re[U^{\ti b}_{12}U^{\ti b*}_{21}+ U^{\ti b}_{11}U^
{\ti b*}_{22}]y - m_b\Im[U^{\ti b}_{12}U^{\ti
  b*}_{21}- U^{\ti b}_{11}U^{\ti b*}_{22}] z &= \Re d_2 \\
\Im[U^{\ti b}_{12}U^{\ti b*}_{22}] x + m_b\Im[U^{\ti b}_{12}U^{\ti
  b*}_{21}+ U^{\ti b}_{11}U^{\ti b*}_{22}]y +
  m_b\Re[U^{\ti b}_{12}U^{\ti b*}_{21}- U^{\ti b}_{11}U^{\ti
  b*}_{22}] z&= \Im d_2 \;,
 \end{align} 
where $(x,y,z)= (\de m_{\ti b_R}^2,\Re \de A_b,\Im \de A_b)$
and\footnote{In the Denner description, $\Re\Sigma^{\tilde{b}}_{i_R
    i_R}$ in \eqref{eq.mbd1} is replaced by
  $\Retilde\Sigma^{\tilde{b}}_{i_R i_R}$.} 
\bea 
d_1&=& \Re\Sigma^{\ti b}_{i_Ri_R}(m_{\ti
    b_{i_R}}^2)- \braket{ U^{\ti b}\de M_{\ti b} U^{\ti
    b\dagger}}_{i_R,i_R}\bigg|_{x=y=z=0}  \label{eq.mbd1}
\\
d_2 &=& \delta Y_{\ti b}-\braket{ U^{\ti b}\de M_{\ti b} U^{\ti
    b\dagger}}_{12}\bigg|_{x=y=z=0} \label{eq.mbd2}\\
\delta Y_{\ti b}&=& \fr 12\Retilde\braket{\Sigma^{\ti b}_{12}(m_{\ti
    b_{1}}^2)+ \Sigma^{\ti b}_{12}(m_{\ti b_{2}}^2)} \,, \label{eq:yb}
\eea
where we have introduced the notation $\delta Y_{\ti b}$ for later use.
The other bottom squark mass gets loop
corrections.\footnote{Note that
      in principle all four masses of 
    the stops and sbottoms can be renormalized on-shell simultaneously
  by adapting the input parameters appropriately.} Its loop-corrected
mass $M_{\ti b_j}$ is obtained by solving iteratively
the following equation\footnote{In the Denner
    description $\Re\hat\Sigma^{\ti b}_{jj}$ is replaced by
    $\Retilde\hat\Sigma^{\ti b}_{jj}$.}  
\be 
M_{\ti b_j}^2 =m_{\ti b_j}^2 -  \Re\hat\Sigma^{\ti b}_{jj}(M_{\ti
  b_j}^2) \;.
\ee
We stop the iteration when the difference between two
consecutive solutions is less than $10^{-5}$. \s

The OS wave function renormalization constants for the squarks are given
by\footnote{In the Denner description, $\Re\Sigma^{\ti q}_{ii}$ in
  \eqref{eq:OSWFR1} is replaced by  $\Retilde\Sigma^{\ti q}_{ii}$ and
  $\Re\Sigma^{\ti q}_{12},
  \Re\Sigma^{\ti q}_{21}$ in
  Eqs.~(\ref{eq:OSWFR2}),\ldots,(\ref{eq:OSWFR5}) are replaced by
  $\Retilde\Sigma^{\ti q}_{12}, \Retilde\Sigma^{\ti q}_{21}$,
  respectively.} 
\begin{align}
\de Z^{\ti q}_{ii} &= - \fr{\pa \Re\Sigma^{\ti q}_{ii}(p^2)}{\pa
                     p^2}\bigg{|}_{p^2 = m_{\ti q_i}^2} \label{eq:OSWFR1} \\
\de Z^{\ti q}_{12} &= 2\fr{\Sigma^{\ti q}_{12}(m_{\ti q_2}^2) -
\delta Y_{\ti q}}{m_{\ti q_1}^2-m_{\ti q_2}^2} \label{eq:OSWFR2}\\ 
\de Z^{\ti q}_{21} &= -2\fr{\Sigma^{\ti q}_{21}( m_{\ti q_1}^2)
                     -\delta Y^*_{\ti q}}{m_{\ti q_1}^2-m_{\ti
                     q_2}^2} \label{eq:OSWFR3}\\
\de \bar{Z}^{\ti q}_{12} &= -2\fr{\Sigma^{\ti q}_{12}( m_{\ti q_1}^2)
                           -\delta Y_{\ti q}}{m_{\ti q_1}^2-m_{\ti
                           q_2}^2}  \label{eq:OSWFR4}\\
\de \bar{Z}^{\ti q}_{21} &= 2\fr{\Sigma^{\ti q}_{21}( m_{\ti q_2}^2)
                           -\delta Y^*_{\ti q}}{m_{\ti q_1}^2-m_{\ti q_2}^2} \,,\label{eq:OSWFR5}
\end{align}
where the $\delta Y_{\ti q}$ are given in
\eqref{eq:deltaY} and \eqref{eq:yb} for $q=t$ and $q=b$, respectively.  
Besides the OS scheme we also provide the option to use the $\DRb$
scheme for all parameters and the wave function renormalization
constants. In the $\DRb$ scheme, all squarks receive loop-corrections
to their masses. We will discuss this issue in \ssect{ssectsquark}.

\subsection{Loop-corrected Higgs Boson Masses and Mixings}
Since we use the mixed OS-$\DRb$ renormalization scheme for the Higgs
sector parameters together with  the $\DRb$ scheme for the Higgs
fields, all Higgs bosons are mixed and receive loop corrections to their masses. 
For the evaluation of the loop-corrected Higgs boson masses and the Higgs
mixing matrix, we use the numerical results obtained from {\tt
  NMSSMCALC}~\cite{Baglio:2013iia}. In this code the two-loop
corrected Higgs boson masses are obtained by determining the zeros of
the determinant of the two-point function $\hat \Gamma^h(p^2)$ with 
\be 
\braket{ \hat{\Gamma}^{h}(p^2)}_{ij} = i\de_{ij}(p^2 - m_{h_i}^2)  + i
\hat\Sigma_{ij}^h(p^2) \;,\quad i,j=1,\ldots,5 \;,
\ee
where $m_{h_i}$ are the tree-level masses and $\hat\Sigma_{ij}^h(p^2)$
is the renormalized self-energy of the $h_i\to h_j$ 
transition at $p^2$. In {\tt NMSSMCALC}, we have included in the
renormalized Higgs self-energies the complete one-loop contributions
with full momentum dependence~\cite{Ender:2011qh,Graf:2012hh} and the
two-loop contributions of ${\cal O}(\al_t\al_s)$
\cite{Muhlleitner:2015dua} and of ${\cal O}(\al_t^2)$
\cite{Dao:2019qaz} in the gaugeless limit at zero
momentum. The loop-corrected masses of the Higgs 
bosons are then sorted by ascending masses\footnote{We
  denote loop-corrected Higgs mass eigenstates by capital letters $H_i$ ($i=1,...,5$).} 
\be 
M_{H_1} \le M_{H_2}\le M_{H_3}\le M_{H_4}\le M_{H_5} \;.
\ee
We have improved the stability of the determination of the Higgs boson
masses in {\tt NMSSMCALC} by  implementing two-point loop integrals
with complex momentum. In the old version of {\tt NMSSMCALC}, in order to
take into account the contribution of the imaginary part of the
complex momentum we expanded the renormalized
Higgs self-energies around the real part of the
complex momentum as
\be 
\hat\Sigma_{ij}^h(p^2) =\hat\Sigma_{ij}^h(\Re p^2 ) + i \Im p^2
\fr{\pa\hat\Sigma_{ij}^h(\Re p^2  )}{\pa\Re p^2} \;. \label{eq:expansionIM}
\ee
Note that this was done only for the one-loop correction with full
momentum dependence. This expansion is not good when $\Re p^2$ is
close to threshold regions in which $ \pa\hat\Sigma_{ij}^h(\Re p^2
)/\pa\Re p^2$ contains threshold singularities. To overcome this  
problem one can use complex masses for the loop particles or complex
momenta. Using complex masses requires the decay
widths of the particles. These have to be obtained in
  an iterative procedure which is very time consuming. We have decided to
use the complex momenta and to keep the masses real. 
We have implemented the two-point loop integrals with complex
momenta and therefore do not use any more the
mentioned approximation.  We have confirmed that the evaluation of
the Higgs masses is stable in the 
singularities region and the differences between Higgs masses using
the complex momentum and the  expansion in
Eq. (\refeq{eq:expansionIM}), defined as
$(M_{H_i}^{\text{expansion}}-M_{H_i}^{\text{complex~}
  p^2})/M_{H_i}^{\text{expansion}}$ ($i=1,\ldots,5$) are of per mille 
level. \s

In processes with external Higgs bosons finite wave-function renormalization
factors $\ZH$ have to be taken into account in order to ensure the
on-shell properties of these Higgs bosons. The wave-function
renormalization factor matrix performing the rotation to the OS states
is given by~\cite{Williams_Weiglein_cMSSM_hdec} 
 \beq
\ZH = \begin{pmatrix}
             \sqrt{\hat Z_{h_1}} & \sqrt{\hat{Z}_{h_1}}
             \hat{Z}_{h_1h_2} &\sqrt{\hat Z_{h_1}}
             \hat Z_{h_1h_3}&\sqrt{\hat Z_{h_1}} \hat
             Z_{h_1h_4}&\sqrt{\hat Z_{h_1}} \hat Z_{h_1h_5} \\
              \sqrt{\hat Z_{h_2}} \hat Z_{h_2h_1}& \sqrt{\hat
                Z_{h_2}}& \sqrt{\hat Z_{h_2}} \hat Z_{h_2h_3}&
              \sqrt{\hat Z_{h_2}} \hat Z_{h_2h_4}& \sqrt{\hat Z_{h_2}}
              \hat Z_{h_2h_5}\\
               \sqrt{\hat Z_{h_3}} \hat Z_{h_3h_1}& \sqrt{\hat
                 Z_{h_3}} \hat Z_{h_3h_2}&\sqrt{\hat Z_{h_3}}&
               \sqrt{\hat Z_{h_3}} \hat Z_{h_3h_4}& \sqrt{\hat
                 Z_{h_3}} \hat Z_{h_3h_5}\\
\sqrt{\hat Z_{h_4}} \hat Z_{h_4h_1}& \sqrt{\hat Z_{h_4}} \hat
Z_{h_4h_2}&\sqrt{\hat Z_{h_4}} \hat Z_{h_4h_3}& \sqrt{\hat Z_{h_4}}&
\sqrt{\hat Z_{h_4}} \hat Z_{h_4h_5}\\
\sqrt{\hat Z_{h_5}} \hat Z_{h_5h_1}& \sqrt{\hat Z_{h_5}} \hat
Z_{h_5h_2}&\sqrt{\hat Z_{h_5}} \hat Z_{h_5h_3}& \sqrt{\hat Z_{h_5}}
\hat Z_{h_5h_4}& \sqrt{\hat Z_{h_5}} \label{eq:mat1}
\end{pmatrix} \;,
\eeq
where 
\be  
\hat Z_{h_i} =\fr 1 {\left(\fr{i}{\Delta^{h}_{i  i}(p^2)}\right)^\prime(M_{H_i}^2)}, \quad
\hat Z_{h_i h_j} = \fr
{\Delta^{h}_{ij}(p^2)}{\Delta^{h}_{ii}(p^2)}\bigg|_{p^2=M_{H_i}^2} \;, \label{eq:mat2}
\ee
 with
\be 
\Delta^{h} =- \sbraket{\hat \Gamma^{h}(p^2)}^{-1} \;. \label{eq:mat3}
\ee
Here prime denotes the derivative with respect to $p^2$.

\subsection{Loop-corrected Neutralino and Chargino Masses}
Within the OS and $\DRb$ schemes defined for the
neutralino and chargino sector, the electroweakinos cannot all be
renormalized OS, and there remain neutralinos and charginos that
receive loop corrections to their masses. In the following we define our
procedure to determine the loop-corrected masses for fermions in the
general case where mixing contributions are present\footnote{In
  the literature there exist many papers that deal with loop-corrected
  masses for neutralinos and charginos in the MSSM such as
  Refs.~\cite{Pierce:1993gj,Eberl:2001eu,Fritzsche:2002bi} to name a few of them. For
  the definition of the loop-corrected masses applied there, OS
  conditions for the field renormalization constants were applied to
  eliminate mixing effects between fermions at tree-level mass values.}. 
This procedure will be used for both the OS and the $\DRb$
scheme. To give an intuitive definition of the loop-corrected fermion
masses, we express the propagator matrix in terms of left- and
right-handed Weyl spinors, $\psi_D=(\psi_L,\psi_R)^T$. In this basis,
the tree-level propagator matrix is given by 
\be 
S(p)= - \Ga(p)^{-1}\,,\quad  \Ga(p)=i\bpmatrix p^\mu \sigma_\mu  & -m  \\ -m
  & p^\mu\bar\sigma_\mu \epmatrix \;, 
\ee 
with
\beq
\sigma^\mu = (\mathbbm{1}, \vec{\sigma}) \quad \mbox{and} \quad 
\bar{\sigma}^\mu = (\mathbbm{1}, -\vec{\sigma}) \;,
\eeq
where $\vec{\sigma} = ( \sigma^1, \sigma^2, \sigma^3)$ denotes the
three Pauli matrices. The mass matrix is given by
\be  
\bpmatrix 0 & m \\ m
 & 0 \epmatrix \;.
\ee 
The loop-corrected propagator matrix in the Weyl basis
$(\psi_L,\psi_R)^T$ is defined as 
\be 
S(p)= i \bpmatrix p^\mu \sigma_\mu (1 + \hat\Sigma^L(p^2)) & -m +
\hat\Sigma^{Ls}(p^2) \\ -m
+ \hat\Sigma^{Rs}(p^2)  & p^\mu\bar\sigma_\mu (1 + \hat\Sigma^R(p^2))
\epmatrix^{-1} \;.
\ee

The loop-corrected mass $M$ is determined from the
real pole $p^2= M^2$ of the propagator matrix satisfying the equation
\be 
\det\bpmatrix p^\mu \sigma_\mu (1 +\Re \hat\Sigma^L(p^2)) & -m +
\Re\hat\Sigma^{Ls}(p^2) \\ -m 
+\Re \hat\Sigma^{Rs}(p^2)  & p^\mu\bar\sigma_\mu (1 +\Re
\hat\Sigma^R(p^2)) \epmatrix=0 \;. 
\label{eq:Realpole1} \ee 
The solution of \eqref{eq:Realpole1} is given by
\be 
p^2 = \fr{m^2(1-\Re
  \hat\Sigma^{Ls}(p^2)/m)(1-\Re\hat\Sigma^{Rs}(p^2)/m)}{(1
  +\Re\hat\Sigma^L(p^2))(1+\Re\hat\Sigma^R(p^2) )}  \;,
\ee
which can be solved iteratively. When the fermion is OS,
$p^2=m^2$, the above relation is nothing else but the OS condition
obtained from \eqref{eq:OS1}. For the case of  $n$ Dirac spinors, the
1PI two-point function  in the basis $(\psi_L^1,\psi_L^2,
  \cdots,\psi_L^n,\psi_R^1,\psi_R^2, \cdots ,\psi_R^n)$ is a $2n\times
2n$ matrix 
\be 
\Ga(p)= i \bpmatrix p^\mu \sigma_\mu a(p^2) & d(p^2)\\ 
  c(p^2) &  p^\mu\bar\sigma_\mu b(p^2)
\epmatrix \;.
\ee
The matrices $a,b,c,d$ are $2\times 2$ matrices in
  case of charginos and $5\times 5$ matrices for neutralinos. With
$m_{\tilde{\chi}_i}$ generically denoting the mass of an
electroweakino with index $i$, the matrices are given by
\bea
a_{ij}(p^2) &=& \de_{ij} +\hat\Sigma_{ij}^L(p^2)\\
b_{ij}(p^2) &=& \de_{ij} +\hat\Sigma_{ij}^R(p^2)\\
 c_{ij}(p^2) &=&- m_{\ti\chi_i}\de_{ij} +\hat\Sigma_{ij}^{Ls}(p^2)\\
 d_{ij}(p^2) &=&- m_{\ti\chi_i}\de_{ij} +\hat\Sigma_{ij}^{Rs}(p^2) \;,
\eea
with $i,j=1,2$ for charginos and $i,j=1,...,5$ for neutralinos. \s

The  poles of the propagator matrix are the solutions of the equation
\be 
\det[\Ga(p)]=0 \;.\label{eq:polepropagator} 
\ee
This is equivalent to~\cite{Martin:2005ch}, 
\be 
\det[p^2 - c(p^2) b^{-1}(p^2) d(p^2) a^{-1}(p^2)] =0 \;.
\ee 
In practice, we solve the equation numerically through iteration
together with the diagonalization of the mass matrix $M_{\tilde{\chi}}=c(p^2) b^{-1}(p^2) 
d(p^2) a^{-1}(p^2)$ to obtain the complex
poles. The loop-corrected masses are
then obtained from the real parts of these complex poles. This procedure is applied 
for the calculation of the loop-corrected masses for neutralinos using
the OS definition of the neutralino WFR constants. 
However, for the chargino sector the  mass matrix $M_{\tilde{\chi}^\pm}$ contains infrared (IR) 
divergences at arbitrary momentum.  We have implemented the 
approximation used in \bib{Fritzsche:2002bi} and calculate the
loop-corrected chargino masses by using the formula 
\bea  
M_{\tilde{\chi}_i^+} &=&m_{\tilde{\chi}_i^+}\braket{1-\fr12\Re
  \Sigma^{\tilde{\chi}_i^+,L}_{ii}(m_{\tilde{\chi}_i^+}^2) -\fr12\Re
  \Sigma^{\tilde{\chi}_i^+,R}_{ii}(m_{\tilde{\chi}_i^+}^2)}-
\fr12\Re\Sigma^{\tilde{\chi}_i^+,Ls}_{ii}(m_{\tilde{\chi}_i^+}^2) \crn
&&-\fr12\Re\Sigma^{\tilde{\chi}_i^+,Rs}_{ii}(m_{\tilde{\chi}_i^+}^2)+\Re \left( U^* \de M_C V^\dagger \right)_{ii} \;.
\label{eq:appMcha}
\eea 

\subsection{Loop-corrected Squark Masses and Mixings}
\label{ssectsquark}
In our $\DRb$ scheme, the counterterms
\be 
\de m_t , \,\de m_b , \, \de m_{\ti Q_3}^2 , \, \de m_{\ti t_R}^2, \, \de m_{\ti
  b_R}^2, \, \de A_t, \, \de A_b
\ee
contain only the UV divergent parts.  For the renormalization of the
squark fields we use a  modified OS scheme. In the following, we will
describe the details of this scheme. We first compute the $\DRb$
squark WFR constants  by taking  the UV-divergent parts of the OS
counterterms, defined in Eqs.~(\ref{eq:OSWFR1}) to (\ref{eq:OSWFR5}),
with tree-level mass values.  
Using these $\DRb$ squark WFR constants, we then compute the
loop-corrected squark masses $M_{\ti q_i}$ that are obtained
by solving iteratively the equations 
\be 
M^2_{\ti q_i} =  m_{\ti q_i}^2 -  \Re \hat\Sigma^{\ti q}_{ii}(M^2_{\ti q_i})\;, \quad i=1,2\;.
\ee
We have assumed here that the off-diagonal renormalized self-energies
vanish and that the residues of the propagators are unity at the
loop-corrected masses. This is equivalent to redefining the diagonal
squark WFR constants at loop-corrected masses as
\be 
\de Z^{\ti q}_{ii}(M_{\ti q_i}^2) = - \fr{\pa
  \Sigma^{\ti q\, \text{div}}_{ii}(p^2)}{\pa p^2}\bigg{|}_{p^2 =
  m_{\ti q_i}^2}- \fr{\pa \hat\Sigma^{\ti q}_{ii}(p^2)}{\pa
  p^2}\bigg{|}_{p^2 = M_{\ti q_i}^2}, 
\quad i=1,2 \;, \label{eq:diagWFRsquark}
\ee
and the off-diagonal WFR constants as
\be  
\de Z^{\ti q}_{ij}(M_{\ti q_k}^2) = \fr{\Sigma^{\ti q \,\text{div}}_{ij}(m_{\ti q_{i}}^2)- \Sigma^{\ti q\, \text{div}}_{ij}(m_{\ti q_{j}}^2)}{m_{\ti q_{j}}^2 -m_{\ti
    q_{i}}^2} -2 \fr{\hat\Sigma^{\ti q}_{ij}(M_{\ti
    q_k}^2)}{M_{\ti q_k}^2- m_{\ti q_i}^2},\quad i,j,k=1,2, \quad i\ne j.
\label{eq:offdiagWFRsquark}
\ee

In the above equations the renormalized self-energies are computed
with the $\DRb$ squark WFR constants.  
We keep also the imaginary part of the two-point loop integrals in the renormalized
self-energies. Note that the WFR constants $\de Z^{\ti q}_{ij}(M_{\ti q_k}^2)$ will enter the evaluation of the decay width. The diagonal WFR constants $\de Z^{\ti q}_{ii}(M_{\ti q_k}^2)$ contain IR divergences
evaluated at the loop-corrected masses. These IR divergences will
cancel exactly with those arising from the virtual part and the real radiation contributions which are also evaluated at the loop-corrected masses.
We have verified that this statement is true for both the EW and the QCD corrections.

\section{Higher-Order Corrections to the Two-Body Decays of the
  Neutral Higgs Bosons \label{sec:oneloopdecays}}
In this section, we present those two-body decay channels that we
have improved by including the missing NLO EW and QCD
corrections. These channels are the decays into OS SM fermion pairs, OS
massive gauge bosons, into a pair of Higgs and gauge bosons, into
chargino or neutralino pairs and into top or bottom squark 
pairs. We will not discuss decays into gluon pairs, photon pairs or $Z
\gamma$ which can be found in \bib{Baglio:2013iia}. For these decays, NLO EW corrections
  are of two-loop order as the leading order (LO) decay widths are already
  loop-induced. The inclusion of the NLO EW corrections 
to Higgs-to-Higgs decays on the other hand have been presented in
\bib{Nhung:2013lpa} and the dominant two-loop corrections of the
${\cal O}(\al_t\al_s)$ have been provided in \bib{Muhlleitner:2015dua}.  \s

For our computation we have used several programs. The generation of the
amplitudes was done by
\texttt{FeynArts}~\cite{Kublbeck:1990xc,Hahn:2000kx} using a model
file created by
\texttt{SARAH}~\cite{Staub:2009bi,Staub:2010jh,Staub:2012pb,Staub:2013tta}. The
output amplitudes were further processed using
\texttt{FeynCalc}~\cite{FeynCalc,Shtabovenko:2016sxi} for the
simplification of the Dirac matrices and for the tensor reduction. The
one-loop integrals were evaluated with the help of
\texttt{LoopTools}~\cite{Hahn:formcalc}.

\subsection{Higgs Boson Decays into Fermion Pairs}
In order to make use of the published results of higher-order
corrections in the literature for CP-even and CP-odd Higgs bosons, it is
convenient to write the interaction vertex of the complex NMSSM Higgs 
boson $h_i$ $(i=1,\ldots,5)$  and quarks as
\be 
\calL_{h_i q\bar q} = - \fr{m_q}{v}\bar q h_i \braket{ \ghqqs -i \ghqqa
  \ga_5  } q \;, 
\ee
where the scalar and  pseudoscalar coupling coefficients for the up-
and down-type quarks at tree-level are given by
\begin{align} 
g_{h_i d\bar{d}}^S &= \fr{\calR_{i1}}{\cbeta} \;, &
g_{h_i d\bar{d}}^P &= \calR_{i4}\tbeta \;, 
\label{eq:bottomHiggsCT} \\  
g_{h_i u\bar{u}}^S &=\fr{ {\cal R}_{i2} }{ \sbeta }\;, & 
g_{h_i u\bar{u}}^P  &=\fr{ {\cal R}_{i4}}{t_\beta}\;,  \label{eq:topHiggsCT}
\end{align}
where ${\cal R}_{ij}$ ($i,j=1,5$) denotes the matrix elements of the
mixing matrix rotating the tree-level Higgs gauge eigenstates to the
mass eigenstates, see \eqref{eq:rotgaugemasstree}.\s 

Following the prescription outlined in our
publication~\cite{Baglio:2013iia}, we improve the widths of the Higgs
boson decays into quark pairs by including the missing SUSY--QCD and
SUSY--EW corrections. We decompose the EW corrections into 
the known QED corrections arising from a virtual photon exchange and a
real photon emission and the remaining unknown EW corrections from
the genuine EW one-loop diagrams. \s

The one-loop SUSY--QCD corrections originate from loop diagrams with the
exchange of a gluino, while the SUSY--EW corrections stem from loop diagrams 
with weak gauge bosons $W,Z$, fermions,  Higgs bosons and their
superpartners in the internal lines. 
They are both IR finite quantities. The computation of the Higgs
boson decays into a bottom quark pair shows that the bottom quark mass
counterterm contains terms proportional to $\tbeta$. This
contribution is large in the large-$\tbeta$ regime and universal. 
In many cases, this contribution is the leading part of the 
SUSY--QCD and SUSY--EW corrections and can be
absorbed into an effective bottom quark Yukawa coupling. 
This can be done by using an effective Lagrangian
formalism~\cite{Carena:1999py,Carena:2002bb,Guasch:2003cv}. In
\bib{Baglio:2013iia}, we have presented the effective bottom Yukawa  
couplings in the real and complex NMSSM. We do not repeat every detail here but
only quote the relevant formulae. In \eqref{eq:bottomHiggsCT} we have 
given the tree-level scalar and  pseudoscalar coupling coefficients
appearing in the Feynman rule for the CP-violating
Higgs bosons $h_i$ to a bottom-quark pair. 
The Feynman rule for the effective
  coupling including the leading SUSY--QCD and
SUSY--EW corrections 
\cite{Carena:1999py,Carena:2002bb,Guasch:2003cv,Hempfling:1993kv,Hall:1993gn,Carena:1994bv,Pierce:1996zz,Carena:1998gk,Noth:2008tw,Mihaila:2010mp} 
(denoted by a tilde to mark the inclusion of the corrections) is also decomposed
  into a scalar and a pseudoscalar part and reads~\cite{Baglio:2013iia}
\beq
-  \frac{im_b}{v} \left[ \tilde{g}_{h_i b\bar{b}}^S -i \gamma_5 \tilde{g}_{h_i b\bar{b}}^P
\right] \;,
\eeq
with
\beq   
\tilde{g}_{h_i b\bar{b}}^S = \mbox{Re}\, \tilde{g}_{b_L}^{h_i} 
\quad \mbox{and} \quad
\tilde{g}_{h_i b\bar{b}}^P = \mbox{Im}\, \tilde{g}_{b_L}^{h_i} \;,
\label{eq:effbotSP}
\eeq
where 
\beq
\tilde{g}_{b_L}^{h_i} = \fr{1}{(1+ \Delta_b) } \left[ \fr{{\cal R}_{i1}}{\cos\beta} 
+  \fr{{\cal R}_{i2}}{\sin\beta} \Delta_b
+  \fr{ {\cal R}_{i3}v}{v_s}\Delta_b + i {\cal
    R}_{i4}\tan\beta \left( 1- \fr{\Delta_b}{\tan^2\beta} \right) -  i \fr{ {\cal
      R}_{i5}v}{v_s}\Delta_b\right] . \label{eq:effbot}
\eeq
The correction $\Delta_b$ including the leading  SUSY--QCD and
SUSY--EW corrections can be cast into the form 
\beq 
\Delta_b &=& \frac{\Delta_b^{\text{QCD}}+\Delta_b^{\text{elw}}}{1+\Delta_1} \;,
\label{eq:deltabcorr} 
\eeq
with the one-loop corrections given by 
\beq
\Delta_b^{\text{QCD}} &=& 
\frac{C_F}{2} \, \frac{\alpha_s (\mu_R)}{\pi} \, M_3^* \,
\mu_{\mbox{\scriptsize eff}}^* \, \tan\beta \, I (m_{\tilde{b}_1}^2,
m_{\tilde{b}_2}^2 , m_{\tilde{g}}^2 ) \;,  \label{eq:deltabcorr_cp1} \\ 
\Delta_b^{\text{elw}} &=& \frac{\alpha_t (\mu_R)}{4\pi} \, A_t^* \,
\mu_{\mbox{\scriptsize eff}}^* \, \tan\beta \, I (m_{\tilde{t}_1}^2, 
m_{\tilde{t}_2}^2 , |\mu_{\mbox{\scriptsize eff}}|^2) \;, \label{eq:deltabcorr_cp2}\\
\Delta_1&=& - \frac{C_F}{2} \, \frac{\alpha_s (\mu_R)}{\pi}\, M_3^*\,
A_b \, I (m_{\tilde{b}_1}^2,
m_{\tilde{b}_2}^2 , m_{\tilde{g}}^2 )\;,
\label{eq:deltabcorr_cp3}
\eeq
where $\alpha_t = y_t^2/(4\pi)$ with $y_t = \sqrt{2} m_t /(v
\sin\beta)$ is the top-Yukawa coupling and $C_F=4/3$. The generic
function $I$ is defined as 
\beq
I(a,b,c) &=& \frac{ab \, \log \frac{a}{b} + bc \, \log \frac{b}{c} +
  ca \, \log \frac{c}{a}}{(a-b)(b-c)(a-c)} \;.
\eeq
Note that the scale of $\alpha_s$ in the SUSY--QCD corrections
has been set equal to
$\mu_R=(m_{\tilde{b}_1}+m_{\tilde{b}_2}+|M_{\tilde{g}}|)/3$, while in
the SUSY--EW corrections it is
$\mu_R=(m_{\tilde{t}_1}+m_{\tilde{t}_2}+|\mu|)/3$. The strong coupling
constant $\alpha_s$ is evaluated with five active flavors. \s

The decay width of the CP-violating NMSSM Higgs bosons
$H_i$ into $q\bar{q}$, including the QCD,  SUSY--QCD, QED, EW and
SUSY--EW corrections, can then be cast into the form 
\beq
\Gamma (H_i \to q\bar{q}) &=& \frac{3 G_F M_{H_i}}{4 \sqrt{2} \pi} \,
\overline{m}_q^2 (M_{H_i})  \, \big[ \braket{1-4 x_q}^{3/2}
\,\Delta^S_{\mbox{\scriptsize QCD}} \deQED^S \Gamma^S_{H_i\to q \bar q }  \crn 
&&+
\braket{1-4 x_q }^{1/2}\,\Delta^P_{\mbox{\scriptsize QCD}} \deQED^P
\Gamma^P_{H_i\to q \bar q }    \big]
\;, \label{eq:correctedgambb}   
\eeq  
where $x_q =m_q^2/M_{H_i}^2$, and 
\beq
\Gamma^S_{H_i \to q \bar{q}} &=& \left( \sum_{j=1}^5 \ZH_{ij}
\tilde{g}_{h_j q\bar{q}}^S \right) \left( \sum_{k=1}^5 \ZH_{ik}
\tilde{g}_{h_k q\bar{q}}^S \right)^* \nonumber \\
&+& 2 \mbox{Re} \left[ \left( \sum_{j=1}^5 \ZH_{ij}
\tilde{g}_{h_j q\bar{q}}^S \right) \left(
\sum_{k=1}^5 \ZH_{ik} \, \delta {\cal M}^{\text{rem},S} (h_k \to q \bar{q})
\right)^* \right] \nonumber \\
&+& 2 \mbox{Re} \left[\left( \sum_{j=1}^5 \ZH_{ij}
\tilde{g}_{h_j q\bar{q}}^S \right) \left( \sum_{k=1}^5 \ZH_{ik}
\, \delta^S_{\text{sub}} (h_k \to q\bar{q})
\right)^* \right] \label{eq:amp1}
\eeq
and
\beq
\Gamma^P_{H_i \to q \bar{q}} &=& \left( \sum_{j=1}^5 \ZH_{ij}
\tilde{g}_{h_j q\bar{q}}^P \right) \left( \sum_{k=1}^5 \ZH_{ik}
\tilde{g}_{h_k q\bar{q}}^P \right)^* \nonumber \\
&+& 2 \mbox{Re} \left[ \left( \sum_{j=1}^5 \ZH_{ij}
\tilde{g}_{h_j q\bar{q}}^P \right) 
\sum_{k=1}^5 (\ZH_{ik})^* \, \left( \delta {\cal M}^{\text{rem},P} (h_k \to q \bar{q})
+ \delta {\cal M}_{GZ,\text{mix}} (h_k \to q\bar{q} )\right)^* \right] \nonumber \\
&+& 2 \mbox{Re} \left[\left( \sum_{j=1}^5 \ZH_{ij}
\tilde{g}_{h_j q\bar{q}}^P \right) \left( \sum_{k=1}^5 \ZH_{ik}
\, \delta^P_{\text{sub}} (h_k \to q\bar{q})
\right)^* \right] \;.\label{eq:amp2}
\eeq
In the numerical analysis presented in Sec.~\ref{sec:numerical} we
will use the quantity $\Gamma^{\text{SEW(+QCD)}}$ for the decays into
fermion pairs to denote the partial decay widths including the SUSY-EW
and (for the quarks) SUSY-QCD corrections, {\it i.e.} exactly the partial decay width as
defined in Eq.~(\ref{eq:correctedgambb}) with the loop-corrected $\Gamma^S$ and
$\Gamma^P$ given in Eqs.~(\ref{eq:amp1}) and (\ref{eq:amp2}),
respectively. In contrast, we will denote by $\Gamma^{\text{tree}}$ the
partial decay widths that only  include the $\Delta_b$ corrections,
{\it i.e.}~Eq.~(\ref{eq:correctedgambb}) but with $\Gamma^S$ and
$\Gamma^P$ given by the first lines in Eqs.~(\ref{eq:amp1}) and (\ref{eq:amp2}),
respectively. \s

Note that in $\delta {\cal M}^{\text{rem},S/P}$,
  $\delta^{S/P}_{\text{sub}}$ and $\delta {\cal
  M}_{GZ,\text{mix}}$ (which will be explained below) we use the
tree-level Higgs couplings $g_{h_k q 
  \bar{q}}$ to the quarks.
We take the occasion to remind the reader that 
  tree-level mass eigenstates are always denoted by $h_i$ and
  loop-corrected ones by $H_i$. Unless stated otherwise, this means
  that we use tree-level couplings for external $h_i$ but with
  loop-corrected masses, and for particles inside loop diagrams we
  always use tree-level masses and tree-level couplings.
We comment on the various terms appearing in
Eqs.~(\ref{eq:correctedgambb}), (\ref{eq:amp1}) and (\ref{eq:amp2})
one by one. The one-loop QED corrections, 
denoted by $\Delta_{\text{QED}}$ have been known in the SM for a long
time, {\it cf.}~Refs.~\cite{Braaten:1980yq,Sakai:1980fa,Inami:1980qp,Drees:1990dq,Bardin:1991dp,Dabelstein:1991ky}. In
the limit $m_{q} \ll M_{H_i}$ they are given by
\bib{Dabelstein:1995js}\footnote{Note that actually in the code we
  have programmed the QCD corrections for the completely massive case at
  next-to-leading order, translated to the $\overline{\mbox{MS}}$
  scheme, and interpolated with the massless expression for large Higgs
masses, according to the implementation in {\tt
  HDECAY}~\cite{Djouadi:1997yw,Djouadi:2018xqq}. \label{foot:comment}},
\beq 
\deQED^S=\deQED^P= 1+ \fr{\al}{\pi}Q_q^2 \braket{\fr94 - 3\log
  \fr{m_q^2}{ M_{H_i}^2} } \;, \label{eq:qedcorr}
\eeq 
where $Q_q$ denotes the electric charge of the quark $q$.
The one-loop SM QCD corrections are similar to the QED corrections, with
the replacement of $ Q_q^2\,\al$  by $(4/3) \al_s( M_{H_i}^2)$. It is
well-known that the SM QCD corrections are rather large. The large
logarithmically enhanced part has been absorbed in
Eq.~(\ref{eq:correctedgambb}) into the running
$\overline{\text{MS}}$ quark mass
$\overline{m}_{q}(M_{H_i}^2)$
at the corresponding Higgs mass scale $M_{H_i}$ to improve the convergence
of the perturbative expansion. The QCD corrections can be
  taken over from the MSSM case by adapting the Higgs
  couplings~\cite{Braaten:1980yq,Sakai:1980fa,Inami:1980qp,Drees:1989du,Drees:1990dq,Gorishnii:1990zu,Gorishnii:1991zr,Kataev:1993be,Gorishnii:1983cu,Surguladze:1994gc,Larin:1995sq,Chetyrkin:1995pd,Chetyrkin:1996sr,Baikov:2005rw}.
After subtracting the enhanced part,
the remaining QCD correction $\Delta_{\text{QCD}}$  reads 
\begin{eqnarray}
\Delta_{\text{QCD}}^S&=&\Delta_{\rm QCD}^P  =  1 + \fr{17}{3}
  \frac{\alpha_s (M_{H_i}^2)}{\pi} +(35.94-1.359
  N_F) \left( \frac{\alpha_s(M_{H_i}^2)}{\pi} \right)^2  \nonumber \\
&& +(164.14-25.77 N_F+ 0.259 N_F^2) \left(
   \frac{\alpha_s(M_{H_i}^2)}{\pi} \right)^3\crn
&& + (39.34 -220.9 N_F + 9.685 N_F^2 - 0.0205 N_F^3)\left(
   \frac{\alpha_s(M_{H_i}^2)}{\pi} \right)^4\;,  \label{eq:deltaqcd}
\end{eqnarray}
where $N_F=5$ active flavors are taken into account.
In the CP-conserving case we also include top quark
induced corrections $\Delta_t^{S/P}$ by adding them to
$\Delta_{\text{QCD}}$. They can be taken over from the MSSM case and
read~\cite{Braaten:1980yq,Sakai:1980fa,Inami:1980qp,Drees:1989du,Drees:1990dq,Gorishnii:1990zu,Gorishnii:1991zr,Kataev:1993be,Gorishnii:1983cu,Surguladze:1994gc,Larin:1995sq,Chetyrkin:1995pd,Chetyrkin:1996sr},
\beq
\Delta_t^S &=& \frac{g_{h_i t\bar{t}}^S}{g_{h_i b\bar{b}}^S} \left(
  \frac{\alpha_s (M_{H_i})}{\pi}\right)^2 
\left[ 1.57 - \frac{2}{3} \log
  \frac{M_{H_i}^2}{m_t^2} + \frac{1}{9} \log^2
  \frac{\overline{m}_{b}^2(M_{H_i})}{M_{H_i}^2}
\right] \\
\Delta_t^P &=& \frac{g_{h_i t\bar{t}}^P}{g_{h_i b\bar{b}}^P} \left(
  \frac{\alpha_s (M_{H_i})}{\pi}\right)^2 
\left[ 3.83 - \log
  \frac{M_{H_i}^2}{m_t^2} + \frac{1}{6} \log^2
  \frac{\overline{m}_{b}^2(M_{H_i})}{M_{H_i}^2}
\right] \;. 
\eeq
In the decay into a bottom quark pair, the large universal corrections
proportional to $\calO(\al_s\tbeta,\al_b\tbeta)$ are resummed into the
effective bottom Yukawa couplings $\tilde{g}_{h_i b\bar{b}}^{S,P}$,
given in Eqs.~(\ref{eq:effbotSP}) and (\ref{eq:effbot}), while in the
decay into top quarks we use the
tree-level values of the effective 
couplings, {\it i.e.}
\beq
\tilde{g}_{h_i t\bar{t}}^{S,P} = g_{h_i t \bar{t}}^{S,P} \;,
\eeq
with $g_{h_i t\bar{t}}$ given in \eqref{eq:topHiggsCT}.
The remaining SUSY--QCD and SUSY--EW corrections are
collected in $\delta {\cal M}^{\text{rem},S}$, $\delta {\cal M}^{\text{rem},P}$ and
$\de {\cal M}_{GZ,\text{mix}}$, where
\beq
\delta {\cal M}^{\text{rem},S/P} = \delta {\cal M}^{\text{rem},S/P}_{\text{SQCD}} + \delta
{\cal M}^{\text{rem},S/P}_{\text{SEW}} + \delta {\cal
  M}^{\text{counter},S/P}  \;,
\eeq
with $\delta {\cal  M}^{\text{counter},S/P}$ denoting the counterterm
contributions. Since in the decay into bottom quarks we have resummed the dominant
corrections into the effective couplings, in the remaining SUSY--QCD
and SUSY--EW correction we have to subtract these corrections to avoid
double counting by adding appropriate counterterms. This is taken care
of by the last terms in Eqs.~(\ref{eq:amp1}) and (\ref{eq:amp2}),
respectively, to which we will come back below. \s

The term $\de {\cal M}_{GZ,\text{mix}}$ is the  sum of
the contributions from the mixing of the CP-odd component of the Higgs
bosons with the neutral Goldstone boson $G$ and with the $Z$  boson,
respectively.  We use the tree-level masses for the Higgs bosons in
the loops in order to ensure the proper cancellation of the
UV-divergent pieces, but we use the loop-corrected Higgs boson masses
for the external particles in the evaluation of the wave-function
renormalization factors, of the amplitudes, and of the decay
widths. It is well-known that the use of loop-corrected Higgs boson
masses for the external particles violates the Slavnov-Taylor identity of the amplitude 
$\de {\cal
  M}_{GZ,\text{mix}}$~\cite{Nhung:2013lpa,Williams_Weiglein_Heidi2011bu,Williams_Weiglein_cMSSM_hdec}. To
 restore the gauge symmetry one should use the tree-level masses for
 both the external and the internal Higgs bosons. This causes a mismatch
with the phase-space factor (where we use the loop-corrected Higgs
boson masses) and with the evaluation of the other amplitudes. We
therefore use the loop-corrected masses for the external particles
also in these contributions, which are then computed in the unitary
gauge. Note that we apply the same method also for the other decays
that contain the contribution $\de {\cal M}_{GZ,\text{mix}}$. In
\bib{Nhung:2013lpa}, $\de {\cal M}_{GZ,\text{mix}}$ was  computed only  
for the $i$th tree-level mass eigenstate of the decay $H_i \to
q\bar{q}$. This may cause instabilities in the case of large mixing
between Higgs boson mass eigenstates at loop-level. We avoid this by
multiplying it also with the WFR factor $\ZH$. 
\s

The remaining SUSY--QCD and SUSY--EW corrections are
computed in the Feynman diagrammatic approach.  
The corrections consist of the contributions from genuine one-loop
diagrams and the counterterms. 
The counterterms are given by
\beq
\delta {\cal M}^{\text{counter},S/P} (h_i \to q\bar{q}) = \delta
\lambda_{h_i q\bar{q}}^{S/P} \;,
\eeq
with the expressions for the scalar and pseudoscalar parts, $\delta
\lambda_{h_i q\bar{q}}^{S/P}$, reading\footnote{Note, that the angle
  $\beta$ in the sense of a mixing angle does not get renormalized.} 
\begin{align}
\de \la_{h_i b \bar b}^S &= \braket{\fr{\de m_b}{m_b}
  -\fr{\de v}{v} -\fr{\de \cbeta}{\cbeta}}\fr{\calR_{i1}}{\cbeta} +\sum_{k=1}^5\fr{\de
  Z_{h_ih_k}}{2} \fr{\calR_{k1}}{\cbeta}+ g_{bL}^{h_i} \fr{\de \widetilde Z_{bL}}{2} +
  g_{bL}^{h_i*} \fr{\de \widetilde Z_{bR}}{2}  \label{eq:hbbCTS}\\ 
\de \la_{h_i b \bar b}^P &= \braket{\fr{\de m_b}{m_b}
  -\fr{\de v}{v} -\fr{\de
    \cbeta}{\cbeta}  }  \calR_{i4} \tbeta 
 +\sum_{k=1}^5\fr{\de Z_{h_ih_k}}{2} \calR_{k4}+ g_{bL}^{h_i} \fr{\de \widetilde Z_{bL}}{2} -
  g_{bL}^{h_i*} \fr{\de \widetilde Z_{bR} }{2}
 \label{eq:hbbCTA}\\ 
\de \la_{h_i t \bar t}^S &= \braket{\fr{\de m_t}{m_t}
  -\fr{\de v}{v} -\fr{\de \sbeta}{\sbeta}}\fr{\calR_{i2}}{\sbeta} +\sum_{k=1}^5\fr{\de
  Z_{h_ih_k}}{2} \fr{\calR_{k2}}{\sbeta} +  g_{tL}^{h_i} \fr{\de \widetilde Z_{tL}}{2} +
  g_{tL}^{h_i*} \fr{\de \widetilde Z_{tR} }{2}  \\ 
\de \la_{h_i t \bar t}^P &=\braket{\fr{\de m_t}{m_t}
  -\fr{\de v}{v} -\fr{\de \sbeta}{\sbeta}}  \fr{\calR_{i4}}{ \tbeta} 
 +\sum_{k=1}^5\fr{\de Z_{h_ih_k}}{2} 
 \fr{\calR_{k4}}{\tbeta} + g_{tL}^{h_i} \fr{\de \widetilde Z_{tL}}{2} -
  g_{tL}^{h_i*} \fr{\de \widetilde Z_{tR} }{2} \;,   
\end{align}
with 
\begin{align}
g_{bL}^{h_i}=&\fr{\calR_{i1}}{\cbeta} +i  \calR_{i4} \tbeta \quad
               \mbox{and} \quad
g_{tL}^{h_i}=\fr{\calR_{i2}}{\sbeta} +i   \fr{\calR_{i4}}{ \tbeta} \;.
\end{align}
The counterterm $\delta v$ for the VEV is given by
\begin{align}
\frac{\delta v}{v}=& -\delta Z_e +\fr{c^2_{\theta_W}}{2 s^2_{\theta_W}} \left(\frac{\delta M_W^2}{M_W^2}-
\frac{\delta M_Z^2}{M_Z^2} \right) +\frac{\delta M_W^2}{2M_W^2} \;.
\end{align}
The electric coupling $e$, the $W$ and $Z$ boson masses and
$\tan\beta$ are renormalized according to the Higgs sector. The top
and bottom quarks are renormalized OS using both the EMT and the
Denner descriptions. The terms being related to left-handed and right-handed OS wave 
function renormalization constants for the quarks ($\de Z_{qL/R} $)
and anti-quarks ($\de \bar Z_{qL/R}$) ($q=t,b$) are\footnote{In the Denner
  description, $\Sigma_q$ is replaced by $\Retilde \Sigma_q$.} 
\begin{align}
\de \widetilde Z_{qL/R} =&\frac{\de Z_{qL/R} + \de \bar Z_{qL/R}}{2}
                           \nonumber \\
 =& -\Sigma_q^{L/R}(m_q^2)- m_q \frac{\partial}{\partial
    p^2}\left[m_q\left(\Sigma_q^L(p^2)+\Sigma_q^R(p^2)\right)
    + \Sigma_q^{Ls}(p^2)+\Sigma_q^{Rs}(p^2) \right]_{p^2=m_q^2}.
\end{align}
The $\DRb$  wave function renormalization constants for the Higgs
bosons are denoted by $\de Z_{h_i   h_k}$. 
We have checked the UV-finiteness of the SUSY--QCD and
SUSY--EW corrections to the decay amplitude. \s

As mentioned above, in the decay into bottom quarks we have to take
care to avoid double counting after resumming the dominant part of the
SUSY--QCD and SUSY--EW corrections into the effective bottom
coupling. To subtract these contributions we add in the decays into a
$b$-quark pair the following
counterterms to \eqref{eq:amp1} and \eqref{eq:amp2},
\beq 
\de^S_{\text{sub}}  (h_i \to b\bar{b}) &=& 
\braket{\fr{\calR_{i1}}{\cbeta} - \fr{\calR_{i2}}{\sbeta}
  - \fr{\calR_{i3}v}{v_s}} \mbox{Re} \De_b 
- \braket{
 \calR_{i4} \tbeta +\fr{\calR_{i4}}{\tbeta} +
 \calR_{i5}\fr{v}{v_s}} \mbox{Im} \De_b, \\
\de^P_{\text{sub}}  (h_i \to b\bar{b}) &=&
\braket{\calR_{i4} \tbeta +\fr{\calR_{i4}}{\tbeta} +
 \calR_{i5}\fr{v}{v_s}} \mbox{Re} \De_b
+ 
\braket{\fr{\calR_{i1}}{\cbeta} - \fr{\calR_{i2}}{\sbeta}
  - \fr{\calR_{i3}v}{v_s}} \mbox{Im} \De_b  , 
\eeq
where $\De_b$ equals $\Delta_b^{\text{QCD}}-\Delta_1$ and
$\Delta^{\text{elw}}_b$ for the SUSY--QCD and SUSY--EW corrections, respectively. 
For the decays into a top-quark pair, these contributions are 
\beq
\de^{S/P}_{\text{sub}} (h_i \to t\bar{t}) = 0 \;.
\eeq

In the decays into strange quarks we also include the one-loop SUSY--QCD
corrections. They are obtained after substituting $\Delta_b$ as given in
\eqref{eq:deltabcorr} with 
\beq
\Delta_s =
\left. \frac{\Delta_b^{\text{QCD}}}{1+\Delta_1} \right|_{b\to s} \;. 
\eeq
The decays into charm quarks are treated as the decays into top
quarks, with the appropriate replacements. \s

The decays into lepton final states $l=e,\mu,\tau$ do not receive QCD
corrections. Their SUSY--EW corrected decay width is given by
\beq
\Gamma (H_i \to l\bar{l}) &=& \frac{ G_F M_{H_i}}{4 \sqrt{2} \pi} \,
m_l^2   \, \big[ \braket{1-4 x_l }^{3/2}\,\deQED^S
  \Gamma^S_{H_i\to l \bar l }  \crn 
&&+
\braket{1-4 x_l }^{1/2}\, \deQED^P \Gamma^P_{H_i\to l \bar l }
\big] \;, \label{eq:correctedgamll}  
\eeq  
where $x_l =m_l^2/M_{H_i}^2$. The $\deQED^{S,P}$ are given by
\eqref{eq:qedcorr} after replacing $(Q_q,m_q)$ with $(Q_l,m_l)$.
Furthermore, we resum the dominant SUSY--EW
corrections into the effective couplings $\tilde{g}_{h_i
  ll}^{S,P}$. They are obtained from the effective couplings
$\tilde{g}_{h_i b\bar{b}}$ in Eqs.~(\ref{eq:effbotSP}) and
(\ref{eq:effbot}) after replacing $\Delta_b$ with
$\Delta_l$, where $\Delta_l$ in the complex case is given by
\beq
\Delta_l =  \frac{e^2}{(4\pi s_W)^2} \,M_1^* \,
\mu_{\mbox{\scriptsize eff}}^* \, \tbeta \, I (m_{\tilde{l}_1}^2, 
m_{\tilde{l}_2}^2 , |M_1|^2) + \fr{e^2}{(4\pi c_W)^2} M_2^* \,
\mu_{\mbox{\scriptsize eff}}^* \, \tbeta  I (m_{\tilde{\nu}_l}^2, 
|M_2|^2 , |\mu_{\mbox{\scriptsize eff}}|^2)\;.
\label{eq:deltabcorrl}
\eeq
The contributions $\Gamma^{S,P}_{H_i \to l\bar{l}}$ are obtained from the
ones given in Eqs.~(\ref{eq:amp1}) and (\ref{eq:amp2}) after replacing
$q\to l$. 

\subsection{Higgs Boson Decays into \boldmath $W^+W^-$ and $ZZ$}
\label{sect:hVV}
We now address the higher-order corrections to the Higgs
  boson decays into gauge boson pairs. We consider corrections only for
  on-shell decays. Off-shell decays are still
  treated at tree-level as done in {\tt NMSSMCALC}~\cite{Baglio:2013iia}.
The one-loop corrected decay amplitude for the decay of a CP-violating
NMSSM Higgs boson $H_i$ ($i=1,...,5$) into a pair of massive gauge
bosons $V=Z,W^\pm$, $H_i(p) \to V(k_1) V(k_2)$, is given by 
\beq 
\sbraket{\sum_{j=1}^5\ZH_{ij}(\calM^{\mu\nu}_{\tree}(h_j\to
  VV) +   \calM^{\mu\nu}_{\text{1L}}(h_j\to VV) )}
\epsilon_\mu(k_1)\epsilon_\nu(k_2) \;,
\eeq
where $\epsilon_{\mu} (k_1)$ and $\epsilon_\nu (k_2)$ are the
polarization vectors of the two gauge bosons with four-momenta $k_1$
and $k_2$, respectively. Note that the $GZ,\text{mix}$ contribution vanishes. 
The tree-level amplitudes for the two final-state pairs are
\beq 
\calM^{\mu\nu}_{\tree}(h_j\to VV)&=& g^{\mu\nu}
\calM_{\tree}(h_j\to VV),\\ 
         &=&g^{\mu\nu} g_{h_j VV} \braket{\cbeta
           \calR_{j1} + \sbeta \calR_{j2} }
\eeq
with
\beq
g_{h_j VV} = \left\{ \begin{array}{ll} \fr{e M_W}{s_{\theta_W}}
                       &\mbox{ for } V=W \\
\fr{e M_Z}{s_{\theta_W}c_{\theta_W}} & \mbox{ for } V= Z \end{array}\right. \;.
\eeq
And the tensor structure of the NLO corrections is given by
 \begin{align}
\calM^{\mu\nu}_{\text{1L}}(h_j\to VV)&= (\calM_{\text{1L}}^{(1)} + \calM_{\text{1L}}^{\text{CT}}
) g^{\mu\nu}+\calM_{\text{1L}}^{(2)} k_1^\nu k_2^{\mu} +i\calM_{\text{1L}}^{(3)}
\varepsilon^{\mu\nu\rho\sigma}k_{1\rho} k_{2\sigma}.
\end{align}
The genuine one-loop triangle diagram contributions are denoted by
$\calM_{\text{1L}}^{(i)}$ ($i=1,2,3$) and the  counterterm contribution by
$\calM_{\text{1L}}^{\text{CT}}$. The term $\calM_{\text{1L}}^{(3)}$
vanishes in the CP-conserving case. \s

The decay width for the decay $H_i \to VV$, including the NLO corrections, is given by
\be 
\Gamma(H_i\to VV) = R\braket{\Gamma^{\tree}(H_i\to
  VV)+\Gamma^{\text{1L}}(H_i\to VV) }, 
\ee
where $R=1/2$ for $V=Z$ and $R=1$ for $V=W$. The improved tree-level
decay width reads
\newcommand{\rv}{r_V}
\beq 
\Gamma^{\tree}(H_i\to VV)= \fr{\sqrt{\rv^2-\rv}M_V^2}{4 \pi M_{H_i}^3}
f_{gg} \left|\sum_{j=1}^5 \, \ZH_{ij} \calM_\tree (h_j\to VV)\right|^2 \;,
\eeq
with 
\beq 
\rv = \frac{M_{H_i}^2}{4 M_V^2}
\eeq
and
\beq
   f_{gg} =4 \rv^2 -4 \rv +3 \;.
\eeq
The NLO partial width for the $Z$-boson pair final state contains
only virtual contributions,
\be  
\Gamma^{\text{1L}}(H_i\to ZZ)= \Gamma^{\text{virt}}(H_i\to ZZ) \;.
\ee 
For the $W$-boson pair final state it contains both virtual and real
radiation contributions, 
\be  
\Gamma^{\text{1L}}(H_i\to W^+W^-)= \Gamma^{\text{virt}}(H_i\to W^+W^-)
+\Gamma^{\text{real}}(H_i\to W^+W^-\gamma) \;. 
\ee 
The virtual part is given by 
\begin{align} 
\Gamma^{\text{virt}}(H_i\to VV)& = \fr{\sqrt{\rv^2-\rv}M_V^2}{2
  \pi M_{H_i}^3} \Re\bigg\{\braket{\sum_{j=1}^5 \,
\ZH_{ij} \calM_{\text{tree}} (h_j\to VV) }
\bigg[f_{gg} \sum_{l=1}^5 \,
  (\ZH_{il})^* \times \crn
& \hspace*{-1cm}
\bigg(\calM_{\text{1L}}^{(1)*}(h_l\to VV)
  +\calM_{\text{1L}}^{\text{CT}*} (h_l\to VV)  \bigg) +M_{H_i}^2 f_{gp}  \sum_{l=1}^5 \,
  (\ZH_{il})^*\calM_{\text{1L}}^{(2)*}(h_l\to VV)\bigg] \bigg\} \;, 
\end{align} 
with 
\beq 
f_{gp} = 2\rv^2 -3 \rv +1 \;.
\eeq
The formulae for ${\cal M}_{\text{1L}}^{(1)}$ and ${\cal
  M}_{\text{1L}}^{(2)}$ are quite lengthy and we do not display them
explicitly here. The counterterm contributions for $V=W$ and $Z$, respectively, read 
\begin{align}
\calM_{\text{1L}}^{\text{CT}}(h_j\to W^+W^-) &= \calM_{\tree}(h_j\to WW)\braket{ \de Z_e + \fr{\de M_W^2}{2M_W^2} -\fr{ \de s_{\theta_W}}{s_{\theta_W}} +\de Z_W} \\
&+
\fr{e M_W}{s_{\theta_W}} \braket{-\sbeta c_\beta^2 \calR_{j1} + c_\beta^3\calR_{j2} }\de\tbeta+\fr12\sum_{l=1}^5 \de Z_{h_jh_l} \calM_{\tree}(h_l\to WW) \crn
\calM_{\text{1L}}^{\text{CT}}(h_j\to ZZ) &= \calM_{\tree}(h_j\to ZZ)\braket{ \de Z_e +
\fr{\de M_Z^2}{2M_Z^2} -\fr{ \de s_{\theta_W}}{s_{\theta_W}} -\fr{ \de
c_{\theta_W}}{c_{\theta_W}} +\de Z_Z} \\
&+
\fr{e M_Z}{s_{\theta_W} c_{\theta_W}}
  \braket{-\sbeta c_\beta^2 \calR_{j1} + c_\beta^3\calR_{j2}
  }\de\tbeta+\fr12\sum_{l=1}^5 \de Z_{h_jh_l} \calM_{\tree}(h_l\to ZZ)
  \;. \nonumber
\end{align}

For the decay $H_i\to W^+W^-$ we have to include to the contribution
from the radiation of a real photon in order to get an
infrared-finite result. This contribution is given by 
\be
\Gamma^{\text{real}}(H_i\to W^+W^-\gamma)= R_3\fr{e^2}{64\pi^3
  M_{H_i}} \Big|\sum_{j=1}^5\ZH_{ij} \calM_{\tree} (h_j\to
  VV) \Big|^2 \;, 
\ee
where 
\begin{align}
R_3= &\fr{1}{M_W^2}\bigg[4 M_W^2 (- 4 r^2_W +4 r_W -3) (I_{1}+I_2+M_W^2 I_{11} +M_W^2 I_{22}) + 4 I_1^2 + 4 I_2^1\crn
& + 2I_{11}^{22}+ 2 I_{22}^{11}+ 4I+ 8 M_W^4 ( 8 r^3_W -12 r_W +10 r_W
  -3 ) I_{12}  \bigg] \;. 
\end{align}
The formula is in agreement with the result given
  in \bib{Kniehl:1991xe}. 
Here, we have neglected the arguments of the Bremsstrahlung integrals
$I_{i_1,\ldots,i_2}^{j_1,\ldots,j_2}(M_{H_i},M_W,M_W)$ for the sake of
readability. In terms of the photon momentum $q$, the $W^+$ momentum
$k_1$ and the $W^-$ momentum $k_2$, these integrals are defined as
\be 
I_{i_1,\ldots,i_n}^{j_1,\ldots,j_m}(M_{H_i},M_W,M_W)=
\fr 1 {\pi^2}\int \fr{d^3k_1}{2k_{10}} 
 \fr{d^3k_2}{2k_{20}} \fr{d^3q}{2q_{0}} \delta(k_0-k_1-k_2-q)
\fr{(\pm 2qk_{j_1})\ldots(\pm 2qk_{j_m})}{(\pm 2qk_{i_1}) \ldots(\pm 2 qk_{i_n})},  
\ee 
where $k_0$ is the four-momentum of $H_i$ and
$i_l,j_k=0,1,2$. The plus signs
  correspond to $k_1$, $k_2$, the minus sign to $k_0$. Their analytic
expressions are given in \bib{Denner:1991kt}. \s

We have checked the UV finiteness of the NLO decay widths of both the
$H_i\to ZZ$ and the $H_i\to W^+W^-$ decays. 
The IR divergence in the decay $H_i\to W^+W^-$, however, is more demanding.
At strict one-loop level, \ie one must use the tree-level Higgs boson
mass for the external Higgs boson and the unity WFR factor matrix, the
IR finiteness if fulfilled. However, the use of the loop-corrected
Higgs masses and the WFR factors $\ZH_{ij}$ breaks the IR
finiteness, because different orders of perturbation theory
are mixed in this case. At tree level there exists a relation between
the  coupling of the neutral Higgs boson with the charged Goldstone
bosons and the coupling of the neutral Higgs boson with the $W$
bosons. Defining the Lagrangian for the interaction between Higgs and
Goldstone bosons by
\beq
{\cal L}_{h_j G^+ G^-} = g_{h_j G^+ G^-} h_j G^+ G^- + {\it h.c.} \;,
\eeq
it is given by
\be 
g_{h_j G^+G^-} = -\fr{m_{h_j}^2}{g v M_W} (c_\beta {\cal R}_{j1} +
s_\beta {\cal R}_{j2} ) g_{h_j W^+ W^-} \;,
\ee
with the tree-level Higgs boson mass $m_{h_j}$. In order to obtain
an IR-finite result while using the loop-corrected mass $M_{H_i}$, we
chose to modify the coupling $g_{h_j G^+G^-}$ as
\be 
g_{h_j G^+G^-} =-\fr{M_{H_i}^2}{g v M_W} (c_\beta {\cal R}_{j1} +
s_\beta {\cal R}_{j2} ) g_{h_j W^+ W^-}\;,
\ee
where the tree-level mass $m_{h_j}^2$ has been replaced by the
loop-corrected mass $M_{H_i}^2$ of the decaying Higgs boson $H_i$.
We verified that the modification of this coupling ensures IR
finiteness while not affecting UV finiteness. The same method has been
used in \bib{Gonzalez:2012mq}. While taking the
  loop-corrected mass for the external decaying Higgs boson ensures
  compatibility with the observation of a 125 GeV SM-like Higgs
  boson, this approach breaks gauge invariance, however. For more
  details on this issue, we refer to an upcoming
  publication~\cite{lukasetal}. \s

For the non-SM-like neutral Higgs
bosons, the tree-level coupling $g_{h_iVV}$ 
is in general suppressed, in particular in case the
  Higgs boson with mass 125.09~GeV behaves very SM-like. In this
case, the one-loop corrected decay width 
$\Gamma^{\text{1L}}(H_i\to VV) $ can be even larger than the
tree-level improved one $\Gamma^{\tree}(H_i\to
VV)$. This becomes a problem when the
one-loop correction is negative, as then the one-loop corrected partial
decay width becomes negative. In this case, we have to include the
one-loop squared term, which is formally of higher order. 
For the decay $H_i\to ZZ$, we will include the one-loop squared
contribution.\footnote{Note, however, that the thus
    obtained result has to be taken with caution. The complete
    two-loop calculation contributes further terms that might cause
    the complete two-loop result to differ considerably from the
    result obtained in the here applied approximation. Moreover, the
    inclusion of (part of) the two-loop corrections explicitly
    includes a dependence on the renormalization scheme chosen at
    one-loop order that would need to be cancelled by transforming the
    input parameters appropriately so as not to become
    inconsistent. We still use this
    approach in order to obtain physical, {\it i.e.}~positive, partial decay widths and hence
    physical branching ratios. Since the partial decay width is suppressed
    here anyway, the effect of the difference between the approximation
    and the full two-loop result on the branching ratio is expected to be
    subleading. Still, the code {\tt NMSSMCALC} will print out a
    warning to make the user aware of this issue.} 
In particular, the decay width now is given by 
\bea 
\Gamma(H_i\to ZZ) = \fr12\braket{\Gamma^{\tree}(H_i\to
  ZZ)+\Gamma^{\text{1L}}(H_i\to ZZ) + \Gamma^{\text{1Ls}}(H_i\to ZZ) }, \eea
where
\bea  \Gamma^{\text{1Ls}}(H_i\to ZZ)&=& \fr{\sqrt{\rv^2-\rv}M_V^2}{4 \pi M_{H_i}^3}\bigg\{
 \braket{\sum_{j=1}^5 \ZH_{ij} \calM_{\text{1L}}^{(1)+\text{CT}}
   (h_j\to ZZ) }\label{eq:olsquared} \\
&&\times\braket{f_{gg} 
\sum_{l=1}^5 \ZH_{il} \calM_{\text{1L}}^{(1)+\text{CT}} (h_l\to ZZ) 
+ M_{H_i}^2f_{gp} \sum_{l=1}^5 \ZH_{il} \calM_{\text{1L}}^{(2)} (h_l\to ZZ)  }^*\crn
&&+M_{H_i}^4(\rv-1)^2 \Big( \sum_{j=1}^5 \ZH_{ij}
  \calM_{\text{1L}}^{(2)} (h_j\to ZZ) \Big)  \braket{\sum_{l=1}^5 \ZH_{il} \calM_{\text{1L}}^{(2)} (h_l\to ZZ)}^*\crn
&& +M_{H_i}^4\fr{\rv-1}{\rv} \Big( \sum_{j=1}^5 \ZH_{ij}
\calM_{\text{1L}}^{(3)} (h_j\to ZZ) \Big) 
\braket{\sum_{l=1}^5 \ZH_{il}
  \calM_{\text{1L}}^{(3)} (h_l\to ZZ)}^*\bigg\} \;, \nonumber
\eea
where $\calM_{\text{1L}}^{(1)+\text{CT}}$ is the sum of $\calM_{\text{1L}}^{(1)}+
\calM_{\text{1L}}^{\text{CT}}$.
For the decay $H_i\to W^+W^-$, the form factor $\calM_{\text{1L}}^{(1)+\text{CT}}$ contains 
IR divergences so that we cannot treat it in the same way as in the decay
$H_i\to ZZ$. Note, however, that 
the 1L decay width $\Gamma^{\text{1L}}(H_i\to WW)$ can always be
divided into three parts that are separately UV and IR
  finite: the (s)fermion contribution arising from loops containing
SM model fermions and their superpartners, the chargino/neutralino
contribution from loops with internal charginos and neutralinos 
and the gauge/Higgs contribution from loops with gauge and Higgs
particles. In many cases the dominant contribution is the (s)fermion
part. That was also observed in the MSSM
case~\cite{Hollik:2011xd,Gonzalez:2012mq}. We therefore add to the
one-loop corrected decay with $H_i \to W^+ W^-$ the one-loop squared
contribution from the (s)fermion part only. This part is IR finite as
it solely involves fermions and sfermions but no
photons. Both for the decays into $ZZ$ and into $WW$
  we include the respective one-loop squared terms in case the one-loop
  contribution is larger than 80\% of the tree-level decay width.

\subsection{Higgs Boson Decays into a \boldmath $Z$ Boson and a Higgs Boson}
The one-loop corrected amplitude for the decay of a heavy Higgs boson
$H_i$ with four-momentum $p$ into a light Higgs boson $H_j$  
and a $Z$ boson, with four-momenta $k_1$ and $k_2$, respectively, $H_i(p)\to H_j(k_1) Z(k_2)$, can be written as 
\be 
 M_{\text{1L}}(H_i\to H_j Z)
 =\epsilon_\mu(k_2)p^\mu \calM_{\text{1L}}(H_i\to
 H_j Z)  \;,
\ee
where 
\be  \calM_{\text{1L}}(H_i\to H_j Z)=
\sum_{k^\prime,l^\prime=1}^5
\ZH_{ik^\prime}\ZH_{jl^\prime}
\braket{ \calM^{(0)}_{h_{k^\prime}
    h_{l^\prime} Z}+
    \calM^{(1)}_{h_{k^\prime}h_{l^\prime} Z} +
    \calM^{GZ,\text{mix}}_{h_{k^\prime}h_{l^\prime} Z}  }\;.
\ee
The tree-level expression $\calM^{(0)}_{h_{i} h_{j} Z}$ reads
\be 
\calM^{(0)}_{h_{i} h_{j} Z} =\fr{e}{s_{\theta_W}c_{\theta_W}}\braket{\sbeta\braket{\calR_{i4}\calR_{j1} -
\calR_{i1}\calR_{j4} } + \cbeta\braket{-\calR_{i4}\calR_{j2} +
\calR_{i2}\calR_{j4} } } \,, \label{eq:treeampHHZ}
\ee
and the one-loop term $\calM^{(1)}_{h_{i}h_{j} Z}$ consists of the
genuine one-loop diagram contribution and the counterterm part 
given by 
\beq 
\calM^{\text{CT}}_{h_{i}h_{j} Z}
& =& \fr12\sum_{i^\prime =1}^5 \braket{ \de
  Z_{h_ih_{i^\prime}}\calM^{(0)}_{h_{i^\prime} h_{j} Z}+ \de
  Z_{h_jh_{i^\prime}}\calM^{(0)}_{h_{i} h_{i^\prime} Z} } +\fr 12
\calM^{(0)}_{h_{i} h_{j} Z}\de Z_Z  
\crn
&& + \calM^{(0)}_{h_{i} h_{j} Z}\braket{\de Z_e - \fr{c_{2\theta_W}
    \de s_{\theta_W} }{s_{\theta_W} c_{\theta_W}^2} } \;.  
\eeq
Also here, the contribution from the one-loop diagrams
with the transition $h_i \to Z(G)$, ${\cal M}^{GZ,\text{mix}}_{h_i h_j Z}$,
is calculated in the unitary gauge. The improved tree-level decay
width is given by
\be 
\Gamma^{\tree}(H_i\to H_j Z)= R_{HHZ} \left|
  \sum_{i^\prime,j^\prime=1}^5
  \ZH_{ii^\prime}\ZH_{jj^\prime}\calM^{(0)}_{h_{i^\prime}
    h_{j^\prime} Z}
\right|^2  
\ee
and the NLO decay width by 
\beq
\Gamma^{\text{NLO}}(H_i\to H_j Z)&=&R_{HHZ}
\abs{\calM_{\text{1L}}  (H_i\to H_j Z)}^2 \;,
\label{eq:1lgaugehiggs}
\eeq
with the 2-particle phase-space factor
\be R_{HHZ}
=\fr{\lambda^{3/2}(M_{H_i}^2,M_{H_j}^2,M_Z^2)}{64 \pi
  M_{H_i}^3 M_Z^2} \;,\quad  
\lambda(x, y, z)= x^2+y^2+z^2-2 xy-2xz-2yz \;.
\ee
Since the formulae for the 1-loop amplitudes are
  quite lengthy we do not display them explicitly here.
Note that, as in the decay into massive gauge bosons, in
\eqref{eq:1lgaugehiggs} we also included, keeping in mind the
  caveat mentioned there, one-loop contributions
squared as the one-loop corrections can be large and negative. 

\subsection{Higgs Boson Decays into Neutralinos and Charginos}
\label{sec:Hchaneu}
The couplings of a neutral Higgs boson $h_i$ with the electroweakinos
can be defined as
\be
-i\fr{e}{2s_{\theta_W}}\braket{g_{h_i\ti\chi_j\ti\chi_k}^L  P_L +
   g_{h_i\ti\chi_{k}\ti\chi_{j}}^R P_R} \;, 
\ee
where $\ti\chi$ stands generically for the neutralinos and
charginos and
  $g_{h_i\ti\chi_{k}\ti\chi_{j}}^R=\braket{g_{h_i\ti\chi_{k}\ti\chi_{j}}^L}^*$. At
tree level, the left- and right-handed coefficients  
for the Higgs-chargino couplings are given in terms of~\cite{Baglio:2013iia} 
\begin{align}
g_{h_i\ti\chi_j^+\ti\chi_k^-}^L& =\fr{\lambda v(\calR_{i3}+
                                 i\calR_{i5})U^*_{j2}V_{k2}^*e^{i\varphi_s}} 
{\sqrt{2}M_Zc_{\theta_W}} + \sqrt{2} U_{j1}^*V_{k2}^{*} e^{-i\varphi_u}(\calR_{i2} -
i c_\beta \calR_{i4}) \nonumber \\
& + \sqrt{2} U_{j2}^* V_{k1}^*(\calR_{i1} - i\sbeta \calR_{i4}) \;,
\end{align}
where $i=1,\ldots,5$, $j,k=1,2$, and for the Higgs-neutralino couplings we
have
\begin{align}
g^L_{h_i\ti\chi_l^0\ti\chi_m^0}&= \bigg[ \fr{1}{c_{\theta_W} }N_{l3}^*\braket{c_{\theta_W} N^*_{m2} - s_{\theta_W} N_{m1}^* } \braket{\calR_{i1} - i\sbeta\calR_{i4}} -\fr{1}{c_{\theta_W}} N_{l4}^*
\braket{c_{\theta_W} N^*_{m2} - s_{\theta_W} N^*_{m1}} e^{-i\varphi_u} \crn
& \braket{\calR_{i2} - i \cbeta 
\calR_{i4} }- \fr{\lambda v N^*_{l5}N_{m3}^*e^{i\varphi_u} \braket{\calR_{i2} +i \cbeta \calR_{i4}} }
{\sqrt{2} M_Z c_{\theta_W} } - \fr{\lambda v N^*_{l5}N_{m4}^* \braket{\calR_{i1} +i \sbeta \calR_{i4}} }
{\sqrt{2} M_Z c_{\theta_W} } \crn
& + \fr{v(\calR_{i3} + i\calR_{i5}) e^{i\varphi_s}(2\kappa N_{l5}^*
N^*_{m5} -\lambda N^*_{l4} N^*_{m3}) }{\sqrt{2} M_Z c_{\theta_W}} +
  l\leftrightarrow m \bigg] \;,  
\end{align}
with $l,m=1,\ldots,5$.
The decay width for the decay of a Higgs boson $H_i$ into a neutralino
pair or a chargino pair including higher order corrections is given by 
\be 
\Gamma(H_i\to \ti\chi_j \ti\chi_k) = R 
\braket{\Gamma^{\tree}(H_i\to \ti\chi_j
  \ti\chi_k)+\Gamma^{\text{1L}}(H_i\to \ti\chi_j \ti\chi_k) } ,
\ee
where $R=1/2$ for identical final states and $R=1$ otherwise. 
The improved tree-level decay width reads
\beq  \Gamma^{\tree}(H_i\to \ti\chi_j \ti\chi_k)&=&  R_2(M_{H_i}^2,M_{\ti\chi_j}^2,M_{\ti\chi_k}^2) 
\bigg[\braket{M_{H_i}^2- M_{\ti\chi_j}^2- M_{\ti\chi_k}^2} \braket{\abs{\calM^{L,0}_{H_i\ti\chi_j \ti\chi_k}}^2 +\abs{\calM^{R,0}_{H_i\ti\chi_j \ti\chi_k} }^2 } \crn
&& - 4 M_{\ti\chi_j}M_{\ti\chi_k}\Re[
\calM^{L,0}_{H_i\ti\chi_j \ti\chi_k} \calM^{R,0*}_{H_i\ti\chi_j \ti\chi_k}]  \bigg],  \eeq
where the 2-body phase space factor is 
\beq 
R_2(x,y,z)&=&  \fr{\lambda^{1/2}(x, y, z)}{16 \pi x^{3/2}} \;, \label{eq:2bodyps}
\eeq
in terms of the improved tree-level amplitude
\be 
\calM^{L/R,0}_{H_i\ti\chi_j \ti\chi_k} = \sum_{i^\prime=1}^5\ZH_{ii\prime}
\fr{e}{2s_{\theta_W}} g_{h_{i^\prime}\ti\chi_j\ti\chi_k}^{L/R} \;. 
\ee
The one-loop decay width for the decay into a neutralino pair is given by
\be 
\Gamma^{\text{1L}}(H_i\to \ti\chi_j^0 \ti\chi_k^0)= \Gamma^{\text{virt}}(H_i\to
\ti\chi_j^0 \ti\chi_k^0)
\ee
and for the decay into a chargino pair it is
\be 
\Gamma^{\text{1L}}(H_i\to \ti\chi_j^+ \ti\chi_k^-)=
\Gamma^{\text{virt}}(H_i\to \ti\chi_j^+ \ti\chi_k^-)+\Gamma^{\text{real}}(H_i\to
\ti\chi_j^+ \ti\chi_k^-) \;,
\ee
where the virtual contribution can be cast into the form
\begin{align}  
\Gamma^{\text{virt}}(H_i\to \ti\chi_j \ti\chi_k)&=  2
R_2(M_{H_i}^2,M_{\ti\chi_j}^2,M_{\ti\chi_k}^2)  \times
\crn
& \bigg[\braket{M_{H_i}^2- M_{\ti\chi_j}^2- M_{\ti\chi_k}^2} \Re\bigg(\calM^{L,0}_{H_i\ti\chi_j
\ti\chi_k} \calM^{L,1*}_{H_i\ti\chi_j \ti\chi_k}
  +\calM^{R,0}_{H_i\ti\chi_j \ti\chi_k} \calM^{R,1*}_{H_i\ti\chi_j
  \ti\chi_k}\bigg)  \crn
& - 2 M_{\ti\chi_j}M_{\ti\chi_k}\Re[\calM^{L,0}_{H_i\ti\chi_j \ti\chi_k}
  \calM^{R,1*}_{H_i\ti\chi_j 
  \ti\chi_k} +\calM^{R,0}_{H_i\ti\chi_j \ti\chi_k} \calM^{L,1*}_{H_i\ti\chi_j \ti\chi_k}] \bigg],  
\end{align}
with the left- and right-handed one-loop amplitudes containing genuine
triangle, counterterm and '$GZ,\text{mix}$' contributions,
\be 
\calM^{L/R,1}_{H_i\ti\chi_j \ti\chi_k} = \sum_{i^\prime=1}^5\ZH_{ii^\prime}
\braket{ \calM^{L/R,\Delta}_{h_{i^\prime}\ti\chi_j
    \ti\chi_k}+\calM^{L/R,\text{CT}}_{h_{i^\prime}\ti\chi_j \ti\chi_k}  
+\calM^{L/R,GZ,\text{mix}}_{h_{i^\prime}\ti\chi_j \ti\chi_k}} . 
\ee

We do not display explicitly here the lengthy
  expressions for the triangle and '$GZ, \text{mix}$' contributions. 
The explicit expressions of the counterterm amplitudes for
the decays into neutralinos are
\begin{align}
\calM^{L,\text{CT}}_{h_{i}\ti\chi_j^0 \ti\chi_k^0}& =\fr{e}{4s_{\theta_W}}\bigg[
\sum_{i^\prime=1,5}g_{h_{i^\prime}\ti\chi_j^0\ti\chi_k^0}^{L}\de Z_{h_i h_{i^\prime}}+
\sum_{j^\prime=1,5}g_{h_{i}\ti\chi_{j^\prime}^0\ti\chi_k^0}^{L}\de Z^{\ti\chi^0}_{L,j^\prime j } 
+\sum_{k^\prime=1,5}g_{h_{i}\ti\chi_{j}^0\ti\chi_{k^\prime}^0}^{L}\de\bar Z^{\ti\chi^0}_{L,kk^\prime }
\bigg]\crn
&+\fr{e}{2s_{\theta_W}}
  g_{h_{i}\ti\chi_{j}^0\ti\chi_{k}^0}^{L}\braket{\de Z_e - \fr{\de
  s_{\theta_W}}{s_{\theta_W}} } +\fr{e}{2s_{\theta_W}} \bigg[
  -N_{l3}^* 
  N_{m1}^*\de t_{\theta_W} \braket{\calR_{i1} - i\sbeta\calR_{i4}}  
\crn
&+N_{l4}^*
  N^*_{m1}
  e^{-i\varphi_u} \de t_{\theta_W} \braket{\calR_{i2} - i \cbeta  
\calR_{i4} } -\fr{\lambda v}{\sqrt{2} M_Z c_{\theta_W} } \big( N^*_{l5}N_{m3}^*e^{i\varphi_u}\crn
&  
  \braket{\calR_{i2} +i \cbeta \calR_{i4}}+ N^*_{l5}N_{m4}^* \braket{\calR_{i1} +i \sbeta \calR_{i4}} +(\calR_{i3} + i\calR_{i5}) e^{i\varphi_s} N^*_{l4} N^*_{m3}\big) \crn
& \times \braket{\fr{\de v}{v} +\fr{\de\lambda}{\lambda} - \fr{\de M_Z^2}{2M_Z^2}
 -\fr{\de c_{\theta_W}}{c_{\theta_W}} }
 + \fr{\sqrt{2}v\kappa(\calR_{i3} + i\calR_{i5}) e^{i\varphi_s}N_{l5}^*
N^*_{m5} }{ M_Z c_{\theta_W}} \crn
& \times \braket{\fr{\de v}{v} +\fr{\de\kappa}{\kappa} - \fr{\de M_Z^2}{2M_Z^2}
 -\fr{\de c_{\theta_W}}{c_{\theta_W}} }  + l\leftrightarrow m \bigg] \;,
\end{align} 
and for the decays into charginos they read
\begin{align}
\calM^{L,\text{CT}}_{h_i\ti\chi_j^+ \ti\chi_k^-}& = \fr{e}{4 s_{\theta_W}}\bigg[ 
\sum_{i^\prime=1,5}g_{h_{i^\prime}\ti\chi_j^+\ti\chi_k^-}^{L}\de Z_{h_i h_{i^\prime}}+
\sum_{j^\prime=1,2}g_{h_{i}\ti\chi_{j^\prime}^+\ti\chi_k^-}^{L}\de Z^{\ti\chi^+}_{L,j^\prime  j } +\sum_{k^\prime=1,2}g_{h_{i}\ti\chi_{j}^+\ti\chi_{k^\prime}^-}^{L}\de \bar Z^{\ti\chi^+}_{L,kk^\prime }\bigg]\crn
&+ \fr{e}{\sqrt{2}s_{\theta_W}}\braket{U_{j1}^*V_{k2}^{*} e^{-i\varphi_u}(\calR_{i2} -
i c_\beta \calR_{i4}) +  U_{j2}^* V_{k1}^*(\calR_{i1} - i\sbeta \calR_{i4})} \braket{\de Z_e - \fr{\de s_{\theta_W}}{s_{\theta_W}} }\crn
&
+\fr{ U^*_{j2}V_{k2}^*e^{i\varphi_s}(\calR_{i3}+ i\calR_{i5})}
{\sqrt{2}}\de\lambda \;.
\end{align} 
The right-handed counterterm amplitudes are equal to the complex
conjugate of the corresponding left-handed parts after
interchanging the indices of the charginos and neutralinos in the
final state. The real photon contribution for the decays into a
chargino pair is expressed in terms of the Bremsstrahlung integrals as 
\begin{align} 
\Gamma^{\text{real}}(H_i\to \ti\chi_j^+ \ti\chi_k^-) &= \fr{e^2}{32\pi^3M_{H_{i}}}\bigg\{ 2  \braket{\abs{\calM^{L,0}_{H_i\ti\chi_j \ti\chi_k}}^2+\abs{\calM^{R,0}_{H_i\ti\chi_j \ti\chi_k}}^2} \braket{ I_2^1+I_1^2+2 I }
\crn &-   \braket{I_1 + I_2 - 2 ( M_{H_i}^2 - 
M_{\ti\chi_j^+}^2 -M_{\ti\chi_k^-}^2) I_{12} + 2 M_{\ti\chi_j^+}^2 I_{11}+  2 M_{\ti\chi_k^-}^2 I_{22} } 
\crn & \times  \bigg[\braket{M_{H_i}^2- M_{\ti\chi_j^+}^2- M_{\ti\chi_k^-}^2} \braket{\abs{\calM^{L,0}_{H_i\ti\chi_j \ti\chi_k}}^2 +\abs{\calM^{R,0}_{H_i\ti\chi_j \ti\chi_k} }^2 } \crn
& - 4 M_{\ti\chi_j^+}M_{\ti\chi_k^-}\Re[
\calM^{L,0}_{H_i\ti\chi_j \ti\chi_k} \calM^{R,0*}_{H_i\ti\chi_j
  \ti\chi_k}]  \bigg]\bigg\} \;,
 \end{align}
where the arguments of the Bremsstrahlung integrals
$I_{i_1,\ldots,i_n}^{j_1,\ldots,j_m}(M_{H_i},M_{\ti\chi^+_j}, 
M_{\ti\chi^-_k})$ have been neglected. Note that we use the
loop-corrected masses for the external Higgs boson and 
the external charginos and neutralinos in the tree-level, virtual and
real contributions. However, for particles inside loops we use the
tree-level masses and tree-level couplings. This does not affect the
UV-finiteness but can break the IR-finiteness in 
the decay into a pair of charginos. We overcome this problem by
replacing the tree-level mass of the chargino in the loop diagrams
with a photon by the corresponding loop-corrected chargino mass. Our
treatment is different from \bib{Goodsell:2017pdq} where the authors define an 
IR divergent counterterm to cancel
the mismatch between the real and virtual contributions. Note finally,
that in case the NLO decay width into neutralino final states becomes
negative, the improved tree-level decay width is calculated instead in
{\tt NMSSMCALCEW}, including the $\ZH$ factor.

\subsection{Higgs Boson Decays into Squark Pairs}
The NLO corrections to the decay of a  neutral Higgs boson into a
squark-antisquark pair consist of the QCD and EW corrections. In the
CP-conserving NMSSM, the NLO corrections to the decay of a CP-odd
Higgs boson into a stop pair have been calculated and discussed in
\bib{Baglio:2015noa}. We extend this computation to the CP-violating
case and include also the decay into a sbottom-antisbottom pair in
this paper. The NLO QCD corrections are positive
and large. They can be larger than 100\% as 
 observed in \bib{Baglio:2015noa} while the EW correction are
 negative and can be of up to $-40\%$. In our calculation,
 we have implemented both the OS and the $\DRb$ scheme.
We have three options here. First, the seven
parameters are renormalized in the OS scheme. Second, the parameters of
the stop sector, $ m_t, m_{\tilde{Q}_3},m_{\tilde{t}_R},A_t$, are
renormalized in the OS scheme while the remaining parameters,
$m_b,m_{\tilde{b}_R},A_b$, are renormalized in the $\overline{\mbox{DR}}$ scheme.
Third, all parameters are renormalized in the $\overline{\mbox{DR}}$ scheme. The
 loop-corrected decay width is decomposed into the  
improved tree-level, one-loop QCD and one-loop EW decay widths,
 \be 
\Gamma(H_i\to \ti q_j \ti q_k^*) = 
\Gamma^{\tree}(H_i\to \ti q_j \ti
q_k^*)+\Gamma_{\text{QCD}}^{(1)}(H_i\to \ti q_j \ti
q_k^*)+\Gamma_{\text{EW}}^{(1)}(H_i\to \ti q_j \ti q_k^*) \;.
\ee
Denoting the color factor by $N_F$, with $N_F=3$, the improved
tree-level decay width is given by 
\beq  \Gamma^{\tree}(H_i\to \ti q_j \ti q_k^*) = N_F R_2(M_{H_i}^2,M_{\ti q_j}^2,M_{\ti q_k}^2) 
\abs{\calM^{0}_{H_i\ti q_j \ti q_k^*}}^2,
\eeq
in terms of the improved tree-level amplitude
\be 
\calM^{0}_{H_i\ti q_j \ti q_k^*} = \sum_{i_1=1}^5
\ZH_{ii'} g_{h_{i'} \ti q_j \ti
  q_k^*} \;.
\ee   
The tree-level Higgs-squark-squark couplings are given by~\cite{Baglio:2013iia}
\begin{align}
g_{h_{i} \ti t_{j} \ti t_{k}^*}&= \fr{eM_Z} {s_{\theta_W}c_{\theta_W}}\bigg[ \fr{m_t^2 \calR_{i2} \braket{U_{j1}^{\ti t*}
U_{k1}^{\ti t} +U_{j2}^{\ti t*} U_{k2}^{\ti t}  } }{\sbeta M_Z^2} +\fr{m_t \braket{ U_{j2}^{\ti t*}
U_{k1}^{\ti t} F_1 +U_{j1}^{\ti t*} U_{k2}^{\ti t} F_1^* } }{2 \sbeta M_Z^2}\crn
&+ \fr 16 \braket{\cbeta \calR_{i1}  - \sbeta \calR_{i2}}\braket{ (4c_{\theta_W}^2 - 1 )U_{j1}^{\ti t*} U_{k1}^{\ti t} +
4 s_{\theta_W}^2 U_{j2}^{\ti t*} U_{k2}^{\ti t} } \bigg],\\
g_{h_{i} \ti b_{j} \ti b_{k}^*}&= \fr{eM_Z} {s_{\theta_W}c_{\theta_W}}\bigg[ \fr{m_b^2 \calR_{i1} \braket{U_{j1}^{\ti b*}
U_{k1}^{\ti b} +U_{j2}^{\ti b*} U_{k2}^{\ti b}  } }{\cbeta M_Z^2} +\fr{m_b \braket{ U_{j2}^{\ti b*}
U_{k1}^{\ti b} F_2 +U_{j1}^{\ti b*} U_{k2}^{\ti b} F_2^* } }{2 \cbeta M_Z^2}\crn
&- \fr 16 \braket{\cbeta \calR_{i1}  - \sbeta \calR_{i2}}\braket{ (2c_{\theta_W}^2 + 1 )U_{j1}^{\ti b*} U_{k1}^{\ti b} +
2 s_{\theta_W}^2 U_{j2}^{\ti b*} U_{k2}^{\ti b} } \bigg],
\end{align}
with
\begin{align}
F_1&=  A_t^* e^{-i\varphi_u} \braket{ \calR_{i2}- i\cbeta\calR_{i4} } -\mueff \braket{\calR_{i1}  + i\sbeta\calR_{i4}}
-\fr{\lambda v \cbeta \braket{\calR_{i3} + i\calR_{i5} }
     e^{i\varphi_s} }{\sqrt{2} } \\ 
F_2&=  A_b^*  \braket{ \calR_{i1}- i\sbeta \calR_{i4} } -\mueff
     e^{i\varphi_u} \braket{\calR_{i2}  +
     i\cbeta\calR_{i4}} 
-\fr{\lambda v \sbeta \braket{\calR_{i3} + i\calR_{i5} }
     e^{i(\varphi_s +\varphi_u )}} {\sqrt{2} } \;.
\end{align}

The one-loop QCD and EW contributions to the decay width  are given by
the sum of the virtual and real contributions, respectively, 
\be  
\Gamma^{(1)}_{\text{QCD/EW}}(H_i\to \ti q_j \ti q_k^*) =
\Gamma^{\text{virt}}_{\text{QCD/EW}}(H_i\to \ti q_j \ti q_k^*)
+\Gamma^{\text{real}}_{\text{QCD/EW}}(H_i\to \ti q_j \ti q_k^*
g/\gamma) \;.
\ee 
For the virtual QCD contribution we have
\begin{eqnarray} 
\Gamma^{\text{virt}}_{\text{QCD}}(H_i\to \ti q_j \ti q_k^*) &=& N_F
R_2(M_{H_i}^2,M_{\ti q_j}^2,M_{\ti q_k}^2) \times \nonumber \\ 
&& 2 \Re\sbraket{\calM^{0*}_{H_i\ti q_j \ti q_k^*} \braket{
    \sum_{i^\prime=1}^5 \ZH_{ii^\prime}(\calM^{\Delta
      ,\text{QCD}}_{h_{i^\prime}\ti q_j \ti q_k^*}  
+\calM^{\text{CT,QCD} }_{h_{i^\prime}\ti q_j \ti q_k^*}) } } \;,
\end{eqnarray}
with the 2-body phase space factor $R_2$ defined in Eq.~(\ref{eq:2bodyps}).
The expression for the virtual EW contribution is different from the
QCD one due to an extra contribution containing the transition $h_i\to
G,Z$. Explicitly, we have
\beq 
\Gamma^{\text{virt}}_{\text{EW}}(H_i\to \ti q_j \ti q_k^*) &=&  N_F
R_2(M_{H_i}^2,M_{\ti q_j}^2,M_{\ti q_k}^2)  \times \nonumber \\
&& 2 \Re\bigg[ \calM^{0*}_{H_i\ti q_j \ti q_k^*}  \sum_{i^\prime=1}^5 \ZH_{ii^\prime}\bigg( \calM^{\Delta,\text{EW} }_{h_{i^\prime}\ti q_j \ti q_k^*} 
+\calM^{\text{CT,EW}}_{h_{i^\prime}\ti q_j \ti q_k^*} 
+\calM^{GZ,\text{mix} }_{h_{i^\prime}\ti q_j \ti q_k^*} \bigg)
\bigg]. 
\label{eq:EWhsqsq}\eeq
The explicit expressions for the counterterm contributions are quite
lengthy and given in \appen{appendix1}. We do not
  display, however, the more cumbersome amplitudes of the 
  virtual QCD and EW contributions, $\calM^{\Delta
      ,\text{QCD}}_{h_{i^\prime}\ti q_j \ti q_k^*} $ and
    $\calM^{\Delta,\text{EW} }_{h_{i^\prime}\ti q_j \ti q_k^*}$, respectively. \s 

The real photon radiation contribution in the EW corrections is
expressed in terms of the Bremsstrahlung integrals as 
\beq 
\Gamma^{\text{real}}_{\text{EW}}(H_i\to \ti q_j \ti q_k \gamma)& =  &\fr{ N_F}{4 \pi^2 M_{H_i} } Q_q^2\al \bigg(
 - I_1 -I_2 -  M_{\ti q_j}^2 I_{11} \crn
&& - M_{\ti q_k}^2 I_{22} +  (M_{H_i}^2 -M_{\ti q_j}^2-M_{\ti q_k}^2 ) I_{12} \bigg)\abs{\calM^{0}_{H_i\ti q_j \ti q_k^*}}^2. \eeq
As usual, we have neglected the arguments of the Bremsstrahlung integrals
$I_{l}(M_{H_i}^2,M_{\ti q_j}^2,M_{\ti q_k}^2) $
  and $I_{lm}(M_{H_i}^2,M_{\ti q_j}^2,M_{\ti q_k}^2) $ ($l,m=1,2$ and $j,k=1,2$).
The real gluon radiation contribution in the QCD corrections can be
obtained from the  EW real photon radiation contribution by replacing  
$Q_q^2\al$ with $C_F\al_s^2$, where $C_F=4/3$ for $SU(3)_C$. We have
checked the UV and IR finiteness of the EW and QCD corrections. We
have compared numerically with the NLO EW and QCD corrections in the
OS scheme for the decay $A_2\to \ti t_1\ti t_2$~\cite{Baglio:2015noa}
using their description in the real NMSSM  and found full agreement. 

\section{Numerical Results \label{sec:numerical}}
To illustrate the importance of the higher-order corrections to the
decays of the light and heavy neutral Higgs bosons and to test the
stability of the NLO results in various regions of the parameter space
we have performed a scan in the NMSSM parameter space. The parameter
points are checked against compatibility with the experimental
constraints from the Higgs data by using the programs {\tt
   HiggsBounds}5.3.2~\cite{Bechtle:2008jh,Bechtle:2011sb,Bechtle:2013wla}
and {\tt HiggsSignals}2.2.3~\cite{Bechtle:2013xfa}. These programs
require as input the effective couplings of the Higgs bosons,
normalized to the corresponding SM values, as well as the masses, the
widths and the branching ratios of the Higgs bosons. These have been
obtained for the SM and NMSSM Higgs bosons from the Fortran code {\tt
  NMSSMCALC}~\cite{Baglio:2013vya,Baglio:2013iia}. One of the neutral
CP-even Higgs bosons is identified with the SM-like Higgs boson
\textendash $\;$it will be called  $h$ from now on \textendash $\;$and
its mass is required to lie in the range
\beq
123 \mbox{ GeV } \le m_h \le 127 \mbox{ GeV} \;.
\eeq

For the SM input parameters we use the following values
~\cite{PhysRevD.98.030001,Dennerlhcnote}  
\begin{equation}
\begin{tabular}{lcllcl}
\quad $\alpha(M_Z)$ &=& 1/127.955, &\quad $\alpha^{\MSb}_s(M_Z)$ &=&
0.1181 \\
\quad $M_Z$ &=& 91.1876~GeV &\quad $M_W$ &=& 80.379~GeV  \\
\quad $m_t$ &=& 172.74~GeV &\quad $m^{\MSb}_b(m_b^{\MSb})$ &=& 4.18~GeV \\
\quad $m_c$ &=& 1.274~GeV &\quad $m_s$ &=& 95.0~MeV \\
\quad $m_u$ &=& 2.2~MeV &\quad $m_d$ &=& 4.7~MeV \\
\quad $m_\tau$ &=& 1.77682~GeV &\quad $m_\mu$ &=& 105.6584~MeV  \\
\quad $m_e$ &=& 510.9989~keV &\quad $G_F$ &=& $1.16637 \cdot 10^{-5}$~GeV$^{-2}$\,.
\end{tabular}
\end{equation} 
Concerning the NMSSM sector, we follow the SUSY Les Houches Accord
(SLHA) format~\cite{Skands:2003cj} in which the soft SUSY breaking
masses and trilinear couplings are understood as $\DRb$ parameters at
the scale
\be 
\mu_R = M_s= \sqrt{m_{\ti Q_3} m_{\ti t_R}} \;. 
\ee
This is also the renormalization scale that we use in the computation
of the higher-order corrections. Note that we chose the charged Higgs
boson mass as an OS input parameter. The computation of the ${\cal
  O}(\alpha_t\alpha_s +\alpha_t^2)$ corrections to the Higgs boson
masses is done in the $\DRb$ renormalization scheme of the top/stop
sector. We have included the contribution of the gauge parameters
$g_1,g_2$ into the conversion from pole to $\DRb$ top masses.
In Table~\ref{tab:nmssmscan} we summarize the ranges applied in our
parameter scan. In order to ensure perturbativity we apply the rough
constraint
\beq
\lambda^2 + \kappa^2 < 0.7^2 \;.
\eeq
The remaining mass parameters of the third generation sfermions that
are not listed in the table are chosen as
\beq
 \quad A_b=A_\tau= 2 \mbox{ TeV},~~ \text{and}~~m_{\tilde{\tau}_R} =
m_{\tilde{L}_3} = m_{\tilde{b}_R} = 3 \mbox{ TeV} \;.
\eeq
The mass parameters of the first and second generation sfermions are
set to
\bea   
m_{\tilde{u}_R,\tilde{c}_R} = 
m_{\tilde{d}_R,\tilde{s}_R} =
m_{\tilde{Q}_{1,2}}= m_{\tilde L_{1,2}} =m_{\tilde e_R,\tilde{\mu}_R}
= 3\;\mbox{TeV} \;.
\eea
\begin{table}
\begin{center}
\begin{tabular}{l|ccc|ccccccccccc} \toprule
& $t_\beta$ & $\lambda$ & $\kappa$ & $M_1$ & $M_2$ & $M_3$ & $A_t$ &
 $m_{\tilde{Q}_3}$ & $m_{\tilde{t}_R}$ & $M_{H^\pm}$
& $A_\kappa$ & $\abs{\mu_{\text{eff}}}$ \\
& & & & \multicolumn{9}{c}{in TeV} \\ \midrule
min & 1 & 0 & -0.7 & 0.5 & 0.5 & 1.8 & -3 &  0.6 &1 & 0.5 &
-2 & 0.2 \\
max & 20 & 0.7 & 0.7 & 1 & 1 & 2.5 & 3 & 3 & 3 & 1.5 & 2 & 1 \\ \bottomrule
\end{tabular}
\caption{Input parameters for the NMSSM scan. All parameters have been 
varied independently between the given minimum and maximum
values. \label{tab:nmssmscan}}
\end{center}
\end{table}
We have performed two scans. In the first (smaller) scan we took care
to select only such scenarios where the lightest CP-even Higgs boson
$H_1$ is singlet-like and the second lightest CP-even Higgs boson is
the SM-like Higgs boson. We refer to this scan as {\it scan1} in the
following. In the second (larger) scan, called {\it scan2} in the
following, we only retained scenarios where the SM-like Higgs boson is
the lightest CP-even Higgs boson. Both scans allow for points that
have a $\chi^2$ computed by HiggsSignals-2.2.3 that is consistent with
an SM $\chi^2$ within $2\sigma$.
All the branching ratios shown in the following have been calculated
by implementing the here presented higher-order corrections to the
various decay widths in {\tt NMSSMCALC}. In this way the new EW
corrections are combined with the state-of-the-art higher-order QCD
corrections already implemented in {\tt NMSSMCALC}. Note, however,
that the EW corrections are only taken into account if the respective
decay is kinematically allowed. Otherwise, the corresponding decay
width without the higher-order corrections discussed in this paper,
which only apply for on-shell decays, are taken into account in the
computation of the total decay width and branching ratios. 
\newcommand{\mmR}{R}
\subsection{Decays into SM Fermion Pairs}
In the old implementation in {\tt NMSSMCALC} the tree-level couplings
entering the various decay widths were improved by including loop
effects in the Higgs mixing matrix elements. Thus, the tree-level
rotation matrix $\calR$ was replaced by the loop-corrected rotation
matrix
\beq
\calR^l = {\bf Z}^H {\cal R} \;, \label{eq:approx1l}
\eeq 
evaluated at zero external momentum both at one-loop and at
two-loop order to ensure unitarity.\footnote{We remind the reader,
  that in contrast the one-loop corrected masses are obtained at
  non-vanishing external momenta and the two-loop corrections at zero
  external momenta.} The implementation here differs by the fact that
in the computation of ${\bf Z}^H$ we include the momentum dependence
at one-loop order and we do not apply the approximation of
\bib{Ender:2011qh} to deduce ${\bf Z}^H$ but proceed as described in
Eqs.~(\ref{eq:mat1})-(\ref{eq:mat3}). In the following, we call the
couplings where we apply $\calR^l$ as obtained from
Eq.~(\ref{eq:approx1l}) with zero external momentum and by applying
the approximation of \bib{Ender:2011qh} 'effective tree-level
couplings' while those with ${\bf Z}^H$ calculated according to
Eqs.~(\ref{eq:mat1})-(\ref{eq:mat3}) including the momentum dependence
at one-loop order are denominated 'improved couplings'.
\s

The decays into SM fermion pairs in the old implementation in {\tt
  NMSSMCALC} were calculated using the loop-corrected rotation matrix,
${\cal R}^{\text{l}}$, evaluated at zero external momenta and by
including the $\Delta_b$ corrections\footnote{For simplicity, we
  collectively call them $\Delta_b$ corrections although we also
  include the corresponding corrections in the decays into strange
  quarks and into leptons.} into the effective tree-level couplings,
as specified in \bib{Baglio:2013iia}. The thus obtained 'effective
couplings' are given by\footnote{We call them 'effective couplings'
  and not 'effective tree-level couplings' as they also contain the
  $\Delta_b$ corrections.}
\be 
\tilde{g}_{h_j q\bar{q}}^{\text{eff},S/P} = 
\tilde{g}_{h_j q\bar{q}}^{S/P}, \label{eq:efftree} \;.
\ee
Beyond the $\Delta_b$ approximation no further SUSY-EW nor SUSY-QCD
corrections were included. To quantify the difference between the
branching ratio computed in this paper and the old implementation in
{\tt NMSSMCALC} we introduce the relative change in the branching
ratio for the decay $H_i \to X_j X_k$ as
\be 
\Delta_{\text{BR}} (H_i X_j X_k) = 
\fr{\text{BR}_{\ZH}^{\text{SEW(+SQCD)}}(H_i\to X_j X_k)-
  \text{BR}_{{\cal R}^l}^{\tree}(H_i\to X_j X_k)} 
{\max(\text{BR}_{\ZH}^{\text{SEW(+SQCD)}}(H_i\to X_j
  X_k),\text{BR}_{{\cal R}^l}^{\tree}(H_i\to X_j X_k) 
  )} \label{eq:difference} \;,
\ee
with $X_j X_k \equiv f\bar{f}$ for the decays into fermions. 
Here the branching ratio $\text{BR}_{\ZH}^{\text{SEW(+SQCD)}}(H_i\to
f\bar f)$ means that we include the SUSY-EW corrections (and SUSY-QCD
corrections for the decays into quarks) together with the
wave-function renormalization factor into the decay width of the decay
$H_i\to f\bar f$. The formulae are given by \eqref{eq:correctedgambb}
for the decays into quarks and by \eqref{eq:correctedgamll} for the
decays into leptons together with the definitions Eqs.~(\ref{eq:amp1})
and (\ref{eq:amp2}). The branching ratio in the old implementation in
{\tt NMSSMCALC} is denoted by $\text{BR}_{{\cal R}^l}^{\tree}(H_i\to
f\bar f)$ (although it also includes the $\Delta_b$ corrections where
applicable). The quantity $\Delta_{\text{BR}}$ hence gives information
on the importance of the improvement of the branching ratios by the
$\ZH$ factor and the SEW(+SQCD) corrections.
This quantity will also be used in the investigations of the decays
into gauge boson pairs and into a pair of $Z$ and Higgs bosons.\s

\begin{figure}[hb!]
  \centering
  \includegraphics[width=0.7\textwidth]{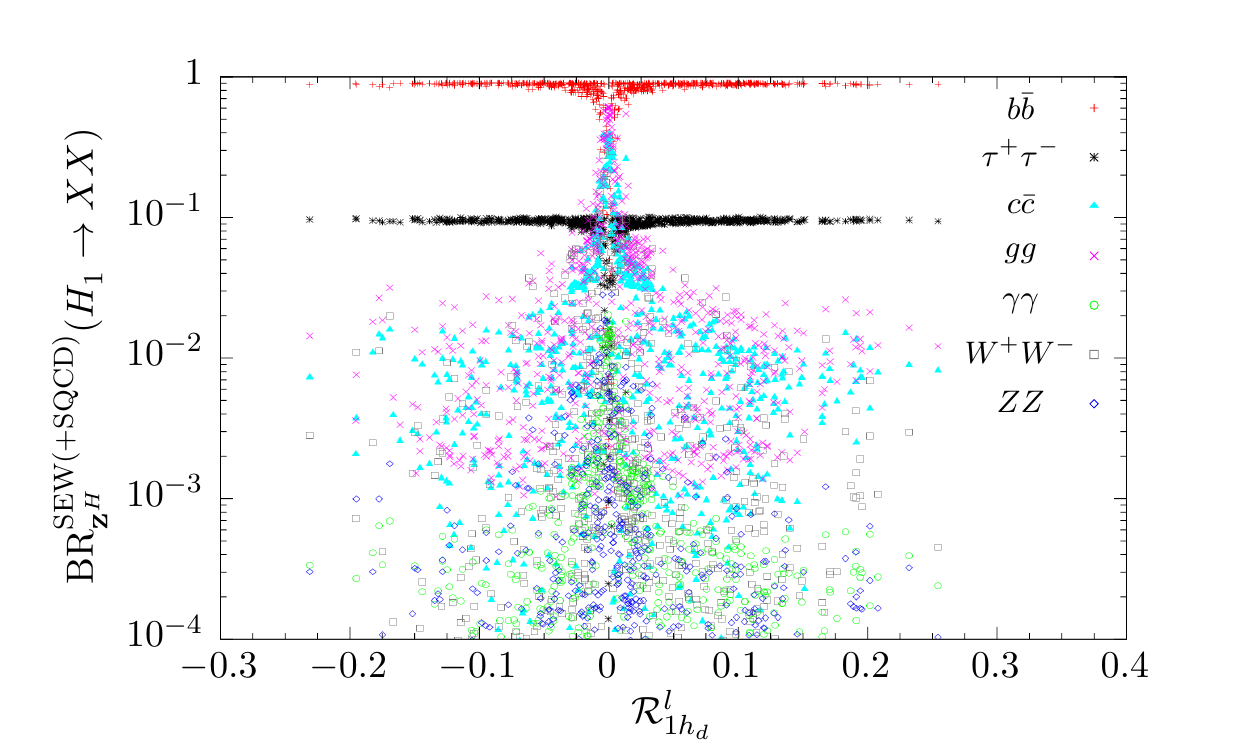}
  \caption{{\it Scan1:} Loop-corrected branching ratios of the
    lightest Higgs boson, which is the singlet-like state, into SM
    particles versus the element ${\cal R}^l_{1 h_d}$ of the
    loop-corrected Higgs rotation matrix.} 
  \label{fig:h1decayscatter1}
\end{figure}
The SM-like Higgs boson is given by the $h_u$-like Higgs
state\footnote{As the SM-like Higgs boson has to comply with the
  experimentally measured Higgs rates and for small values of
  $\tan\beta$, as preferred by the NMSSM, is dominantly produced
  through gluon fusion it needs a substantial coupling to top quarks
  so that it is the $h_u$-dominated Higgs state that turns out to be
  SM-like.}, and in our scans we found valid scenarios where this can
be the lightest or the second lightest of the CP-even Higgs bosons. We
first consider only the parameter points where the lightest CP-even
Higgs boson $H_1$ is the singlet-like state, {\it i.e.}~has a large
$h_s$ component. These are points obtained in the above described {\it
  scan1}. Here and in the following we denote a Higgs boson $H_i$ to
be dominantly $x$-like ($x=h_s,h_u,h_d,a_s,a_d$) if the corresponding
matrix element squared $|{\cal R}^l_{i x}|^2$ exceeds 80\%. When $H_1$
is $h_s$-like, the question which final state has the largest decay
width strongly depends on the amount of admixture of $h_d$ and $h_u$
components to the singlet-state. In \figref{fig:h1decayscatter1} we
show, for all parameter points that pass our constraints, the scatter
plot of the $H_1$ branching ratios into SM particles against its $h_d$
component represented by the element ${\cal R}^l_{1h_d}$ of the
loop-corrected Higgs rotation matrix\footnote{Note, that ${\cal
    R}^l_{1h_d}$ is the $(1,h_d)$-component of the mixing matrix
  ${\cal R}^l$ given by Eq.~(\ref{eq:approx1l}), evaluated at zero
  external momentum both at one- and at two-loop order and where for
  the computation of ${\bf Z}^H$ the approximation of
  \bib{Ender:2011qh} is used. In the computation of the loop-corrected
  branching ratios, however, we of course use the new implementation
  described at the beginning of this subsection.}. The mass of $H_1$
lies between 70 and 118 GeV for these points.
As can be inferred from the plot, the dominant decays are those into
$b\bar b$, $c\bar c$, $\tau\bar\tau$ and $gg$. In most cases the
branching ratio into a bottom-quark pair is dominant followed by the
decay into $\tau\tau$. However, when the $h_d$ component of $H_1$ is
very small the branching ratios into $gg$ and $c\bar c$ become
competitive and can even be larger than those for the decay into
$b\bar b$ with values beyond 60\% for the $gg$ final state and of up
to 35-39\% for $c\bar{c}$ in some of the scenarios. In this case, {\it
  i.e.}~for $|{\cal R}^l_{1 h_d}| \lsim 0.02$, also the branching
ratios into $\gamma\gamma$ and into the off-shell final state $W^{+*}
W^{*-}$ increase and can reach up to about 30\% in the latter and
about 2\% in the former case. The branching ratio into the off-shell
$Z^*Z^*$ final state, which also increases then, is about one order of
magnitude smaller than the one into $W^{+*}W^{-*}$. But already for
$|{\cal R}^l_{1 h_d}| \gsim 0.02$ the decay into $b\bar{b}$ takes over
again and reaches branching ratio values of up to 90\% followed by the
branching ratio into $\tau\tau$ with values of up to 10\%. \s

\begin{figure}[t!]
  \centering
  \includegraphics[width=0.6\textwidth]{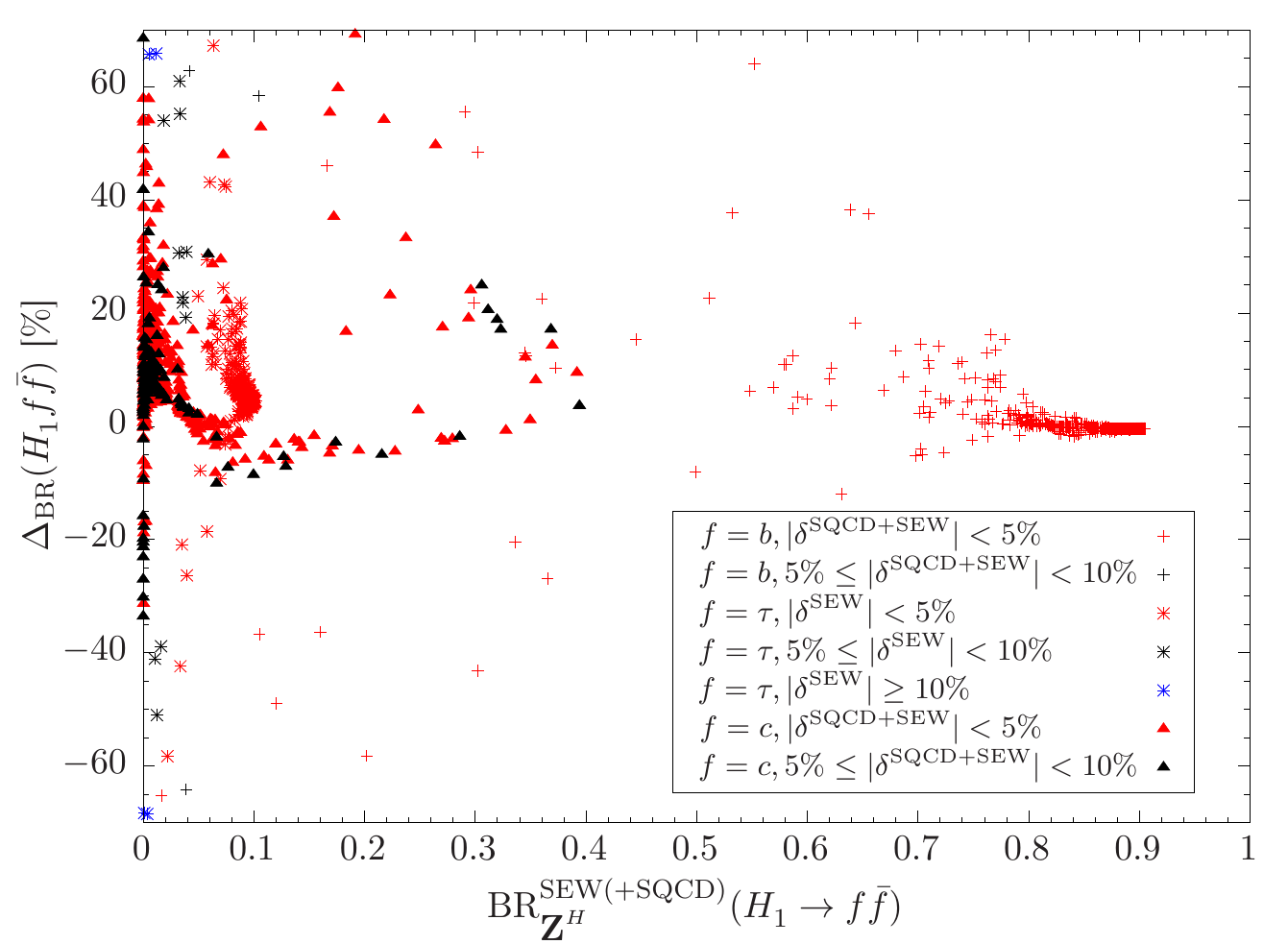}
  \caption{{\it Scan1:} Relative difference $\Delta_{\text{BR}}$ 
    (see text, for definition) in percent for the $H_1$ decays into
    $b\bar b, \tau\bar \tau, c\bar 
    c$. Red: relative corrections $\delta$ (see text, for definition) in percent of the
    SUSY-EW(+SUSY-QCD) corrections to the decay widths with $|\delta| <
    5$\%; black: $5 \le |\delta| < 10$\%; blue: $|\delta| \ge 10$\%.
    \label{fig:h1decayscatter2}}  
  \label{fig:h1decayscatterDelta}
\end{figure}

In order to investigate the importance of the higher-order corrections
we define for our new implementation the relative correction of the
partial width for the decay $H_i \to X_j X_k$ as 
\be 
\delta^{\text{SEW(+SQCD)}} (H_i X_j X_k)= \fr{\Gamma^{\text{SEW(+SQCD)}}_{\ZH}
  (H_i \to X_j X_k) 
  -\Gamma^{\tree}_{\ZH} (H_i \to X_j X_k) }{\Gamma^{\tree}_{\ZH} (H_i
  \to X_j X_k)} \;,  
\label{eq:correction}
\ee
with the higher-order decay widths for the decays $H_i \to q \bar{q}$
into quarks given in \eqref{eq:correctedgambb} and the higher-order
decay widths for the decays $H_i\to l\bar l$ into leptons given in
\eqref{eq:correctedgamll} and with the tree-level decay width
$\Gamma^{\tree}_{\ZH} $ including only the $\Delta_b$ corrections. The
tree-level and higher-order decay widths are both evaluated with the
new implementation of $\ZH$. Note that the quantity $\delta$ gives
information on the importance of the SEW(+SQCD) corrections in the
decay width alone as the factor $\ZH$ cancels in the ratio. In
\figref{fig:h1decayscatter2} we show the scatter plot of the relative
change of the branching ratios, $\Delta_{\text{BR}}(H_1f\bar f)$,
$f=b,c,\tau$, for all the parameter points passing the constraints,
against $\text{BR}_{\ZH}^{\text{SEW(+SQCD)}}(H_1\to f\bar f)$. \s

\begin{figure}[t!]
  \centering
  \includegraphics[width=0.6\textwidth]{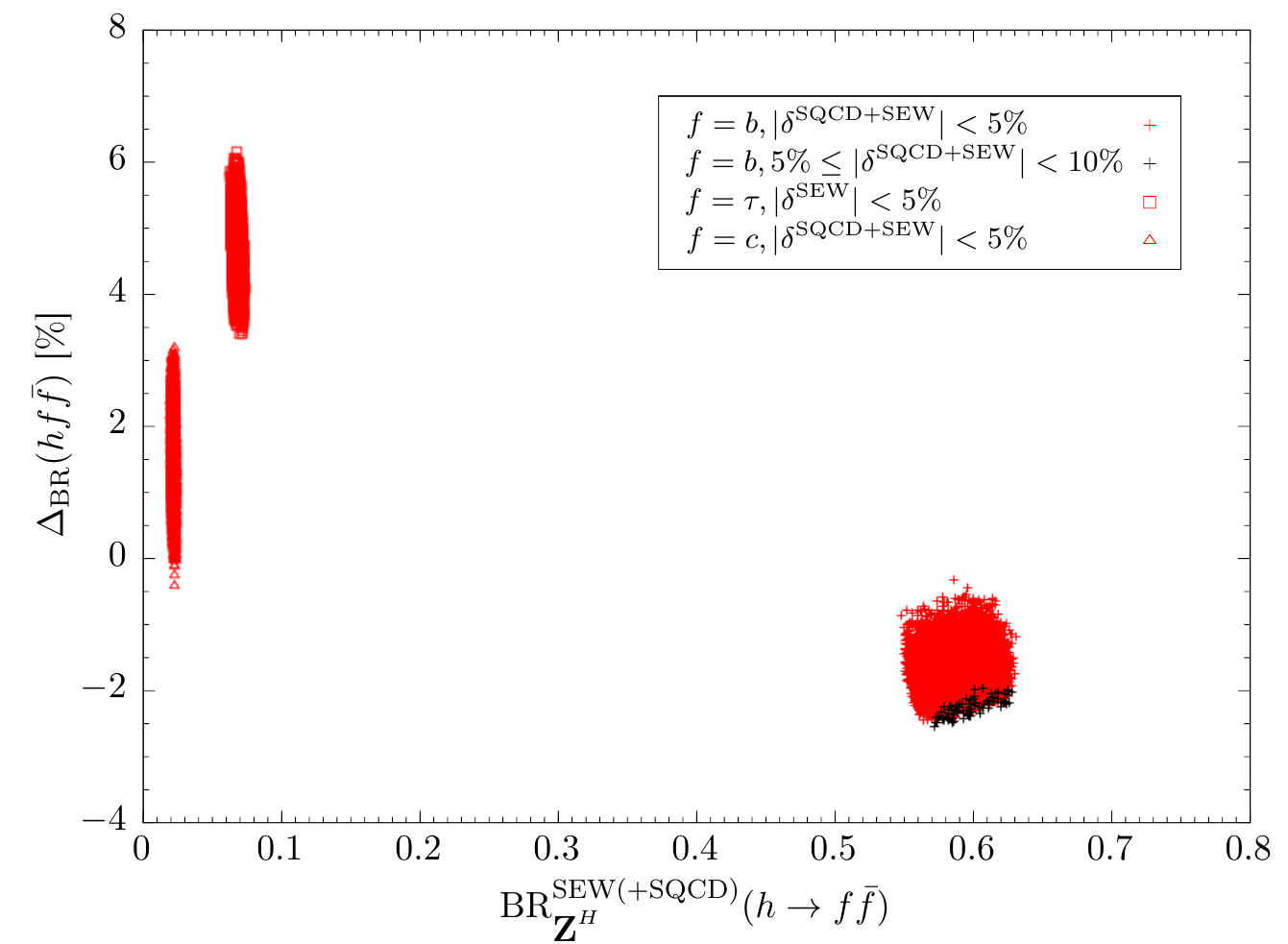}
  \caption{{\it Scan2:} Same as
    \protect\figref{fig:h1decayscatterDelta} but for the SM-like Higgs
    boson $h$ that is $h_u$-like.} 
  \label{fig:hsmdecayscatterDelta}
\end{figure}

\begin{figure}[t!]
  \centering
  \subfloat[ ]{\includegraphics[width=0.48\textwidth]{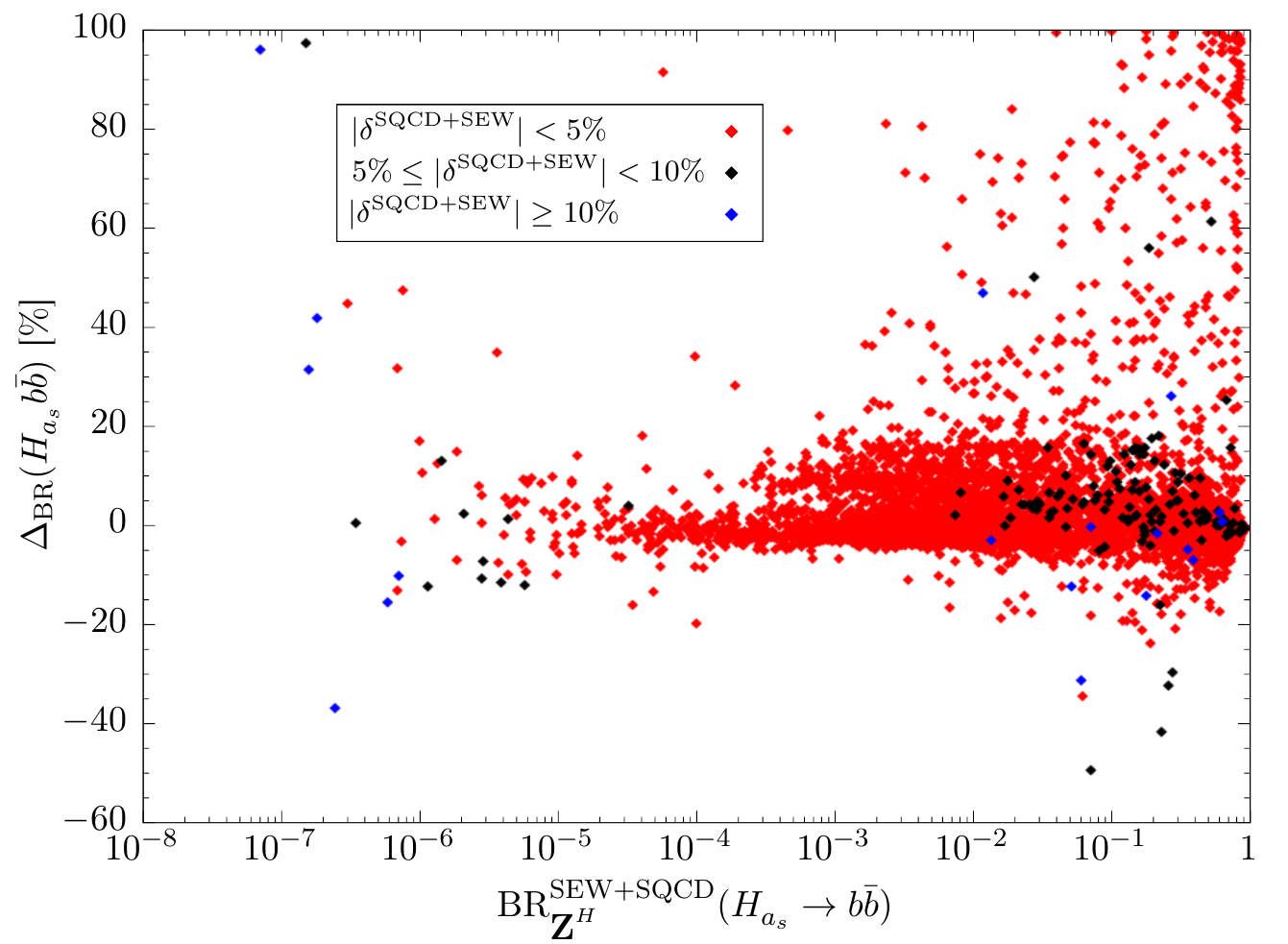}\label{fig:asdecayscatter3}}\quad
  \subfloat[]
  {\includegraphics[width=0.48\textwidth]{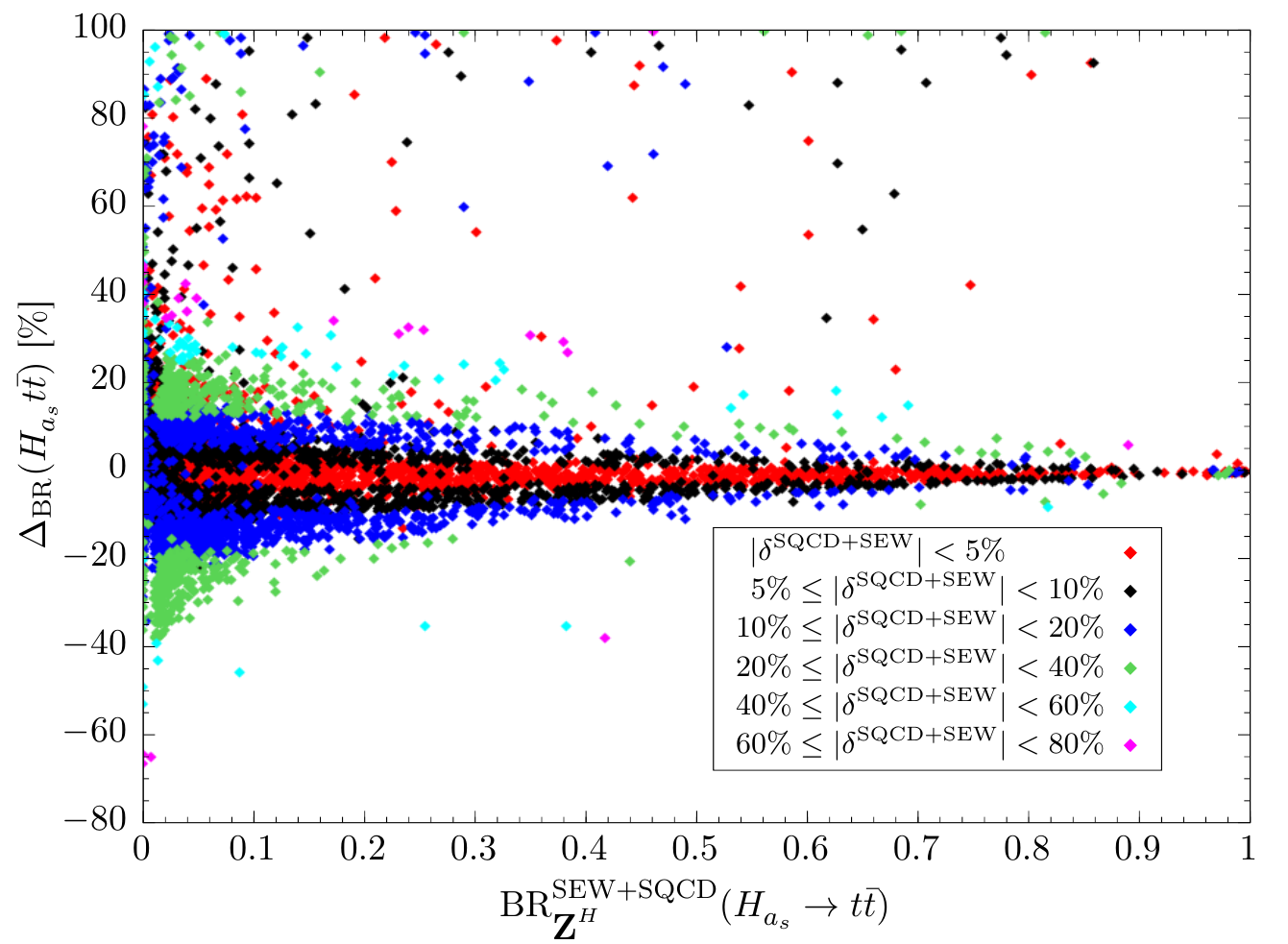}\label{fig:asdecayscatter5}}
  \caption{{\it Scan2:} Same as \protect\figref{fig:h1decayscatterDelta} but for
    the heavier $a_s$-like Higgs boson $H_{a_s}$ decaying into $b\bar
    b$ (left) and $t\bar t$ (right).} 
\label{fig:asdecayscatterDelta}
\end{figure}

\begin{figure}[t!]
  \centering
  \subfloat[ ]{\includegraphics[width=0.48\textwidth]{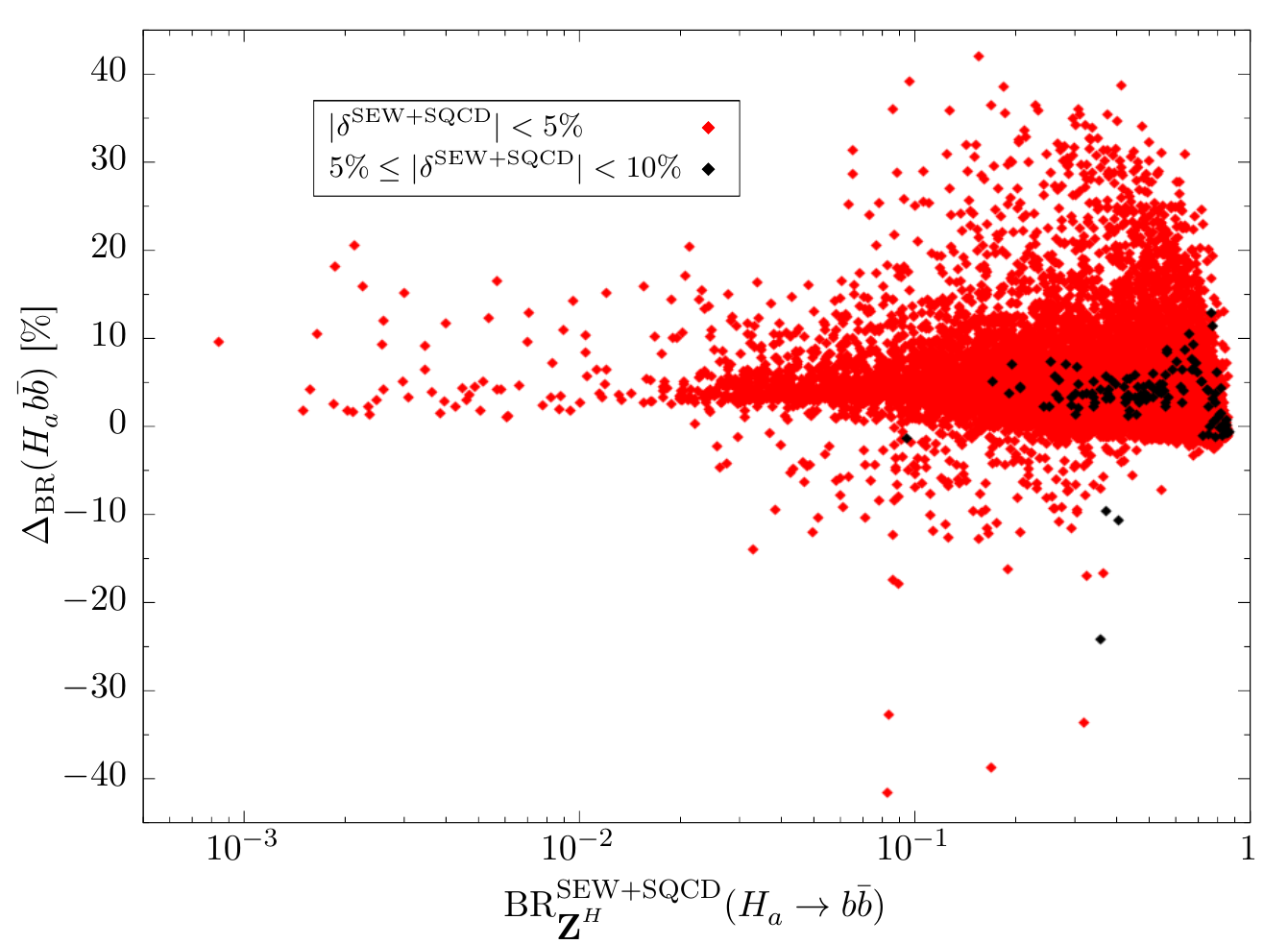}\label{fig:hadecayscatter3}}\quad
  \subfloat[]{\includegraphics[width=0.48\textwidth]{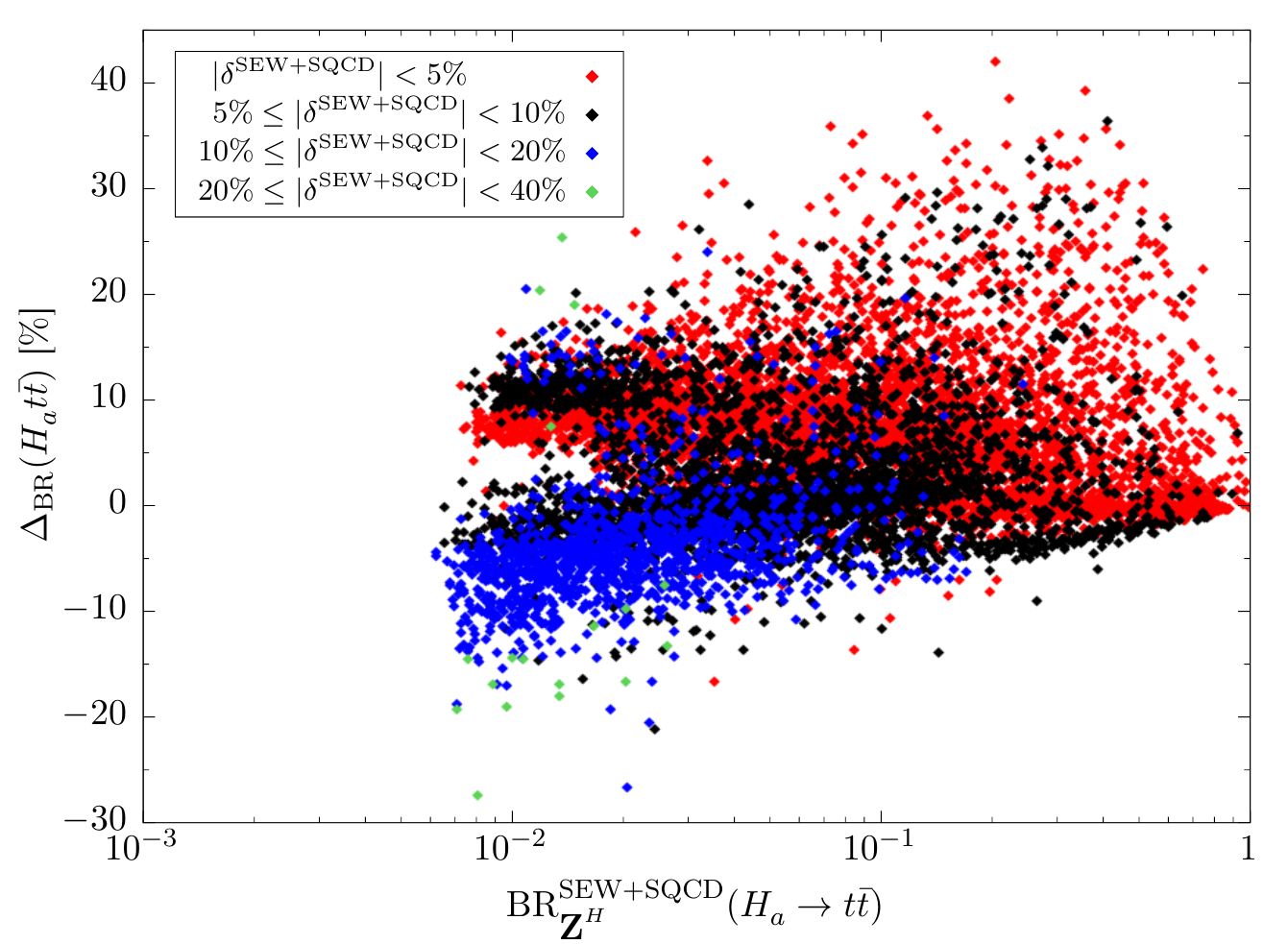}\label{fig:hadecayscatter5}}
  \caption{{\it Scan2:} Same as \protect\figref{fig:h1decayscatterDelta} but for
    the heavier $a$-like Higgs boson $H_{a}$ decaying into $b\bar
    b$ (left) and $t\bar t$ (right).} 
\label{fig:hadecayscatterDelta}
\end{figure}

\begin{figure}[t!]
  \centering
  \subfloat[ ]{\includegraphics[width=0.48\textwidth]{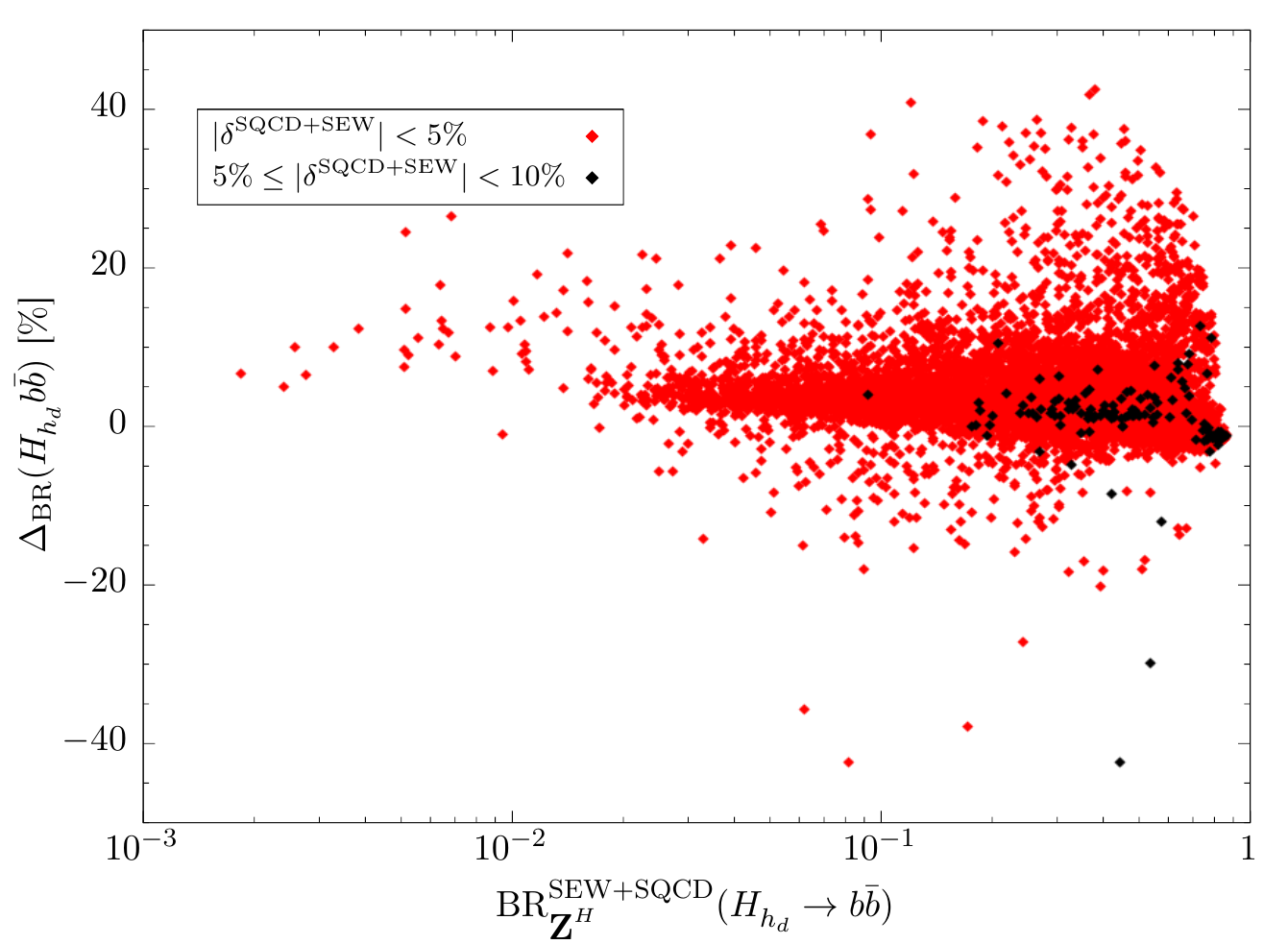}\label{fig:hddecayscatter3}}\quad
  \subfloat[]
  {\includegraphics[width=0.48\textwidth]{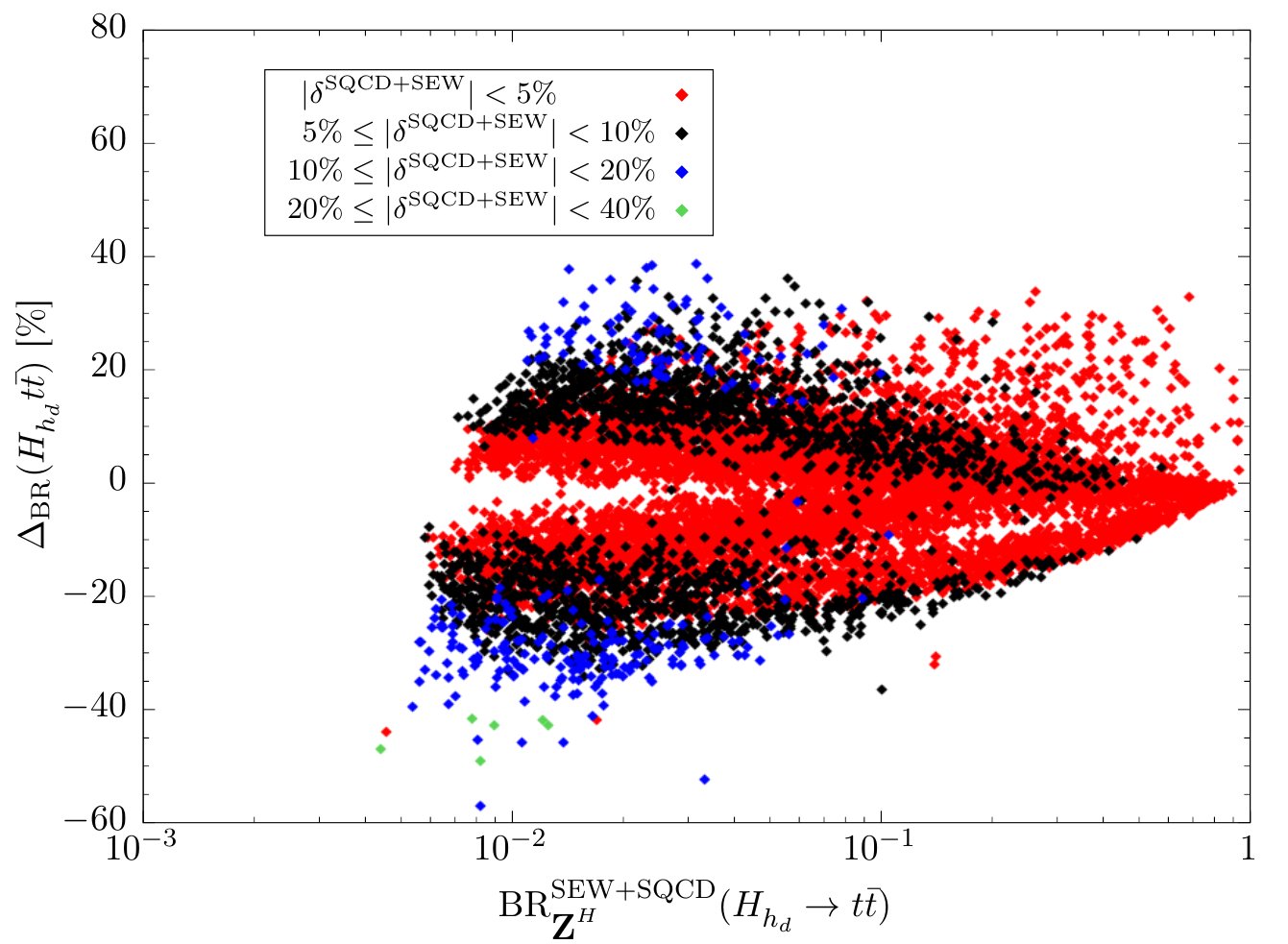}\label{fig:hddecayscatter5}}\quad
  \caption{{\it Scan2:} Same as \protect\figref{fig:h1decayscatterDelta} but for
    the heavier $h_d$-like Higgs boson $H_{h_d}$ decaying into $b\bar
    b$ (left) and $t\bar t$ (right).} 
\label{fig:hddecayscatterDelta}
\end{figure}

The color code in \figref{fig:h1decayscatter2} as well as in
Figs.~\ref{fig:hsmdecayscatterDelta}--\ref{fig:hddecayscatterDelta} 
denotes the sizes of the relative corrections of the
partial decay widths. The points where the absolute value of
$\delta^{\text{SEW(+SQCD)}}$ exceeds 10\% are marked in blue, those with
$\left|\delta^{\text{SEW(+SQCD)}}\right|$ in the [5,10]\% range in black
and those with relative corrections less than 5\% in red. For
Figs.~\ref{fig:hadecayscatter5} and \ref{fig:hddecayscatter5} we
distinguish two regimes for the larger corrections, in blue where
$\left|\delta^{\text{SEW+SQCD}}\right|$ is in the [10,20]\% range
and in green where $\left|\delta^{\text{SEW+SQCD}}\right|$ is in the
[20,40]\% range; in Fig.~\ref{fig:asdecayscatter5} we also add two
other categories of points, in cyan where
$\left|\delta^{\text{SEW+SQCD}}\right|$ is in the [40,60]\% range and
in pink where $\left|\delta^{\text{SEW+SQCD}}\right|$ is in the
[60,80]\% range. Note that for the $\tau$ decays there are no
SUSY-QCD corrections. The ballpark of the relative change
$\Delta_{\text{BR}} ({H_1 b\bar b})$ in the branching ratios between
the old and the new implementation ranges below about 30\% with vertex
corrections $|\delta|$ smaller than 5\%. There are some very rare
scenarios where $\left|\Delta({H_1b\bar b})\right|$ exceeds 50\% and
where at the same time the relative vertex corrections are between 5
and 10\%. We investigated these cases and observed that there is an
accidental cancellation either in the effective tree-level or in the
improved couplings. These parameter points lead to similar results for
the $\tau$ final states, {\it i.e.}~$|\Delta_{\text{BR}}({H_1
  \tau\bar{\tau}})| > 50$\% and at the same time $|\delta|$ between 5
and 10\%. The cancellation results in a suppression of the branching
ratio, to less than 4\% for the $\tau\bar{\tau}$ final state and 10\%
at maximum for the $b\bar{b}$ final state. In most of the cases, the
large $\Delta_{\text{BR}} ({H_1f\bar f})$ is due to the use of the
wave-function renormalization factor $\ZH$, however.
There are also cases with a cancellation between the SUSY-EW/SUSY-QCD
corrections and the wave-function renormalization factor $\ZH$
correction. This results in  $\Delta_{\text{BR}}({H_1c\bar c})$ being
less than 1\%. \s

We have performed the same analysis for the heavier Higgs
bosons, using the full set of points from our {\it scan2}. 
Figure~\ref{fig:hsmdecayscatterDelta} is the scatter plot of the
$\Delta_{\text{BR}}({h f\bar f})$, $f=b,\tau, c, t$, against
$\text{BR}_{\ZH}^{\text{SEW(+SQCD)}}(h\to f\bar f)$ for an SM-like
Higgs boson $h$, while Figures~\ref{fig:asdecayscatterDelta},
\ref{fig:hadecayscatterDelta}, and \ref{fig:hddecayscatterDelta} are
the scatter plots of $\Delta_{\text{BR}}({H_if\bar f})$, $f=b, t$,
against $\text{BR}_{\ZH}^{\text{SEW(+SQCD)}}(H_i\to f\bar f)$ for a
heavy $a_s$-, $a$- and $h_d$-like Higgs boson {\it i.e.}
$H_i=H_{a_s}$, $H_a$, $H_{h_d}$, respectively.
As before, the SM-like Higgs boson $h$ is always $h_u$-like 
and decays dominantly into a bottom-quark pair with a branching
ratio of about 60\%, as expected, followed by the decay into a $\tau$ pair
and the decay into a $c$-quark pair. As can be inferred
from Fig.~\ref{fig:hsmdecayscatterDelta}, the relative changes between
the old and the new implementation are much smaller than for the
singlet-like lightest Higgs boson and amount only to a few
percent. The relative vertex corrections $|\delta|$ are below 10\% for
the $b$-quark pair final state and below 5\% for the decays both into
$\tau\bar{\tau}$ and $c\bar{c}$. \s


For the heavy Higgs bosons, the decay into a top quark pair can become
kinematically possible. 
We start by discussing the decay pattern of the heavy singlet state
$H_{a_s}$, with a mass between 120~GeV and
1.7~TeV, into  the $b$-quark and $t$-quark final states, 
presented in Fig.~\ref{fig:asdecayscatterDelta}.\footnote{Since we
  will not gain much new information, for $H_{a_s}$, $H_a$ and
  $H_{h_d}$ we do not show the corresponding plots into
  $\tau\bar{\tau}$ and $c\bar{c}$.} 
For the $b\bar{b}$ final states the relative change in
the branching ratios due to the new implementation is mostly 
between -20\% and 20\%. We also find points where the relative
change is close to 100\%, in particular for branching ratios close to
100\%.
Most points exhibit small relative vertex corrections (see, red points
in \figref{fig:asdecayscatter3}), so that the large changes of
$\Delta_{\text{BR}}(H_{a_s} b\bar b)$ are due to the implementation of
$\ZH$. This is especially the case for large branching ratios
close to 100\%. There are a few points where the relative vertex
corrections lie between 5 and 10\% (black) and even above 10\%
(blue). This happens for the cases where the effective tree-level
couplings $H_{a_s}f\bar f$ are suppressed. The relative change
$\Delta_{\text{BR}}$ can still be very small when the effects from the
$\ZH$ factor and the vertex corrections cancel.
The decay pattern for the $t\bar t$ channel, finally, is
displayed in Fig.~\ref{fig:asdecayscatter5}. The branching ratio takes
all values between almost 0 and 100\%. The relative changes
$\Delta_{\text{BR}}(H_{a_s} t\bar t)$ are mostly between -20\% and
20\% and close to 0\% for large branching ratios above about 60\%. We
also observe large $\Delta_{\text{BR}}$, in particular for branching
ratios close to zero. This is mainly due to very suppressed effective
tree-level $H_{a_s}t\bar t$ couplings corresponding to the regions where
the branching ratio $\mathrm{BR}(H_{a_s}\to b\bar b)$
is enhanced. These 
regions correspond to large values of $\tan\beta$ close to the upper
bound in our scan, or to smaller mass values of $H_{a_s}$ with not
sufficient phase space to decay into an on-shell top-quark pair. In
these regions the relative corrections $|\delta|$ are most of the time
below 40\%, and for cases where $|\delta|<5\%$ the large changes in
$\Delta_{\text{BR}}$ are mostly due to the use of the wave-function
renormalization factor $\ZH$. For larger branching ratios the relative
corrections $|\delta|$ are mostly below 10\% (red and black
points). Some rare scenarios display corrections above 40\% and up to
80\% (in cyan and in pink), again mostly in regions with lower
branching ratios.\s

Similar observations can be made for the other heavier Higgs states
$H_a$ (with a mass between 539~GeV and 2~TeV) and $H_{h_d}$ (with a
mass between 548~GeV and 2~TeV), with the notable exception that the
relative changes $\Delta_{\text{BR}}({H_{a/h_d} } X_i X_j)$ are more
reduced and never reach 100\%. The relative changes
$\Delta_{\mathrm{BR}}(H_{a/h_d}b\bar b)$ are most of the time positive
and below 40\% as seen in 
Figs.~\ref{fig:hadecayscatter3} and \ref{fig:hddecayscatter3}.
The decays into top-quark pairs can be dominant 
where the decays into $b\bar{b}$ are suppressed,
and the relative changes $\Delta_{\text{BR}}$ between the old and new
implementation are close to zero when
$\mathrm{BR}(H_{a/h_d}\to t\bar t)\to 100\%$ as seen in
Figs.~\ref{fig:hadecayscatter5} and \ref{fig:hddecayscatter5}. For
some rare scenarios the relative vertex corrections $|\delta|$ can reach 40\%,
depicted in green in the figures. Note that $\mathrm{BR}(H_{a/h_d}\to
b\bar b)$ can reach 90\%, corresponding to regions
where the effective tree-level coupling $H_{a/h_d}b\bar b$ is strongly enhanced due
  to large values of $\tan\beta$ while at the same time the effective
tree-level coupling $H_{a/h_d}t\bar t$ is strongly suppressed. 

\subsection{Decays into a Massive Gauge Boson Pair}
In the CP-conserving case, the heavy Higgs boson that can decay into
two on-shell massive gauge bosons is $h_d$-like. The tree-level
coupling of a Higgs boson $H_i$ to $VV$ ($V=W,Z$) is proportional to
\beq
\calR_{i,1}\cbeta +\calR_{i,2}\sbeta \;.
\eeq
Due to the SM-like ({\it i.e.}~$h_u$-like) Higgs boson coupling with almost
SM-strength to the massive gauge bosons the tree-level coupling of the
$h_d$-like heavy Higgs boson to $VV$ is almost zero because of sum
rules. This leads to very suppressed tree-level partial decay
  widths $\Gamma(H_{h_d}\to VV)$. \s

\begin{figure}[b!]
  \centering
  \subfloat[]{\includegraphics[width=0.48\textwidth]{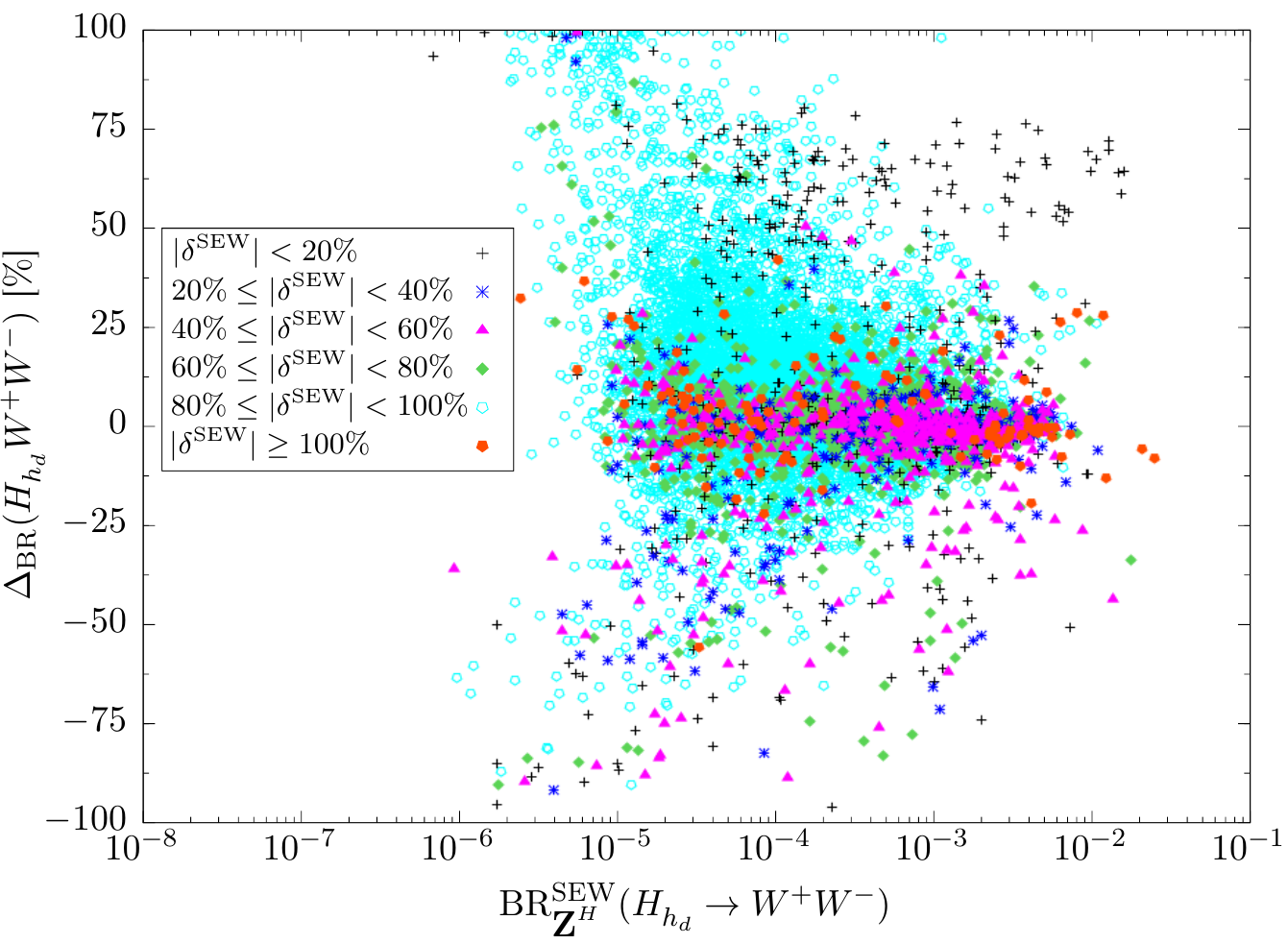}\label{fig:ScatterPlotHdWW}}\quad
  \subfloat[]
  {\includegraphics[width=0.48\textwidth]{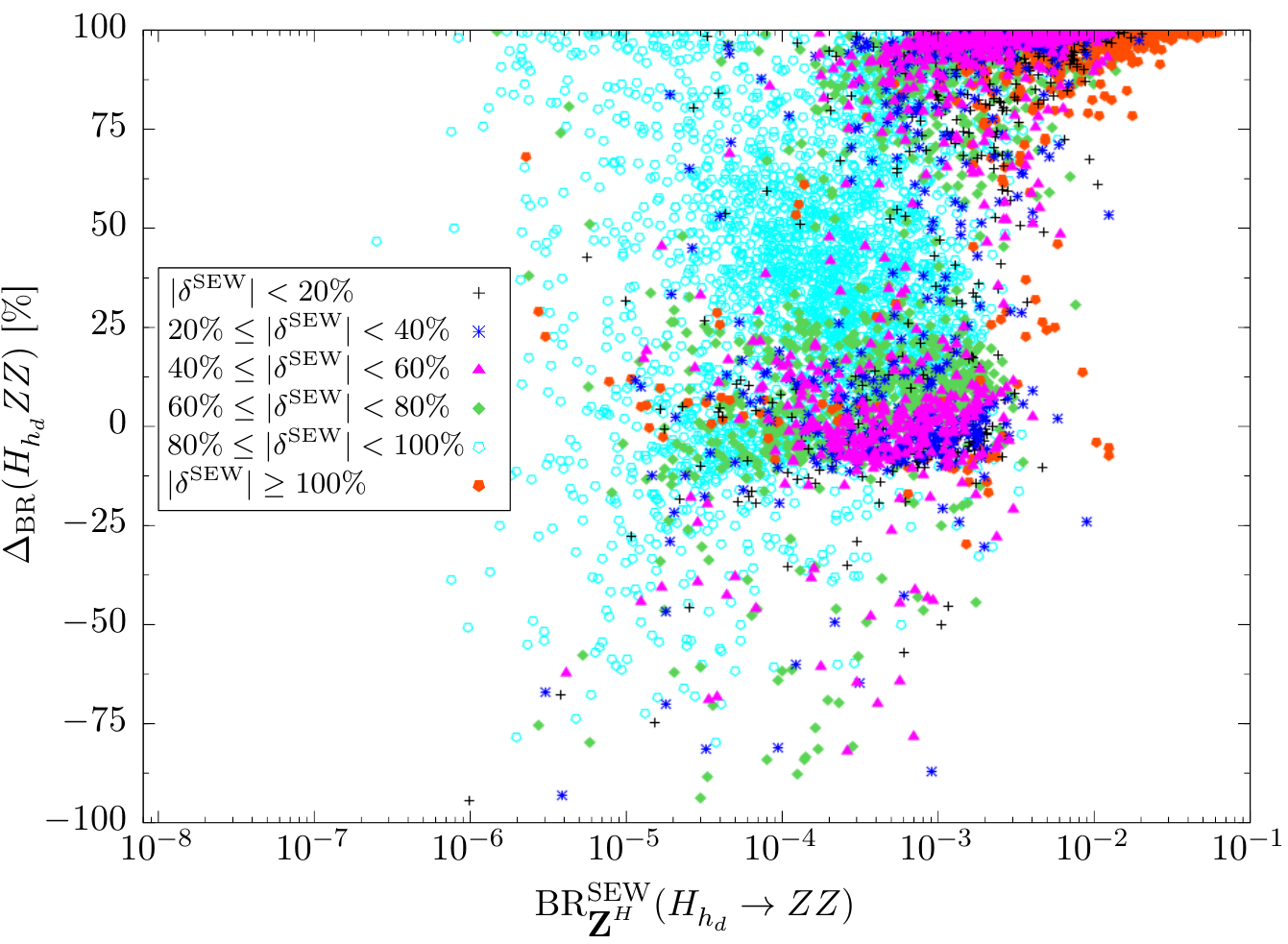}\label{fig:ScatterPlotHdZZ2}}
  \caption{{\it Scan2:} Relative difference $\Delta_{\text{BR}}$ 
    in percent for the $h_d$-like Higgs boson $H_{h_d}$ into a $W$ boson
    pair (left) and into a  $Z$ boson pair (right) as
    a function of the corresponding loop-corrected branching ratio. 
    Black: relative corrections $\delta^{\text{SEW}}$ in percent of the
    SUSY-EW corrections to the decay widths with $|\delta^{\text{SEW}}| <
    20$\%; blue: $20 \le |\delta^{\text{SEW}}| < 40$\%; pink: $40 \le |\delta^{\text{SEW}}| <
    60$\%; green: $60 \le |\delta^{\text{SEW}}| < 80$\%; cyan: $80 \le |\delta^{\text{SEW}}| <
    100$\%; red: $|\delta^{\text{SEW}}|\ge 100$\%.}
  \label{fig:ScatterPlotHdVV}
\end{figure}

In order to compare the results obtained in this paper with the old
implementation in {\tt NMSSMCALC}  using the tree-level coupling
together with the loop-corrected rotation matrix $\calR^{l}$, we show
in \figref{fig:ScatterPlotHdWW} the relative change $\Delta_{\text{BR}}
(H_{h_d} WW)$ of the branching ratio into $WW$ between the old and the
new implementation including the NLO-EW vertex corrections as described in
Sec.~\ref{sect:hVV} and the improvement with the $\ZH$ factor, as a function of the
loop-corrected branching ratio ${\rm BR}_{\ZH}^{\rm SEW}(H_{h_d}\to
W^+W^-)$. The plotted points are those of our scan that pass the
constraints we have applied. We display in
\figref{fig:ScatterPlotHdZZ2} the same but for the decay into
$ZZ$. The color and symbol code denotes the magnitude of the relative
NLO electroweak vertex corrections alone, with $\delta$ defined in
\eqref{eq:correction}. As can be inferred from the plots  both the
$\ZH$ factor and the NLO electroweak corrections can be responsible
for the large relative changes in the branching ratios. This is in
particular reflected by the black points for which the vertex
corrections are below 20\% and at the same time the relative changes
$\Delta_{\text{BR}}$ can reach up to 100\%. In some cases there is a
cancellation between the two contributions (the vertex corrections and
the $\ZH$ factor) leading to relatively small relative changes in the
branching ratios. These cases are the pentagon-marked full (red)
points in \figref{fig:ScatterPlotHdWW} for the decay
  $H_{h_d}\to W^+W^-$, which are mostly located in regions where
$|\Delta_{\text{BR}}(H_{h_d}WW)|\lsim 25\%$ while the relative vertex correction
$|\delta|$ is at least 100\%. In the case of the decay into a $Z$
boson pair, however, the bulk of these pentagon-marked full (red)
points, indicating again a relative vertex correction $|\delta|$ of
  at least 100\%, induces large relative changes of $\Delta_{\text{BR}}(H_{h_d} Z Z)$
close to 100\%. These points correspond to a region which is
  discussed in more detail in the next paragraph. \s

We also note that there are two regions concentrating many
points for the decay into a $Z$ boson pair, the region for which
$\Delta_{\text{BR}}(H_{h_d}ZZ)\simeq 0$\% and the one for which
$\Delta_{\text{BR}}(H_{h_d}ZZ)\simeq 100$\%. This is in contrast to the decay into
a $W$ boson pair which is mostly centered around
$\Delta_{\text{BR}}(H_{h_d}WW)\simeq 0$\%
for vertex corrections $|\delta|<80$\% and much more scattered for
the points where $80 < |\delta| < 100$\%, displayed with
cyan-pentagon-marked points (for $|\delta|\ge 100$\% the above
described cancellation takes place in the decay $H_{h_d}\to
W^+W^-$). This presence of the second 
region in the $Z$ boson final state, for which $\Delta_{\text{BR}}$ is close to
100\% can be explained by the occurrence of many parameter points
having a very suppressed tree-level coupling $H_{h_d}ZZ$. They also
correspond to regions where the loop-corrected partial decay width
$\Gamma(H_{h_d}\to ZZ)$ is higher, up to 1 GeV, while the decay width
is at most 5 MeV for the region centered around
$\Delta_{\text{BR}}(H_{h_d}ZZ)=0$. Note, that while the tree-level couplings
$H_{h_d}ZZ$ and $H_{h_d} WW$ are the same, the loop-corrected decay
widths differ by the fact that the decay into $WW$ bosons receives
real corrections and that in the one-loop squared contributions to the
decay width $H_{h_d} \to W^+W^-$ we
only include the (s)fermion contributions in contrast to the decay
$H_{h_d} \to ZZ$.\footnote{We remind the
    reader that we take into account this part of the two-loop
    corrections in case the one-loop corrected partial decay width
    becomes negative, see also \eqref{eq:olsquared}.}

\subsection{Decays into a \boldmath $Z$ Boson and a Higgs Boson}
\begin{figure}[t!]
  \centering
  \subfloat[]{\includegraphics[width=0.48\textwidth]{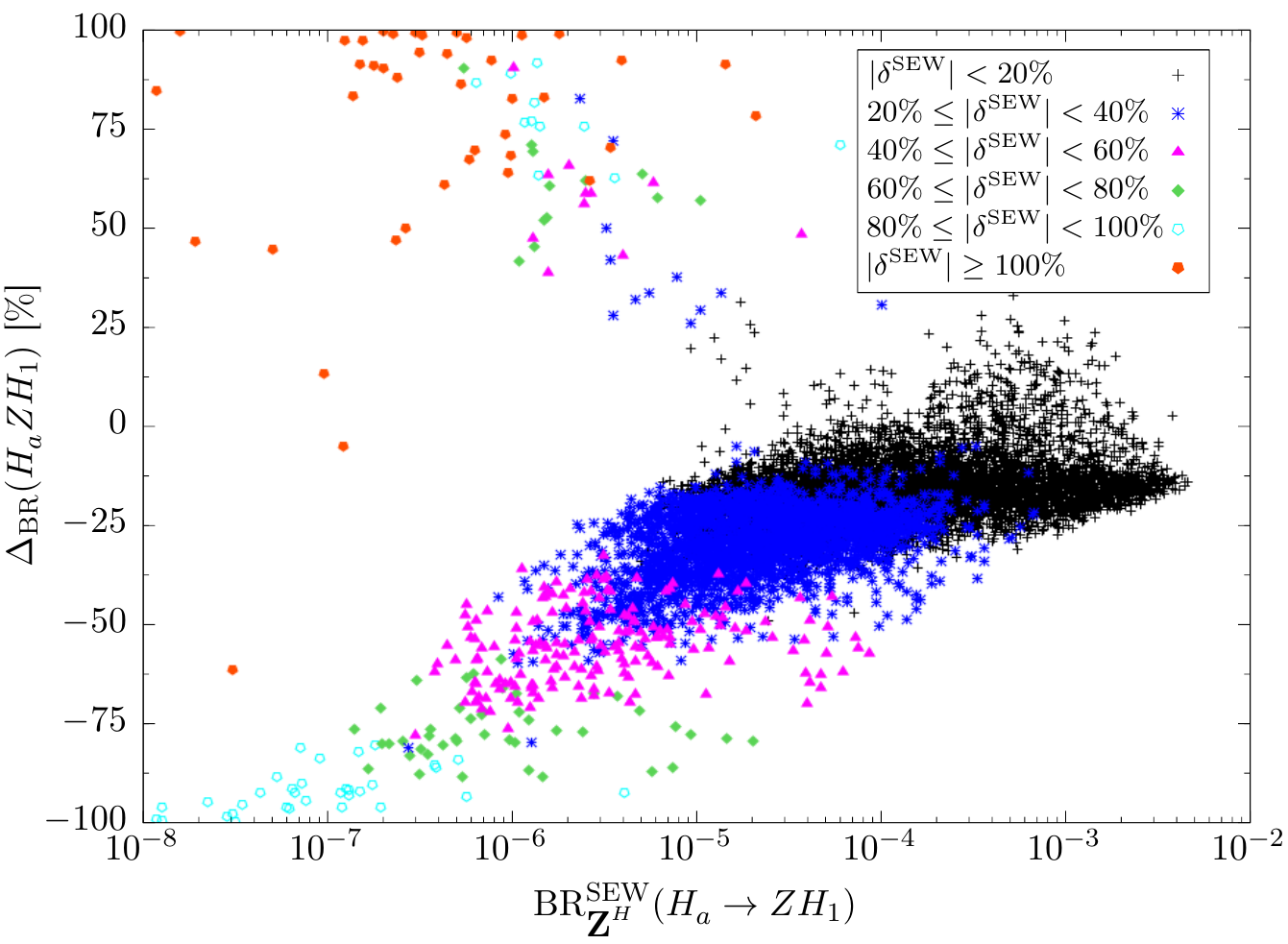}\label{fig:ScatterPlotH3aH1Z}}\quad
  \subfloat[]
  {\includegraphics[width=0.48\textwidth]{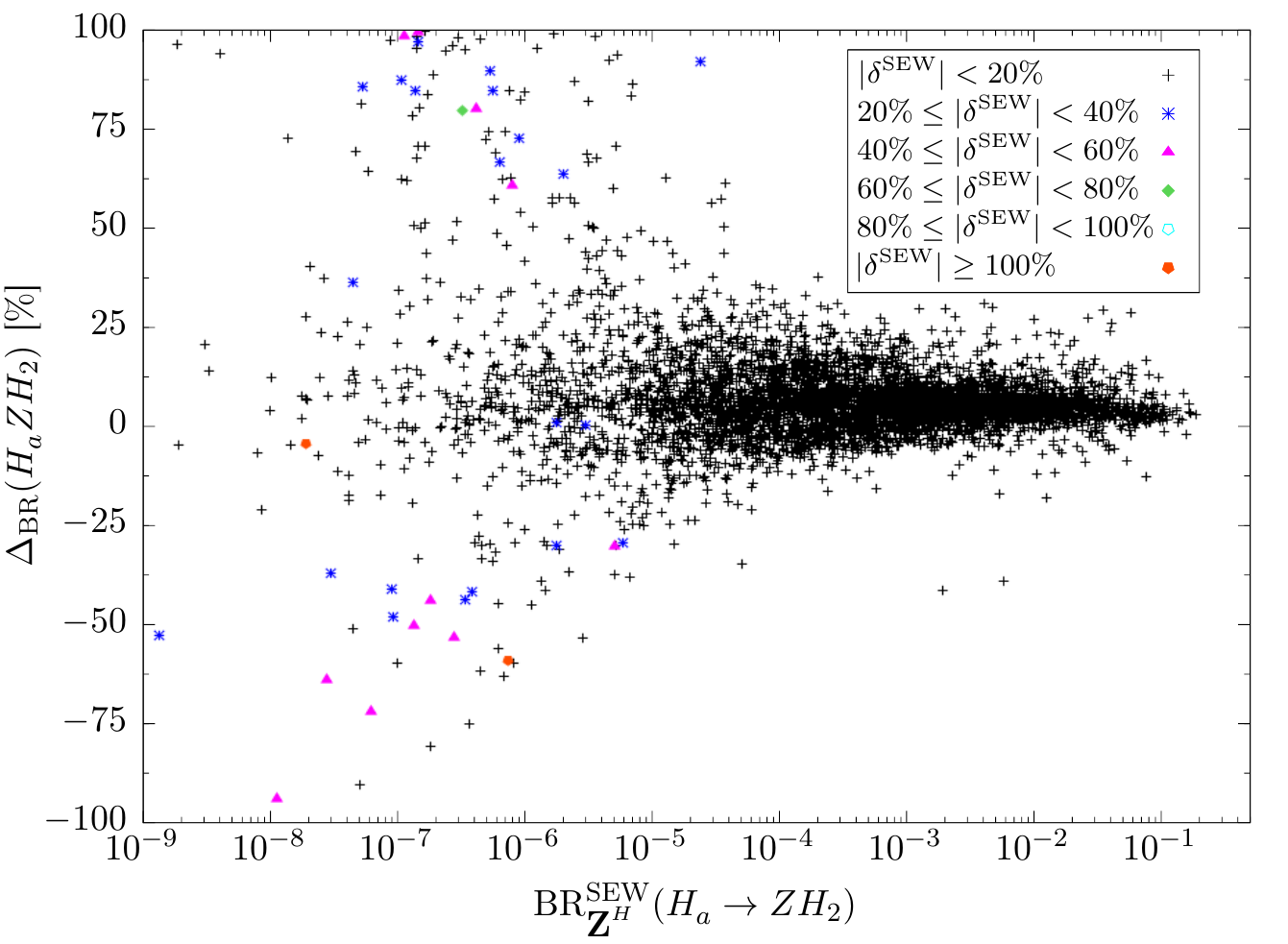}\label{fig:ScatterPlotH3aH2Z}}\quad
  \caption{{\it Scan2:} Relative difference $\Delta_{\text{BR}}$ 
    in percent for the $H_a$-like Higgs boson decay into $ZH_1$
    (left) and into $ZH_2$ (right) as
    a function of the corresponding loop-corrected branching ratio. 
    Black: relative corrections $\delta^{\text{SEW}}$ in percent of the
    SUSY-EW corrections to the decay widths with $|\delta^{\text{SEW}}| <
    20$\%; blue: $20 \le |\delta^{\text{SEW}}| < 40$\%; pink: $40 \le |\delta^{\text{SEW}}| <
    60$\%; green: $60 \le |\delta^{\text{SEW}}| < 80$\%; cyan: $80 \le |\delta^{\text{SEW}}| <
    100$\%; red: $|\delta^{\text{SEW}}|\ge 100$\%.}
  \label{fig:ScatterPlotH3aHZ}
\end{figure}

In the searches for heavy pseudoscalars, this decay can be an
important search
channel~\cite{Khachatryan:2015tha,Khachatryan:2016are}. We are
interested here in how large the branching ratio can be and how
important are the newly included higher-order corrections, in
  the case for which on-shell decays are possible. With the
obtained valid set of parameter points, we present in
\figref{fig:ScatterPlotH3aH1Z} scatter plots of the relative changes of the branching
ratios between the old and new implementation for the decay of a heavy
pseudoscalar-like Higgs boson $H_a$ into $Z H_1$ and in
\figref{fig:ScatterPlotH3aH2Z} for 
the decay into $Z H_2 $, against the respective 
  loop-corrected branching ratios. 
The color and marker codes denote the relative sizes of the one-loop
vertex corrections within specific ranges, identical to the 
ranges used in the previous sub-section for the decays into gauge
boson pairs. The mass values of the individual involved CP-even Higgs bosons
in the final state range in $123~\text{GeV} \le
m_{H_1} \le 127~\text{GeV}$ and $463~\text{GeV} \le m_{H_2} \le
1.73~\text{TeV}$, while the mass of the decaying Higgs boson ranges in
$539~\text{GeV} \le m_{H_a} \le 2.0~\text{TeV}$ for the on-shell decay
into $Z H_1$ pairs and in $713~\text{GeV} \le m_{H_a} \le
2.0~\text{TeV}$ for the on-shell decay into $Z H_2$ pairs. As can
already be inferred from the mass ranges, the $H_1$ state is the
SM-like Higgs boson $h$, while the $H_2$ state is the
singlet-like scalar Higgs boson $H_{h_s}$. \s

We observe that the branching ratios into the $Z H_1$
final state remain very small, below 0.4\%, while those of the decay
into $Z H_2 $ can reach 11\%. This is due to the nature of the $H_1$
Higgs boson that is SM-like, with very suppressed tree-level
$H_1 H_a Z$ couplings. 
The relative changes $\Delta_{\text{BR}}$ of
the branching ratio for the decay $H_a\to Z H_1$ are mostly
between 0 and -75\% corresponding to relative vertex corrections
$|\delta^{\text{SEW}}|$ 
being at most 60\% (black, blue, and pink points), while a few points
corresponding to higher vertex corrections up to more than 100\% (green,
cyan, and red points) can reach $\Delta_{\text{BR}} = \pm 100$\%. These
extreme points correspond to very small values for the branching
ratios themselves which explains in turn the very large relative
corrections $|\delta|$ that we observe. Note that in these decays we
take into account the one-loop squared term as described in
\eqref{eq:1lgaugehiggs} which makes up for the main contribution to the very large
relative corrections. 
As for the decay $H_a\to Z H_2$ the relative correction $\Delta_{\text{BR}}$ is
most of the time between 0 and $\pm 25$\%, corresponding to points
where the relative vertex corrections $|\delta|$ are below 20\%. A few points
display larger $\Delta_{\text{BR}}$ values, and also larger relative vertex
corrections $|\delta|$ that can reach 100\% and even beyond, again for points that 
display very small branching ratios, below about $10^{-4}\,\%$. 
Note that there are points for which the correction $|\Delta_{\text{BR}}|$ is rather limited,
below 25\%, while the relative vertex correction can reach
40\% (for one scenario even more than 100\%). This
can be explained by a sign compensation between the $\ZH$ factor and
the vertex correction. \s

\begin{figure}[t!]
  \centering
  \subfloat[]{\includegraphics[width=0.48\textwidth]{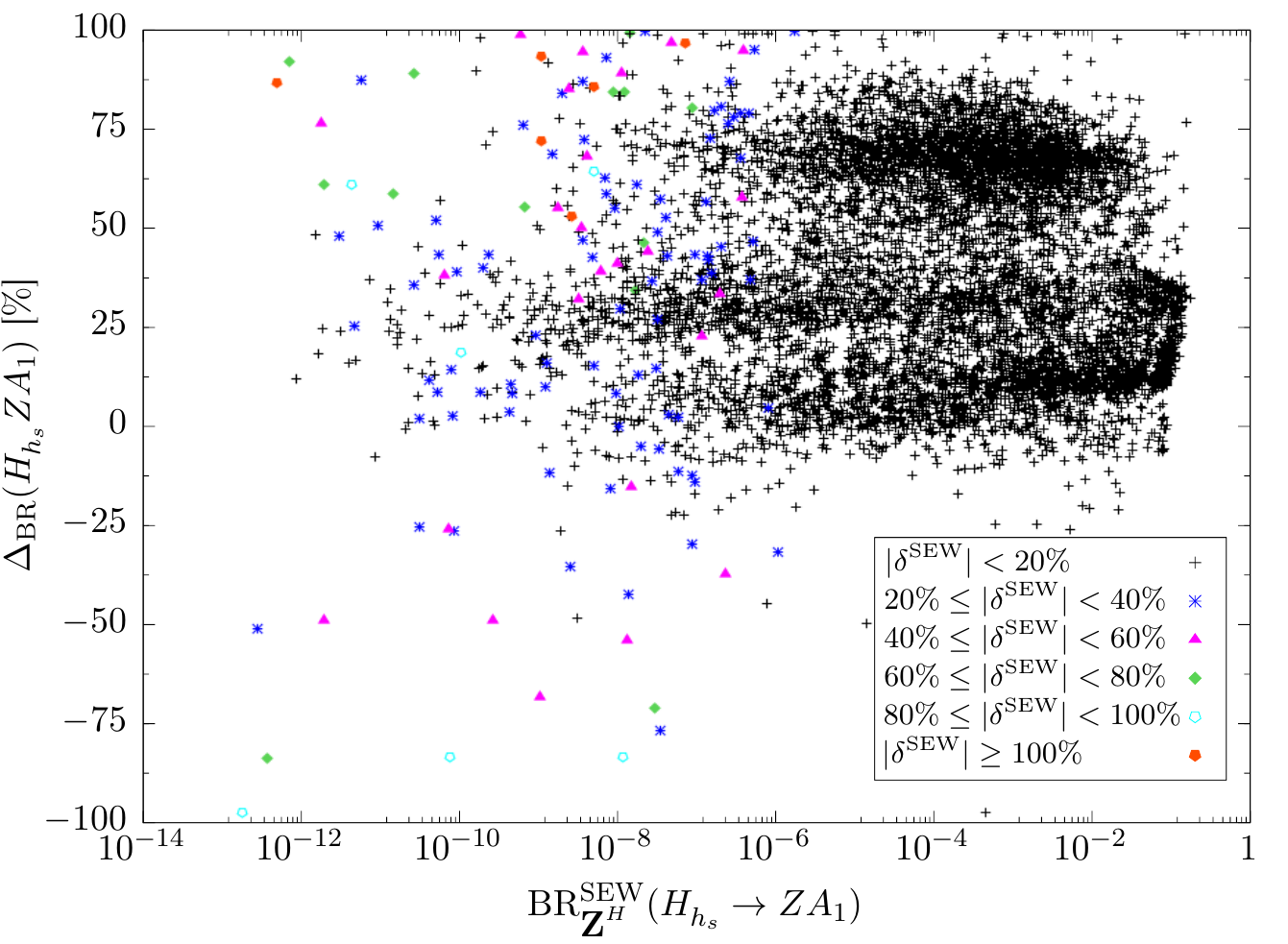}\label{fig:ScatterPlotH3SA1Z}}\quad
  \subfloat[]
  {\includegraphics[width=0.48\textwidth]{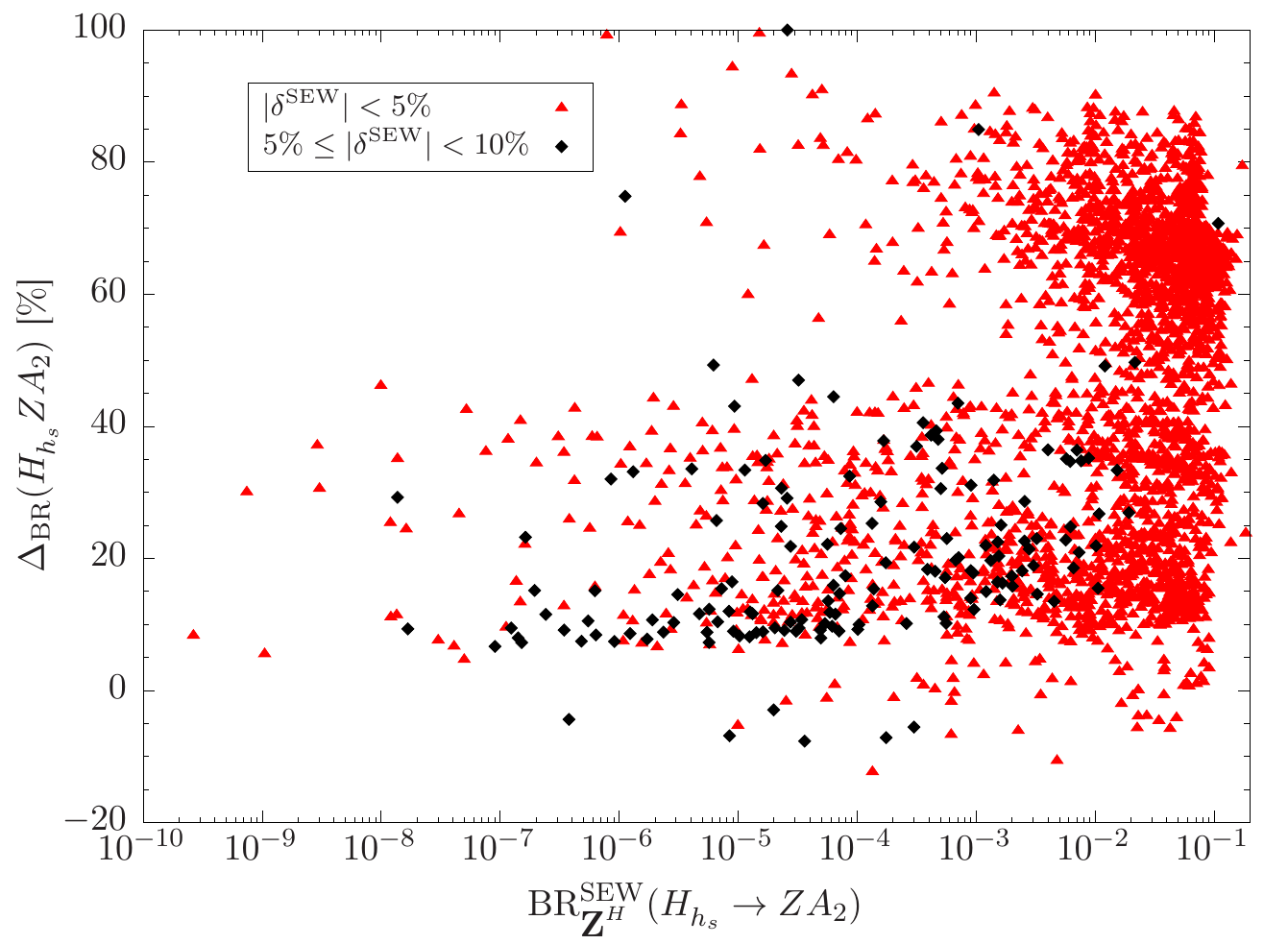}\label{fig:ScatterPlotH3SA2Z}}\quad
  \caption{{\it Scan2:} Same as in \protect\figref{fig:ScatterPlotH3aHZ} but for a
    heavy singlet-like Higgs boson $H_{h_s}$ decaying into $Z A_1$ and
    $Z A_2$ pairs. In the right plot, however, red:
      $|\delta^{\text{SEW}}|< 5\%$; black: $5 \le |\delta^{\text{SEW}}|< 10\%$.}
  \label{fig:ScatterPlotH3sAZ}
\end{figure}

The corresponding results for the heavy singlet-like Higgs boson
$H_{h_s}$ decaying into $Z A_1$ and $Z A_2 $ are shown in
Figs.~\ref{fig:ScatterPlotH3SA1Z} and Figs.~\ref{fig:ScatterPlotH3SA2Z},
respectively. We see that the maximum achieved branching ratios for
the decay $H_{h_s}\to Z A_1$ are below 20\% and for a large number of
parameter points are tiny. In the case of the decay $H_{h_s}\to Z A_2$ the
branching ratio can reach around 15\%. In most of the cases $A_1$ is
singlet-like, corresponding to points where the branching ratio is
small (below 10\%), while $A_2$ is doublet-like. The mass values of the
individual involved Higgs bosons in the final state range in
$120~\text{GeV} \le m_{A_1} \le 1.50~\text{TeV}$ and $562~\text{GeV}
\le m_{A_2} \le 1.63~\text{TeV}$, while the mass of the decaying Higgs
boson ranges in $464~\text{GeV} \le m_{H_{h_s}} \le 1.75~\text{TeV}$ for
the on-shell decay into $Z A_1$ pairs and in $696~\text{GeV} \le
m_{H_{h_s}} \le 1.75~\text{TeV}$ for the on-shell decay into $Z H_2$
pairs. The cases with larger branching ratios for the decay $H_{h_s}\to Z A_1$
(larger than 10\%) correspond mostly to the few $A_1$ pseudoscalar
Higgs bosons with doublet-like admixture and mass values
above $400$~GeV. \s

The relative changes $\Delta_{\mathrm{BR}}$ in the branching
ratios are mostly positive, and can reach values of 100\%. For some
very rare scenarios we get $\Delta_{\mathrm{BR}}(H_{h_s} Z A_1)$ close
to -100\%, while $\Delta_{\mathrm{BR}}(H_{h_s} Z A_2)$ is not below
-15\%. The relative vertex corrections are moderate for the decay
$H_{h_s}\to Z A_1$, mostly $|\delta| \le 20$\% (black points). This
means that the large changes in $\Delta_{\mathrm{BR}}(H_{h_s} Z A_1)$
are mostly due to the $\ZH$ factor. For very small branching fractions
below $10^{-4}\%$ larger vertex corrections are possible, mainly
because the denominator in the definition of $\delta$ is very small in
these regions and can lead to sharp changes in $\delta$. Note that the
bulk of the changes between the old and the new implementation in
these cases stems from the vertex corrections. In the case of the
decay $H_{h_s}\to Z A_2$ the relative vertex corrections are mostly
small, with values $|\delta|<5$\% (red triangle-marked points). For
large relative changes $\Delta_{\text{BR}} (H_{h_s} Z A_2)$ the $\ZH$
factor is responsible then.

\subsection{Decays into  Charginos and  Neutralinos}
We start by investigating the loop corrections to the masses of the
charginos and neutralinos using the three renormalization schemes OS1,
OS2 and $\DRb$, imposed on the two gaugino masses $M_1$ and $M_2$, as
defined in \sect{sssect:NCrenor}. According to the SLHA format that we
apply in our code, $M_1, M_2$ are 
$\DRb$ parameters given at the scale
$M_{\text{SUSY}} = \sqrt{m_{\tilde{Q}_3} m_{\tilde{t}_R}}$. When we
use the OS schemes we have to 
translate the $\DRb$ input parameters to the OS values by applying
the approximate transformation formulae
\bea 
M_1^{\text{OSi}} &=& M^{\DRb}_1 -\de M_1^{\text{finOSi}}\crn
M_2^{\text{OSi}} &=& M^{\DRb}_2 -\de M_2^{\text{finOSi}}  \;,
\eea
 where $\de M_{1/2}^{\text{finOSi}}$ are the finite parts of the
 $M_{1/2}$ counterterms computed in the $\text{OSi}$ ($i=1,2$) renormalization
 scheme.  Since the finite parts $\de M_{1/2}^{\text{finOSi}}$ should
 be computed with OS input parameters we have used an iterative method
 to obtain these. For all parameter points in our scan, the size of the
 loop corrections to the neutralino and charginos masses, quantified
 by $\Delta M^{\tilde{\chi}}_i =
 M_{\tilde{\chi}_i}^{\text{loop}}-
 m_{\tilde{\chi}_i}^{\text{tree}}$, with $i=1,...,5$ for the
 neutralinos and $i=1,2$ for the charginos, never exceeds 46 GeV. 
\s

We exemplary present here a particular point, called {\it scenario1},
with the soft SUSY breaking masses and trilinear couplings given by
\beq
&&  
m_{\tilde{t}_R}=1384\,\gev \,,\;
 m_{\tilde{Q}_3}=1743\,\gev\,,\; m_{\tilde{b}_R}=
m_{\tilde{L}_3}= m_{\tilde{\tau}_R}=3000\,\gev \,,
 \non\\ 
&& |A_{u,c,t}| = 3594\,\gev\, ,\; |A_{d,s,b}|=2000\,\gev\,,\; |A_{e,\mu,\tau}| =2000\,\gev\,,\; \\ \non
&& |M_1| = 560\,\gev,\; |M_2|= 684\,\gev\,,\; |M_3|=2494\,\gev \,,\\ \non
&&  \varphi_{A_{e,\mu,\tau}}=\varphi_{A_{d,s,b}}=0\,,\; 
\varphi_{A_{u,c,t}}=\varphi_{M_1}=\varphi_{M_2}=\varphi_{M_3}=0
 \;, \label{eq:param4scen2}
\eeq
and the remaining input parameters set to\footnote{The
  imaginary part of $A_\kappa$ is obtained from the tadpole
  condition.}
\beq
&& |\lambda| = 0.307 \;, \quad |\kappa| = 0.517 \; , \quad \Re A_\kappa = 361\,\gev\;,\quad 
|\mu_{\text{eff}}| = 272\,\gev \;, \non \\ 
&&\varphi_{\lambda}=\varphi_{\kappa}=\varphi_u=0\;,
\quad \varphi_{\mu_{\text{eff}}}=\pi \;,  
\quad \tan\beta = 9.38 \;,\quad M_{H^\pm} = 1393 \,\gev \;.
\eeq
The Higgs boson masses and their main composition in terms of
singlet/doublet and scalar/pseu\-do\-sca\-lar components at  two-loop
order ${\cal O}(\alpha_t \alpha_s + \alpha_t^2)$ 
for $\DRb$ renormalization in the top/stop sector
computed by {\tt NMSSMCALC}, are summarized in
Table~\ref{tab:massvalues1}. \s
\begin{table}[t!]
\begin{center}
 \begin{tabular}{|l||c|c|c|c|c|}
\hline
$\DRb$ & ${H_1}$ & ${H_2}$ & ${H_3}$ & ${H_4}$ & ${H_5} $  \\ \hline \hline 
two-loop ${\cal O}(\alpha_t \alpha_s+ \alpha_t^2)$ & 125.14 &  698.05 &  813.53 &  1391.6 &  1392.56\\ \hline  
main component & $h_u$ &  $a_s$ &  $h_s$ &  $a$ &  $h_d$\\ \hline  
\end{tabular}
\caption{{\it Scenario1:} Masses and main components of the neutral
  Higgs bosons at two-loop order ${\cal O}(\alpha_t \alpha_s+
  \alpha_t^2)$, using $\DRb$ renormalization in the top/stop sector.}
\label{tab:massvalues1}
\end{center}
\end{table}

For {\it scenario1}, we present in \tab{tab:massneucha1} the
tree-level and loop-corrected 
masses of the neutralinos and charginos in the three different  
renormalization schemes and for the Denner description. 
As expected the wino-like neutralino and  
the wino-like chargino which couple to the electroweak
  gauge bosons, get significant loop corrections in the $\DRb$ 
scheme. The one-loop corrected masses themselves, however, barely
differ in the three renormalization schemes so that the remaining
theoretical error due to missing higher-order corrections is very
small. \s

\begin{table}[t!]
\begin{center}
 \begin{tabular}{|ll||c|c|c|c|c|c|c|c|c|c|c|c|c|c|}
\hline
~~ & &${M_{\tilde{\chi}_1^0}}$&${M_{\tilde{\chi}^0_2}}$&${M_{\tilde{\chi}_3^0}}$&${M_{\tilde{\chi}_4^0}}$&${M_{\tilde{\chi}_5^0}}$&${M_{\tilde{\chi}_1^+}}$&${M_{\tilde{\chi}_2^+}}$ \\ \hline \hline
 \multirow{2}{*} {OS1}  & tree-level & 265.97 &  276.05 &  565.2 &  730.92 &  920.76 &  270.72 &  730.83\\  
                       & one-loop & 273.22 &  282.48 &  565.2 &  730.78 &  914.39 &  278.02 &  730.83\\ \hline 
\multirow{2}{*} {OS2}  & tree-level & 265.97 &  276.05 &  565.2 &  730.78 &  920.76 &  270.72 &  730.69\\  
                       & one-loop & 273.22 &  282.48 &  565.2 &  730.78 &  914.39 &  278.02 &  730.83\\ \hline 
\multirow{2}{*} {$\DRb$}  & tree-level & 265.5 &  276.2 &  563.47 &  694.32 &  920.76 &  270.38 &  694.18\\  
                       & one-loop & 273.21 &  282.44 &  565.21 &  730.01 &  914.39 &  278.01 &  729.86\\ \hline 
& main component & $\tilde H_d^0$ &  $\tilde H_u^0$ &  $\tilde{B}$ &  $\tilde W_3$ &  $\tilde S $ &  $\tilde H^+$ &  $\tilde{W}^+$\\ \hline  
\end{tabular}
\caption{{\it Scenario1:} Masses and main components of the
  neutralinos and charginos at  
  tree and one-loop level in the three renormalization schemes OS1,
  OS2 and $\DRb$.}
\label{tab:massneucha1}
\end{center}
\vspace*{-0.9cm}
\end{table}

\begin{figure}[t!]
\includegraphics[width=0.45\textwidth]{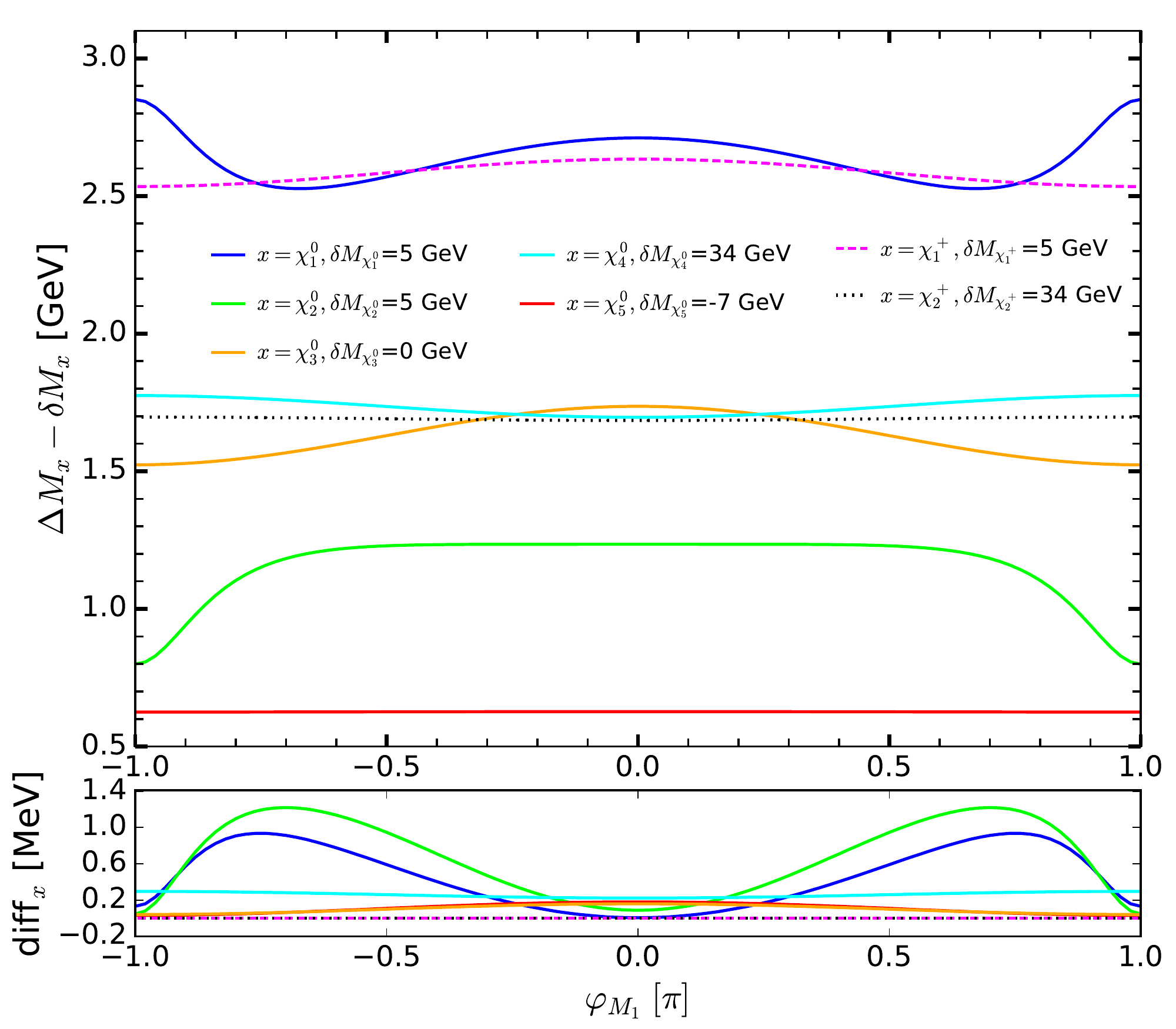}
\includegraphics[width=0.45\textwidth]{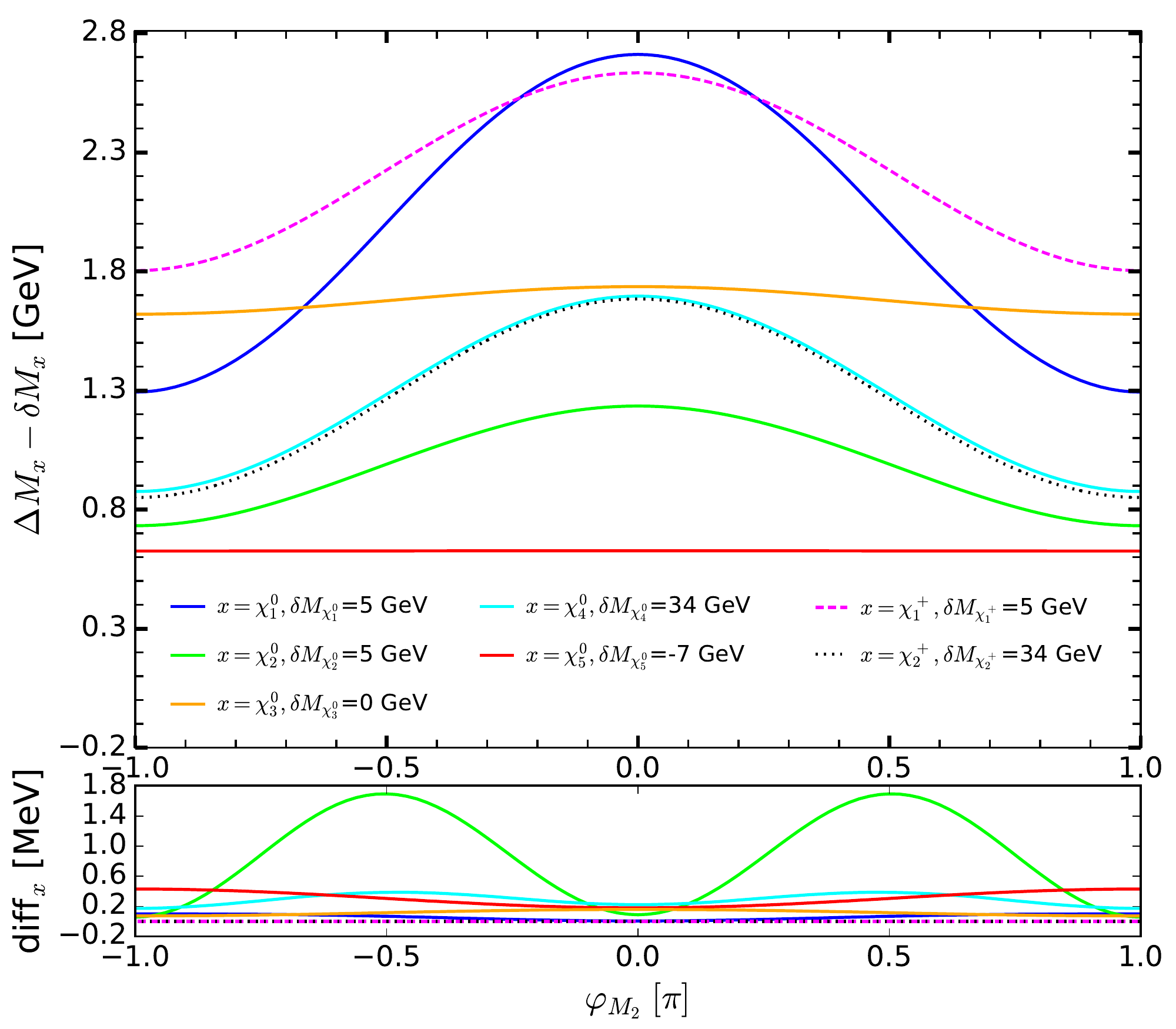}
  \caption{{\it Scenario1}: Upper: loop corrections to the electroweakino masses, 
$\Delta M_{x}$ (see text, for the definition), $x= \tilde{\chi}_i^0,
\tilde{\chi}_j^\pm$ ($i=1,...,5$, $j=1,2$), 
    as a function of $\varphi_{M_1}$ (left) and $\varphi_{M_2}$
    (right) in the $\DRb$ scheme using the Denner description. The
    subtraction values $\delta M_x$ are specified in the legends. 
Lower: differences $\text{diff}_x $ (see text, for the definition)
between the loop-corrected masses using  the Denner and the EMT descriptions.
 \label{fig:plotMchimass}}
\end{figure} 

We vary the phases of the gaugino mass parameters $M_1$ and $M_2$,
$\varphi_{M_1}$ and $\varphi_{M_2}$, in order to study their effect on
the loop-corrected neutralino and chargino masses. Note that these
complex phases have negligible impact on the Higgs sector
\cite{Graf:2012hh,Dao:2019qaz}. We use the $\DRb$ scheme for this analysis and 
show in the upper plots of \figref{fig:plotMchimass} the loop corrections to the
electroweakino masses,
\beq
\Delta M_{x} = M_{x}^{\text{loop}} - m_x^{\text{tree}}\;, \qquad
x= \tilde{\chi}_i^0, \tilde{\chi}_j^\pm \; (i=1,...,5, j=1,2) \;,
\label{eq:diffinomass}
\eeq
as function of $\varphi_{M_1}$ (left) and $\varphi_{M_2}$ (right). We apply a
subtraction $\delta M_x$ to the mass corrections $\Delta M_x$ of the
different electroweakinos that allows us to show all corrections,
which can be very different in size, in one plot. In the lower plots of
\figref{fig:plotMchimass} 
we show the differences
\beq
\text{diff}_x =
M_{x}^{\text{loop,\;D}}-M_{x}^{\text{loop,\;EMT}} \;,
\eeq 
with $x$ denoting any electroweakino, 
between the loop corrected electroweakino masses computed using the
Denner and the EMT 
descriptions presented in \sect{sssect:NCrenor}. As can be inferred
from the upper plots, the wino-like neutralino $\tilde\chi_4^0$ (cyan
line) and wino-like chargino $\tilde{\chi}_2^+$ (black line) receive
the largest loop corrections of about 35~GeV in absolute values. The
corrections to the Higgsino-like neutralinos $\tilde{\chi}_1^0$ (blue
line) and $\tilde{\chi}_2^0$ (green line), the singlino-like neutralino
$\tilde{\chi}_0^5$ (red line) and the Higgsino-like chargino
$\tilde{\chi}_1^+$ (pink line) range around 6-8~GeV. The correction to
the bino-like neutralino $\tilde{\chi}_3^0$ (orange line) is somewhat
smaller with values around 1.5-1.8~GeV. A small difference between the
Denner and the EMT descriptions of about 1-2~MeV is observed for
the Higgsino-like neutralinos, {\it cf.}~lower plots.  We do not see
any difference, however, for the
chargino masses. This is because we have used the approximation in
\eqref{eq:appMcha} for the loop-corrected chargino masses. There is a
compensation between 
$\Re\Sigma^{\tilde{\chi}^+,Ls}_{ii}(m_{\tilde{\chi}_i^\pm}^2)$ and
$\Re\Sigma^{\tilde{\chi}^+,Rs}_{ii}(m_{\tilde{\chi}_i^\pm}^2)$ that kills the
effect of the imaginary part of the loop integral function. \s
 
\begin{table}[t!]
\begin{center}
 \begin{tabular}{|ll||c|c|c| |l||c|c|c|}
\hline
~~ & &$\Ga^l [\mev]$&$\de [\%]$&BR$^l$ [$\%$]& ~~ &$\Ga^l [\mev]$&$\de [\%]$&BR$^l$ [$\%$] \\ \hline \hline
 \multirow{2}{*} {$H_2\to\ti \chi_1^0\ti \chi_1^0$}  & OS1     &405.67 & 0.08 & 25.14 & \multirow{2}{*} {$H_2\to\ti \chi_1^0\ti \chi_2^0$}     &0.43 & -0.08 & 0.03 \\
                                                   &  $\DRb$   &407.74  &  8.4  &  25.15&                                                     &0.4  &  -16.34  &  0.02 \\ \hline
\multirow{2}{*} {$H_2\to\ti \chi_2^0\ti \chi_2^0$}  & OS1     &404.44 & 0.07 & 25.06 & \multirow{2}{*} {$H_2\to\ti \chi_1^+\ti \chi_1^-$}  &802.3 & 0.08 & 49.72 \\
                                                   &  $\DRb$   &404.84  &  7.23  &  24.98 &                                                   &  807.18  &  9.03  &  49.8\\ \hline
\multirow{2}{*} {$H_3\to\ti \chi_1^0\ti \chi_1^0$}  & OS1     &321.53 & 0.08 & 22.01& \multirow{2}{*} {$H_3\to\ti \chi_1^0\ti \chi_2^0$}   &0.78 & -0.11 & 0.05 \\
                                                   &  $\DRb$   &322.45  &  8.13  &  22.04 &                                                    &0.73  &  -16.24  &  0.05 \\\hline
\multirow{2}{*} {$H_3\to\ti \chi_2^0\ti \chi_2^0$}  & OS1     &308.94 & 0.07 & 21.15 &\multirow{2}{*} {$H_3\to\ti \chi_1^+\ti \chi_1^-$}  & 633.52 & 0.08 & 43.38 \\ 
                                                   &  $\DRb$   &309.0  &  7.08  &  21.12 &                                                    & 635.37  &  8.79  &  43.42 \\\hline
\multirow{2}{*} {$H_4\to\ti \chi_1^0\ti \chi_1^0$}  & OS1     &102.86 & 0.02 & 0.8 & \multirow{2}{*} {$H_5\to\ti \chi_1^0\ti \chi_1^0$}     &151.31 & -0.01 & 1.18 \\
                                                   &  $\DRb$   &89.4  &  -11.31  &  0.69 &                                                  & 132.77  &  -13.56  &  1.04 \\\hline
\multirow{2}{*} {$H_4\to\ti \chi_1^0\ti \chi_2^0$}  & OS1     &17.87 & 0.08 & 0.14 &\multirow{2}{*} {$H_5\to\ti \chi_1^0\ti \chi_2^0$}      &9.29 & 0.24 & 0.07 \\
                                                   &  $\DRb$   &14.59  &  -11.68  &  0.11&                                                  & 7.65  &  1.94  &  0.06  \\\hline
\multirow{2}{*} {$H_4\to \ti\chi_1^0\ti\chi_3^0$}  & OS1     &395.9 & -5.5 & 4.59&\multirow{2}{*} {$H_5\to\ti \chi_1^0\ti \chi_3^0$}      &423.04 & 0.02 & 3.3 \\
                                                    & $\DRb$ & 451.86  &  6.59  &  3.5&                                                  &448.9  &  7.83  &  3.5 \\\hline
\multirow{2}{*} {$H_4\to\ti \chi_1^0\ti \chi_4^0$}  & OS1     &1448.21 & -0.03 & 11.22 &\multirow{2}{*} {$H_5\to\ti \chi_1^0\ti \chi_4^0$}     &1161.18 & -0.02 & 9.05 \\
                                                    & $\DRb$ &1438.6  &  -3.68  &  11.15&                                                  &1157.98  &  -2.81  &  9.03  \\\hline
\multirow{2}{*} {$H_4\to\ti \chi_1^0\ti \chi_5^0$}  & OS1     &194.66 & -0.02 & 1.51 &\multirow{2}{*} {$H_5\to\ti \chi_1^0\ti \chi_5^0$}    &338.35 & -0.02 & 2.64\\
                                                    & $\DRb$ &195.02  &  -1.63  &  1.51&                                                  & 341.12  &  -1.58  &  2.66 \\\hline
\multirow{2}{*} {$H_4\to\ti \chi_2^0\ti \chi_2^0$}  & OS1     &27.86 & -0.13 & 0.22 & \multirow{2}{*} {$H_5\to\ti \chi_2^0\ti \chi_2^0$}     &34.6 & -0.1 & 0.27  \\
                                                    & $\DRb$ &25.54  &  -20.08  &  0.2  &                                                  &32.72  &  -15.35  &  0.26\\\hline
\multirow{2}{*} {$H_4\to\ti \chi_2^0\ti \chi_3^0$}  & OS1     &425.3 & 0.02 & 3.29 &\multirow{2}{*} {$H_5\to\ti \chi_2^0\ti \chi_3^0$}     &413.74 & 0.01 & 3.22 \\
                                                    & $\DRb$ & 450.13  &  7.77  &  3.49 &                                                  &438.11  &  6.48  &  3.42  \\\hline
\multirow{2}{*} {$H_4\to\ti \chi_2^0\ti \chi_4^0$}  & OS1     &1125.46 & -0.02 & 8.72 & \multirow{2}{*} {$H_5\to\ti \chi_2^0\ti \chi_4^0$}     &1379.31 & -0.03 & 10.75 \\
                                                    & $\DRb$ & 1116.5  &  -3.34  &  8.66 &                                                &1366.13  &  -3.91  &  10.66  \\\hline
\multirow{2}{*} {$H_4\to\ti \chi_2^0\ti \chi_5^0$}  & OS1     &382.4 & -0.02 & 2.96 &\multirow{2}{*} {$H_5\to\ti \chi_2^0\ti \chi_5^0$}     &180.21 & -0.02 & 1.4  \\
                                                    & $\DRb$ & 381.89  &  -2.35  &  2.96 &                                               &    180.34  &  -1.86  &  1.41 \\\hline
\multirow{2}{*} {$H_4\to\ti \chi_3^0\ti \chi_3^0$}  & OS1     &1.2 & 0.19 & 0.01  & \multirow{2}{*} {$H_5\to\ti \chi_3^0\ti \chi_3^0$}    &1.19 & 0.18 & 0.01 \\
                                                    & $\DRb$ &1.29  &  28.87  &  0.01 &                                                  & 1.29  &  27.81  &  0.01  \\\hline
\multirow{2}{*} {$H_4\to\ti \chi_3^0\ti \chi_4^0$}  & OS1     &3.18 & 0.05 & 0.02 &\multirow{2}{*} {$H_5\to\ti \chi_3^0\ti \chi_4^0$}    &1.48 & 0.06 & 0.01 \\
                                                    & $\DRb$ &2.95  &  -2.79  &  0.02 &                                                  &  1.43  &  0.89  &  0.01  \\\hline
\multirow{2}{*} {$H_4\to\ti \chi_1^+\ti \chi_1^-$}  & OS1     &393.59 & -0.01 & 3.05 & \multirow{2}{*} {$H_5\to\ti \chi_1^+\ti \chi_1^-$}    &257.68 & -0.03 & 2.01 \\
                                                    & $\DRb$ & 343.95  &  -13.61  &  2.67 &                                                  &215.16  &  -19.09  &  1.68  \\\hline
\multirow{2}{*} {$H_4\to\ti \chi_1^+\ti \chi_2^-$}  & OS1     &2440.51 & -0.01 & 18.91 & \multirow{2}{*} {$H_5\to\ti \chi_1^+\ti \chi_2^-$}     &2482.01 & -0.01 & 19.34  \\
                                                    & $\DRb$ & 2461.41  &  -0.61  &  19.0 &                                                  &2498.8  &  -0.9  &  19.49 \\\hline
\multirow{2}{*} {$H_4\to\ti \chi_2^+\ti \chi_1^-$}  & OS1     &2440.51 & -0.01 & 18.91 & \multirow{2}{*} {$H_5\to\ti \chi_2^+\ti \chi_1^-$}   &2482.01 & -0.01 & 19.34  \\
                                                    & $\DRb$ & 2461.41  &  -0.61  &  19.0 &                                                  &2498.8  &  -0.9  &  19.49 \\\hline
 \hline
\end{tabular}
\caption{{\it Scenario 1:} Loop-corrected decay widths $\Gamma^l$,
  relative loop corrections $\delta$ and loop-corrected branching
  ratios BR$^l$ of all kinematically allowed decays into chargino and
  neutralino pairs of heavy Higgs bosons in the OS1 and $\DRb$
  renormalization schemes. The results in the OS2 scheme are nearly the
  same as in OS1 scheme.} 
\label{tab:correctionneucha1}
\end{center}
\end{table}

In order to study the loop corrections on the decay widths, we
computed the tree-level and loop-corrected decay widths, defined
in \sect{sec:Hchaneu}, for the three different renormalization schemes
OS1, OS2 and $\DRb$ for the {\it scenario1}. Note that we use the
loop-corrected masses for external Higgs bosons, charginos and
neutralinos not only in the loop-corrected decay widths but also in
the tree-level ones. For illustration, we present in
\tab{tab:correctionneucha1} for the decays of all Higgs bosons in all
possible electroweakino final states the loop-corrected decay widths
$\Gamma^{l} \equiv \Gamma_{\ZH}^{\text{SEW}} (H_i \to \tilde{\chi}_j \tilde{\chi}_k)$,
the relative corrections $\delta (H_i \tilde{\chi}_j \tilde{\chi}_k)$ as defined in
Eq.~(\ref{eq:correction}), and the loop-corrected branching ratios
$\mbox{BR}^l \equiv \mbox{BR}_{\ZH}^{\text{SEW}} (H_i \to \tilde{\chi}_j \tilde{\chi}_k)$
using {\it scenario1}, for the OS1 and $\DRb$
  renormalization schemes. We found that $\Ga^{l}$ is
almost identical in the two OS schemes.  
 The relative size of the loop corrections is
below 10\% in the OS scheme. The relative corrections in the
$\DRb$ scheme are always larger than 
in the OS schemes. For some channels with small decay widths, we see 
significant corrections in the $\DRb$ scheme. For example  in the
decay channel $H_4\to \ti\chi_2^0 \ti\chi_2^0$, 
the relative loop correction is $-0.13\%$ in the OS scheme while it is
$-20\%$ in the $\DRb$ scheme. 
Based on our investigation, we conclude that it is better to use the
OS scheme in the decays of the neutral Higgs bosons into
electroweakinos. The largest uncertainty due to missing higher-order
corrections that we 
estimate from the variation of the renormalization schemes is found to
be 17\% in the decay $H_5\to \ti \chi_1^0\ti\chi_2^0$.
We also studied the difference between the Denner and EMT descriptions and
did not observe any significant difference. Defining the difference as
$(\Gamma^{\text{D}}_x -\Gamma^{\text{EMT}}_x)/\Gamma^{\text{D}}_x$
with $\Gamma_x$ being the loop-corrected decay width for some Higgs
decay into an electroweakino final state, we see that
the differences are of per mille level for all investigated decays. 

\subsection{Decays into a Squark Pair}
We start by discussing the top and bottom squark masses in the OS and
$\DRb$ renormalization schemes defined in \sect{sect:Resquark}. We
follow the SLHA convention where the input parameters $m_{\ti Q_3},
m_{\ti t_R}, m_{\ti b_R},  A_t,  A_b$ are $\DRb$ parameters
at the scale $M_{\text{SUSY}}$. When we apply the
OS scheme, these parameters must be translated into OS parameters by
applying the approximate transformation formula ($i=1,2$)
\be 
X^{\text{OSi}} = X^{\DRb} -\de X^{\text{finOSi}} \;,\label{eq:DROSconversion}
\ee
with $X=m_{\ti Q_3}, m_{\ti t_R}, m_{\ti b_R},  A_t,  A_b$ and the
finite part of their OS counterterms denoted as 
 $\de X^{\text{finOSi}}$. We have used an iterative method 
to obtain a stable value of $\de X^{\text{finOSi}}$. Note that we
include both the NLO QCD and the full NLO EW contribution in the
conversion \eqref{eq:DROSconversion}.\footnote{This is a bit different
  from the Higgs mass calculation in {\tt NMSSMCALC} where we include
  the NLO QCD correction and the 
NLO Yukawa correction of order ${\cal O}(\alpha_t)$ to the conversion
in the OS renormalization scheme of the top/stop sector.} Using 
these OS parameters together with the OS top mass in the tree-level
mass matrices, we obtain the top and bottom squark masses. When we
apply the $\DRb$ renormalization scheme, the top pole mass has to be
translated to the $\DRb$ top mass for which we follow the description
in appendix C of  \bib{Dao:2019qaz}. The $\DRb$ top and bottom masses
together with the $\DRb$ parameters $m_{\ti Q_3},
m_{\ti t_R}, m_{\ti b_R},  A_t,  A_b$ are then used in the tree-level
mass matrices to get the tree-level rotation matrices. They are subsequently
used in the computation of the renormalized self-energies of the top and
bottom squarks to  obtain the loop-corrected squark masses as
described in \sect{ssectsquark}. In principle, we would expect that 
the loop-corrected masses computed in the $\DRb$ scheme are closer to
the OS masses in the OS description if one includes 
more higher order corrections. We consider here a parameter point
({\it scenario2}) given by 
the following soft SUSY breaking masses and trilinear
couplings\footnote{This parameter point 
is allowed by {\tt HiggsBounds}5.3.2 and its $\chi^2$ computed by
{\tt HiggsSignals}2.2.3 is consistent with an SM $\chi^2$ less than 1$\sigma$.} 
\beq
&&  m_{\tilde{u}_R,\tilde{c}_R} = 
m_{\tilde{d}_R,\tilde{s}_R} =
m_{\tilde{Q}_{1,2}}= m_{\tilde L_{1,2}} =m_{\tilde e_R,\tilde{\mu}_R} = 3\;\mbox{TeV}\, , \;  
m_{\tilde{t}_R}=623\,\gev \,,\; \non \\ \non
&&  m_{\tilde{Q}_3}= 1180\,\gev\,,\; m_{\tilde{b}_R}=
m_{\tilde{L}_3}= m_{\tilde{\tau}_R}=3 3\;\mbox{TeV} \,,
 \\ 
&& |A_{u,c,t}| = 1760\,\gev\, ,\; |A_{d,s,b}|=2000\,\gev\,,\; |A_{e,\mu,\tau}| = 2000\,\gev\,,\; \\ \non
&& |M_1| = 1000\,\gev,\; |M_2|= 1251\,\gev\,,\; |M_3|=2364\,\gev \;,
\label{eq:param1}
\eeq
with the CP phases given by 
\beq
&&  \varphi_{A_{u,c,t}}=\varphi_{A_{d,s,b}}=
\varphi_{A_{e,\mu,\tau}}=\varphi_{M_1}=\varphi_{M_2}=\varphi_{M_3}=0 
\;. \label{eq:param2}
\eeq
The remaining input parameters have been set to
\beq
&& \lambda = 0.106 \;, \quad \kappa = -0.238 \; , \quad \mbox{Re}(A_\kappa) = -647\,\gev\;,\quad 
\mu_{\text{eff}} = -603\,\gev \;, \non \\ 
&&\varphi_u=0\;,
\quad  \tan\beta = 17.5 \;,\quad M_{H^\pm} = 1867 \,\gev \;.
\eeq
With the given $\DRb$ parameters of the squark sector, we obtain their
corresponding OS parameters  
\beq 
&& m_{\ti Q_3}=1120\,\gev \; , \quad
 m_{\ti t_R}=402\,\gev \; , \quad  m_{\ti b_R}=2997\,\gev \; ,\crn
&& \quad A_t=1720\,\gev \; , \quad A_b=-581\,\gev\:.
\eeq
The tree-level and loop-corrected masses of the stops and sbottoms in
the OS and $\DRb$ scheme 
are shown in \tab{tab:StopSbottomMasses}. We see that for the $\DRb$
scheme there are large changes between the tree-level and one-loop
masses, in particular for the lightest stop $\tilde{t}_1$. The
loop-corrected masses, however, are then closer to each other in both
schemes, as expected. The maximum difference if found for the light
sbottom $\tilde{b}_1$ mass, where the OS and $\DRb$ results differ by
27~GeV at one-loop order (compared to 59~GeV at tree level). 
The two-loop corrected neutral Higgs boson masses at ${\cal O}(\alpha_t
\alpha_s + \alpha_t^2)$ together with their respective main component
are displayed in Table~\ref{tab:massPSTOP}. The SM-like Higgs boson mass is around
124~GeV while the remaining Higgs spectrum is quite heavy with masses
above 1.6~TeV. \s

\begin{table}[t!]
\begin{center}
 \begin{tabular}{|ll||c c c c|}
 \hline
    ~~&  & $m_{\ti t_1} \,[\mbox{GeV}] $ &$m_{\ti t_2} \,[\mbox{GeV}]$
   & $m_{\ti b_1}\,[\mbox{GeV}]$ & $m_{\ti b_2}\,[\mbox{GeV}]$\\ 
\hline
\multirow{2}{*}  {OS} & tree &  334  & 1166 & 1122  & 2297 \\
                      & 1loop & 334 & 1166& 1125 & 2297 \\ \hline
\multirow{2}{*}  {$\DRb$} & tree &  585  & 1216 & 1181  & 3000 \\
                      & 1loop & 341 & 1152& 1152 & 2294 \\ \hline
\end{tabular}
\caption{{\it Scenario2}: The tree-level and one-loop corrected stop
  and sbottom masses in the $\DRb$ and OS schemes.} 
\label{tab:StopSbottomMasses}
\end{center}
\vspace*{-0.5cm}
\end{table}

\begin{table}[t!]
\begin{center}
 \begin{tabular}{|l||c|c|c|c|c|}
\hline
 &${H_1}$&${H_2}$&${H_3}$&${H_4}$&${H_5}$\\ \hline \hline
two-loop ${\cal O}(\alpha_t \alpha_s+ \alpha_t^2)$
 &123.63 & 1621.65 & 1865.39 & 1895.83 & 2538.29\\  
main component&$h_u$&$a_s$&$h_d$&$a$&$h_s$\\ \hline
\end{tabular}
\caption{{\it Scenario2}: Mass values in GeV and main components of the neutral Higgs
  bosons at  two-loop order ${\cal O}(\alpha_t \alpha_s +
  \alpha_t^2)$ obtained by using the $\DRb$ renormalization in the top/stop sector.}
\label{tab:massPSTOP}
\end{center}
\vspace*{-0.7cm}
\end{table}

\begin{figure}[b!]
\includegraphics[width=0.45\textwidth]{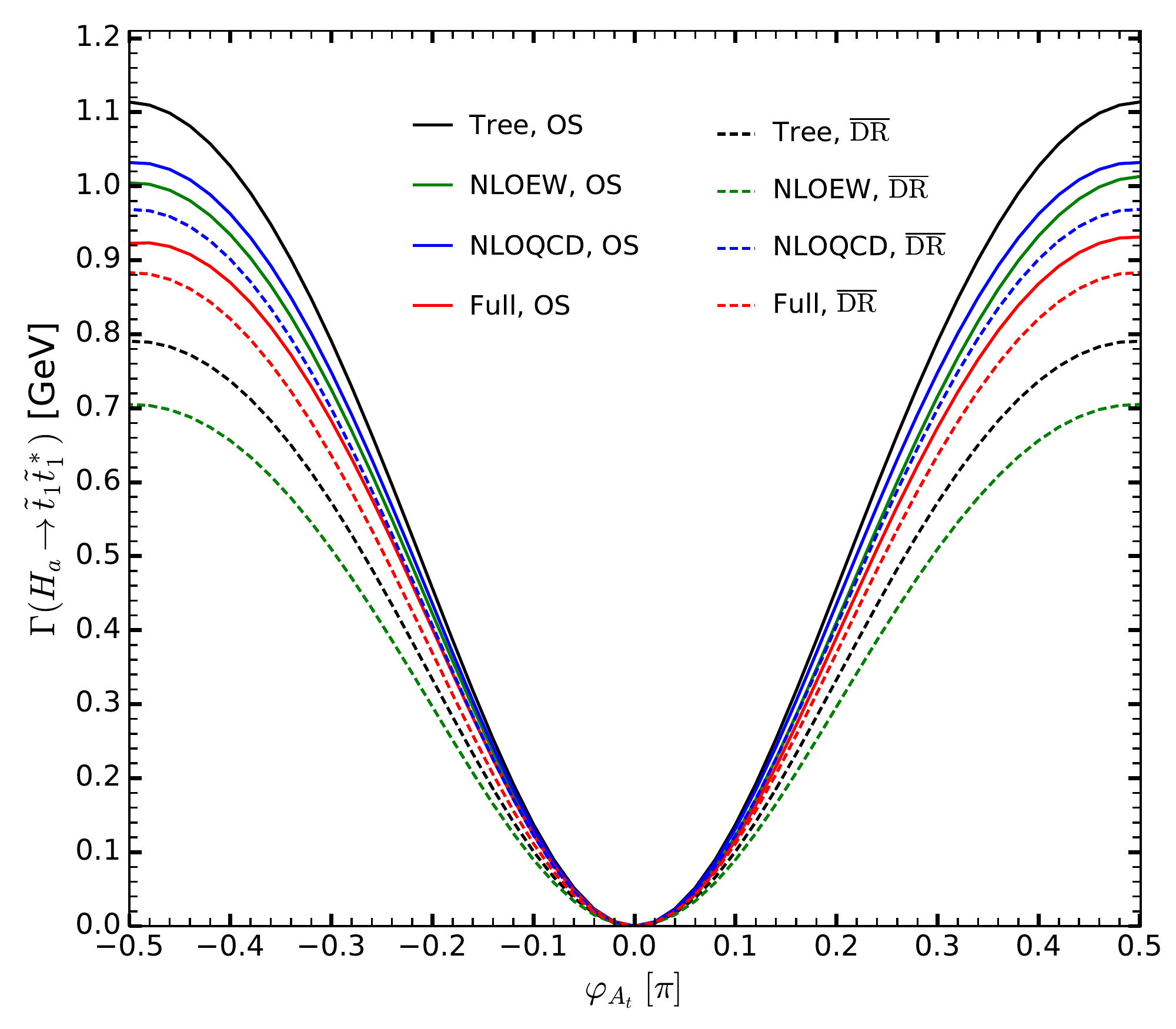}
\includegraphics[width=0.45\textwidth]{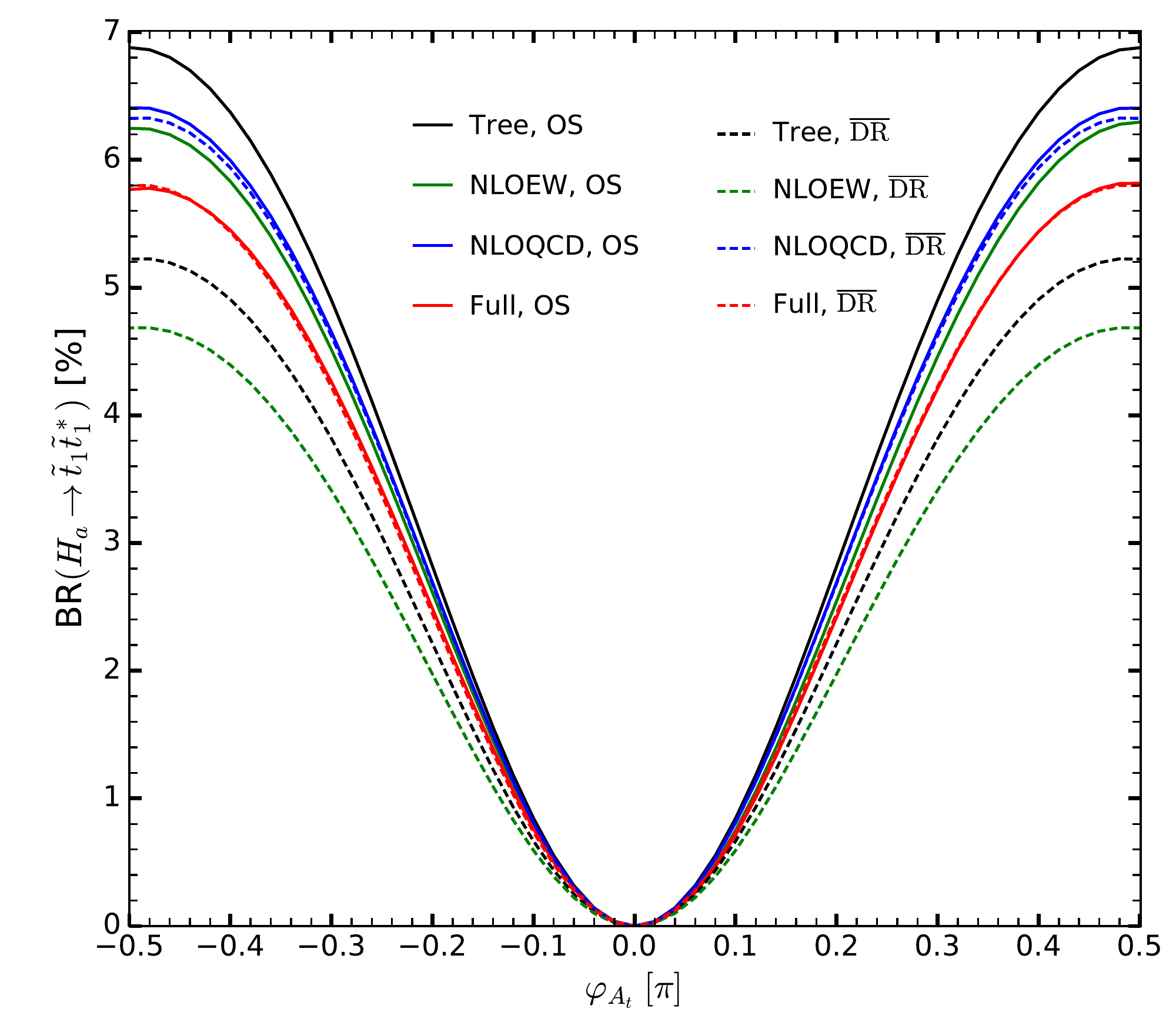}
  \caption{{\it Scenario2}: Tree-level (black), NLO EW (green),
    NLO QCD (blue) and full (red) partial width (left) and branching
    ratio (right) of the decay $H_4\to \ti t_1\ti
    t_1^*$ ($H_4$ is $a$-like) as function of the complex 
    phase $\varphi_{A_t}$. They are shown for the OS (full) and $\DRb$ scheme (dashed).}
 \label{fig:plotH4st1st1}
\end{figure}

We now turn on the complex phase of $A_t$. In the left plot of
\figref{fig:plotH4st1st1}, we show the tree-level (black), NLO EW
(green), NLO QCD (blue), and full, {\it i.e.}~NLO
  QCD+EW, (red) corrections to the
partial width of the decay $H_4\to \ti t_1 \ti t_1^*$ as function of
the phase $\varphi_{A_t}$ in both the OS 
(full lines) and the $\DRb$ (dashed lines) 
schemes while their corresponding branching ratios are
depicted in the right plot. The decay $H_4\to \ti t_1 \ti t_1^*$
vanishes in the CP-conserving limit where $H_4$, which
  is $a$-like in {\it scenario2}, is a CP-odd Higgs 
boson. (Note that CP-odd Higgs bosons at tree-level only couple to two
  different stops.)
In the OS scheme, the relative EW corrections $\delta$ (see Eq.~(\ref{eq:correction}) for
the definition) vary in the range (-6\%,-10 \%) and the
QCD corrections in the range $(-4\%, -8\%)$ depending on the phase
$\varphi_{A_t} $ that is varied from zero to
$\pm\pi/2$. In the $\DRb$ scheme,  the relative EW  
corrections are of order $-10\%$ and the relative QCD
corrections of $22\%$ and depend slightly on the phase $\varphi_{A_t} $.
We define the relative differences between
the OS and $\DRb$ decay widths and branching ratios, as 
\be 
\Delta_\Gamma =\left| \fr{\Ga^{\text{OS}}_{\ZH} -\Ga^{\DRb}_{\ZH}
}{\Ga^{\text{OS}}_{\ZH}} \right|\;, \label{eq:diffosdr} 
\ee
and
\be
\Delta_{\text{BR}} = \left|\fr{\mathrm{BR}^{\text{OS}}_{\ZH} -\mathrm{BR}^{\DRb}_{\ZH}
}{\mathrm{BR}^{\text{OS}}_{\ZH}} \right| \;, \label{eq:diffosdrbr}
\ee
respectively. For $\varphi_{A_t}= -\pi/2$, the relative difference
$\Delta_\Gamma $ of the partial decay widths between the OS and $\DRb$
schemes is then about 40\%  at tree-level while 
it reduces to $4\%$ when both QCD and EW corrections are
included, so that at one-loop level we clearly see a reduction of the
theoretical error due to missing higher-order corrections For the
relative error in the branching ratios we find values
  between 32\% and 27\% at tree level and between  0.3\% and 3\% at
  one-loop order including both the EW and QCD corrections while the
  phase $\varphi_{A_t} $ is varied from $\pm
      \pi/2$ to zero. \s

\begin{figure}[t!]
\includegraphics[width=0.45\textwidth]{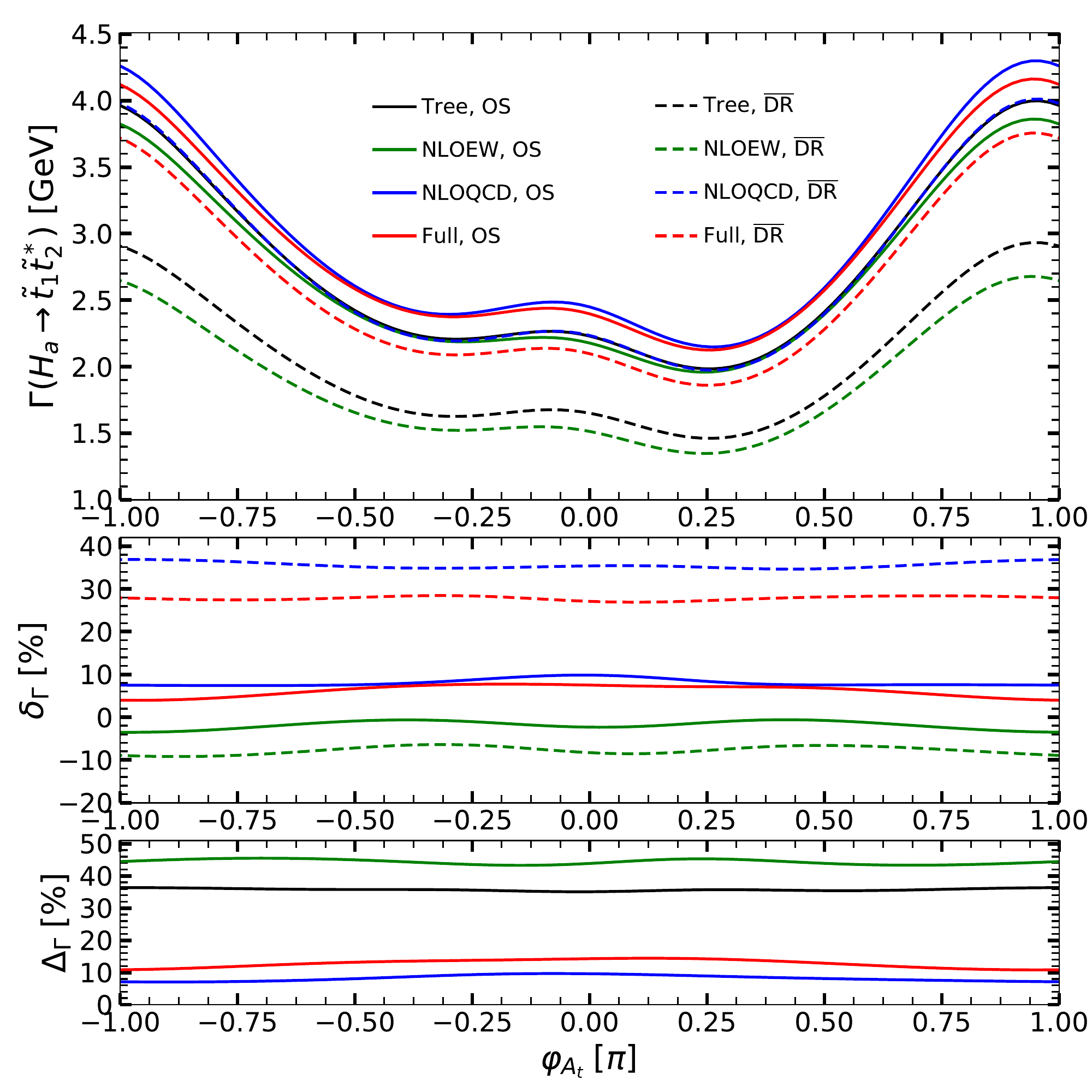}
\includegraphics[width=0.45\textwidth]{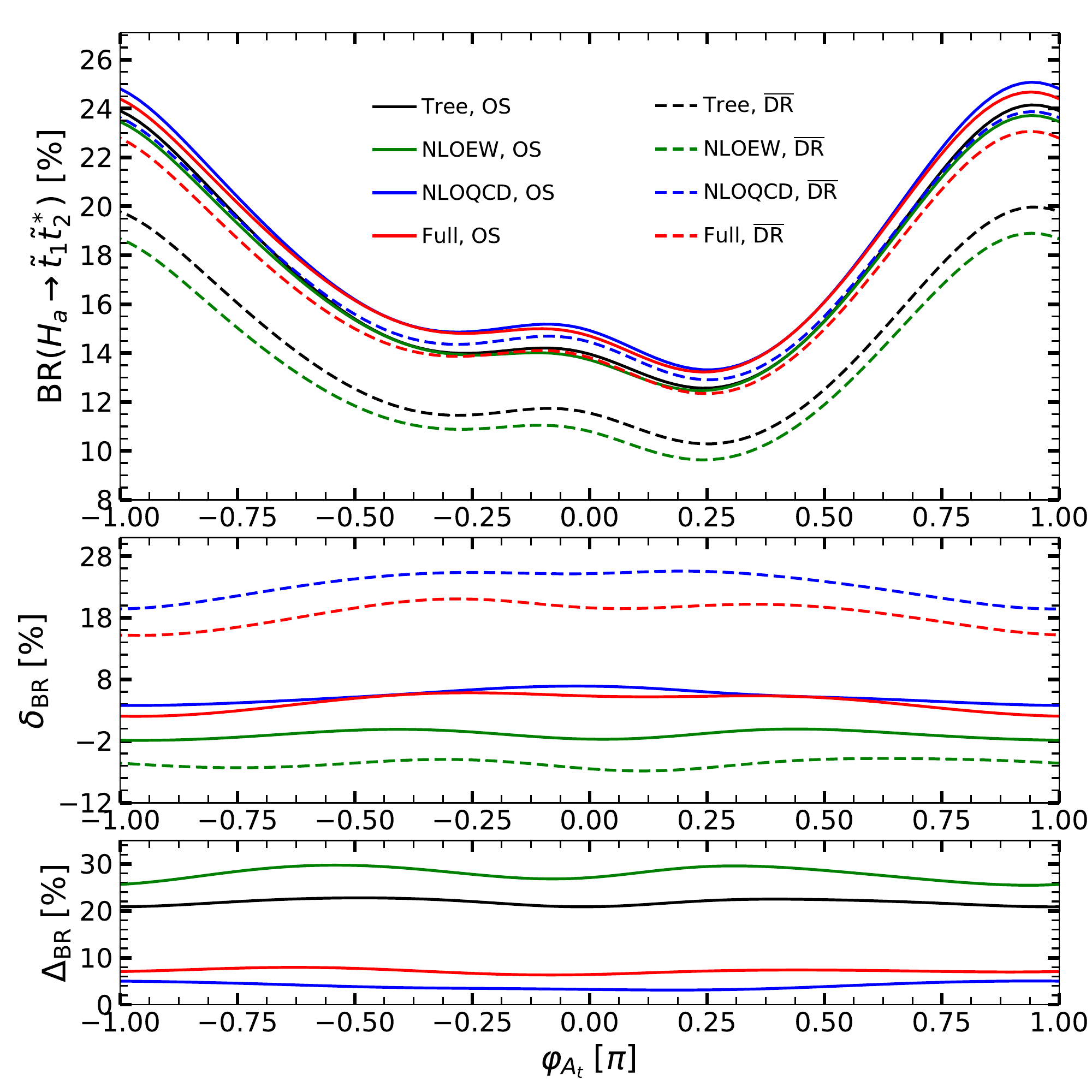}
  \caption{{\it Scenario2}: Upper: Tree-level (black), NLO EW (green),
    NLO QCD (blue) and full (red) partial width (left) and branching
    ratio (right) of the decay $H_4 (=H_a) \to \ti t_1\ti t_2^*$ as function
    of the complex phase $\varphi_{A_t}$. They are shown for the OS
    (full) and $\DRb$ (dashed) scheme. Middle: Relative EW, QCD and
    EW+QCD corrections $\delta$  (see text, for definition)
    for the decay width (left) and branching ratio (right) in the OS
    (full) and $\DRb$ (dashed) scheme.  Lower:
   Relative differences $\Delta$ between the OS and $\DRb$ scheme (see
   text, for definition) for the decay width (left) and branching
   ratio (right).}
 \label{fig:plotHast1st2}
\end{figure}

For the decay $H_4 (\equiv H_a) \to \ti t_1 \ti t_2^*$, we show in the upper panels of
\figref{fig:plotHast1st2} the partial decay width (left) and branching
ratio (right)  at tree-level
(black), NLO EW (green), NLO QCD (blue), and NLO EW+QCD (red) as a
function of $\varphi_{A_t}$, both for the OS (full) and $\DRb$
(dashed) scheme . 
In the middle panels we show the relative NLO EW, NLO QCD and NLO EW+QCD
corrections which are defined as 
\be 
\delta_\Gamma = \fr{\Ga^{\text{EW/QCD/EW+QCD}}_{\ZH} -\Ga^{\text{tree}}_{\ZH}
}{\Ga^{\text{Tree}}_{\ZH}}
\ee
and
\be 
\delta_{\text{BR}} = \fr{\mathrm{BR}^{\text{EW/QCD/EW+QCD}}_{\ZH}
  -\mathrm{BR}^{\text{tree}}_{\ZH} }{\mathrm{BR}^{\text{Tree}}_{\ZH}} \;,
\ee
respectively. The lower panels display the relative differences between
the OS and $\DRb$ decay widths and branching ratios, $\Delta_\Gamma$
and $\Delta_{\text{BR}}$, as defined in Eq.~(\ref{eq:diffosdr}) and
Eq.~(\ref{eq:diffosdrbr}), respectively. The corrections vary slightly with the phase
  $\varphi_{A_t}$. The EW corrections are negative in both schemes while the QCD
corrections are positive and of the same order of magnitude. This
shows the importance to include both types of corrections to make
reliable predictions. Overall, the relative corrections $\delta$ in the $\DRb$ scheme
are larger than in the OS scheme. 
As can be inferred from the bottom left panel of 
\figref{fig:plotHast1st2}, for $\varphi_{A_t}=0$, the
relative difference in the partial width, $\Delta_{\Gamma}$, between the OS and $\DRb$ 
scheme at tree level is about $37\%$ and decreases dramatically to
less than $12\%$ when both the EW and the QCD corrections are
included.\footnote{When we only include the EW
    corrections the scheme dependence increases when going from tree-
    to one-loop level. Overall, the behavior is as expected, however,
  when the full set of corrections is included. This shows that care
  has to be taken, when estimating the theoretical uncertainty due to
  missing higher-order corrections based on a change of the
  renormalization schemes, if not all corrections of a given loop
  order are included. See also Ref.~\cite{Dao:2019qaz} for a similar discussion.}
Similar results are found for the branching ratios,
presented on the right plots of \figref{fig:plotHast1st2}, with
smaller values of $21\%$ at tree level and $7\%$ at full one-loop
order. Note that in the right hand side plots we treated the decays
$H_a\to \ti t_1 \ti t_2^*$ and $H_a\to \ti t_2 \ti t_1^*$ at the same
level of precision while all other decays are computed at the highest
possible precision. \s 

\begin{figure}[t!]
\bc
\includegraphics[width=0.45\textwidth]{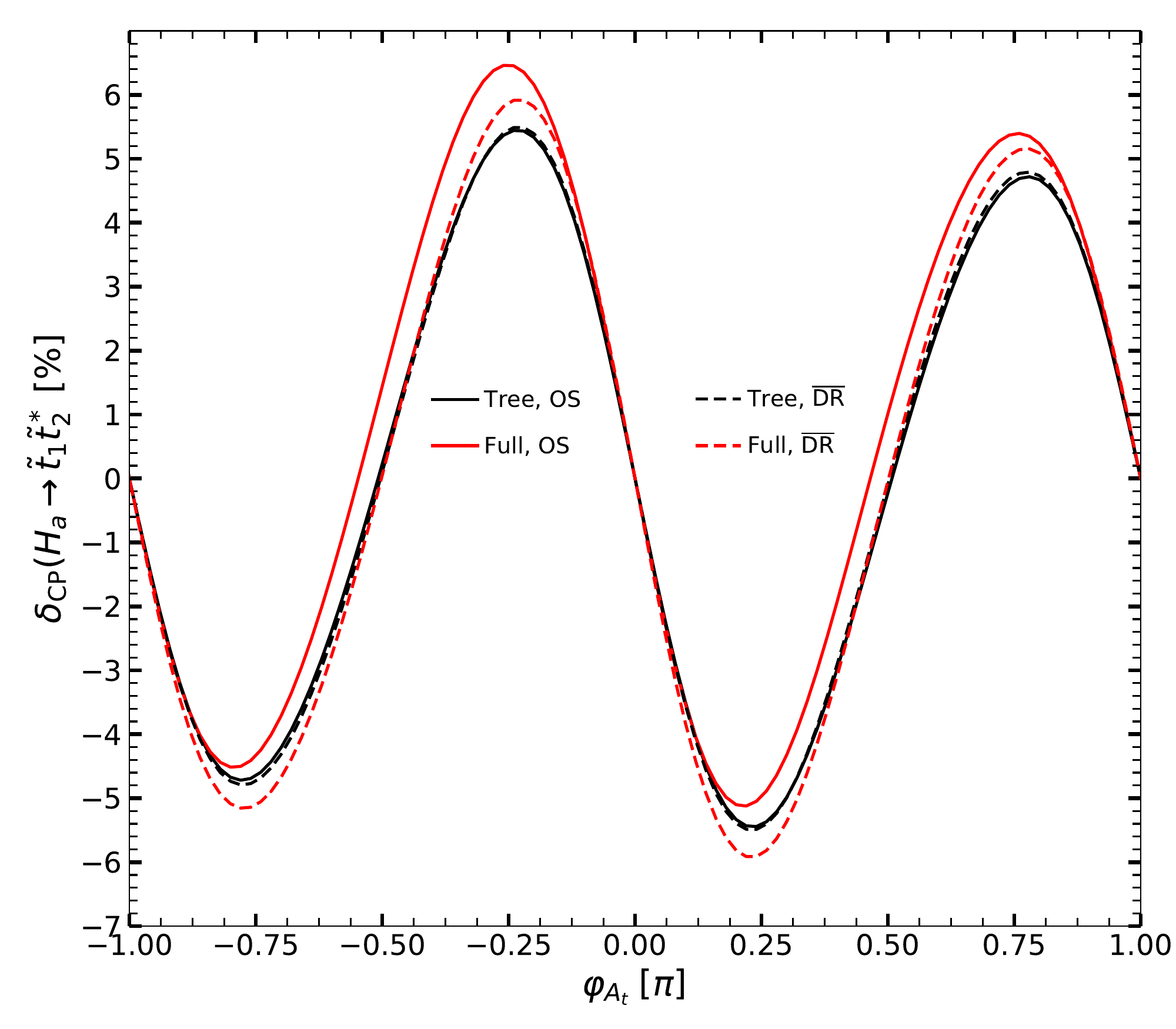}
\caption{CP-asymmetry $\delta_{\text{CP}}$ as function of the
  complex phase $\varphi_{A_t}$ at tree level (black) and including both
  the EW and the QCD corrections (red) in the $\DRb$ (dashed) and OS
  (full) scheme.}
 \label{fig:CPasyH4st1st2}
\ec
\end{figure}

In the CP-invariant scenario the decay width of decay $H_{a}\to \ti
t_1 \ti t_2^*$ is equal to the one of its charge conjugate decay
$H_{a} \to \ti t_2 \ti t_1^*$. For non-vanishing $\varphi_{A_t}$,
however, the CP asymmetry, defined as 
\be 
\delta_{\text{CP}} = \fr{ \Ga(H_a\to \ti t_1 \ti t_2^*) -\Ga(H_a\to \ti t_2
  \ti t_1^*)}{\Ga(H_a\to \ti t_1 \ti t_2^*) +\Ga(H_a\to \ti t_2 \ti
  t_1^*)} \;,
\ee
is non-zero. In \figref{fig:CPasyH4st1st2} we show the CP asymmetry as
a function of  $\varphi_{A_t}$. We see that the CP asymmetry appears 
already at tree-level, which results from the imaginary part of the
WFR factor $\ZH$ and the imaginary part of the tree-level couplings
$g_{h_i\ti q_j\ti q_k^*}$. The relative change of the
asymmetry due to loop corrections ranges between 18\% and -9\% in the OS
scheme while in the $\DRb$ scheme it is about 8\% when the
phase $\varphi_{A_t}$ is varied from $-\pi/4$ to $\pi/4$. \s

\begin{figure}[t!]
\includegraphics[width=0.45\textwidth]{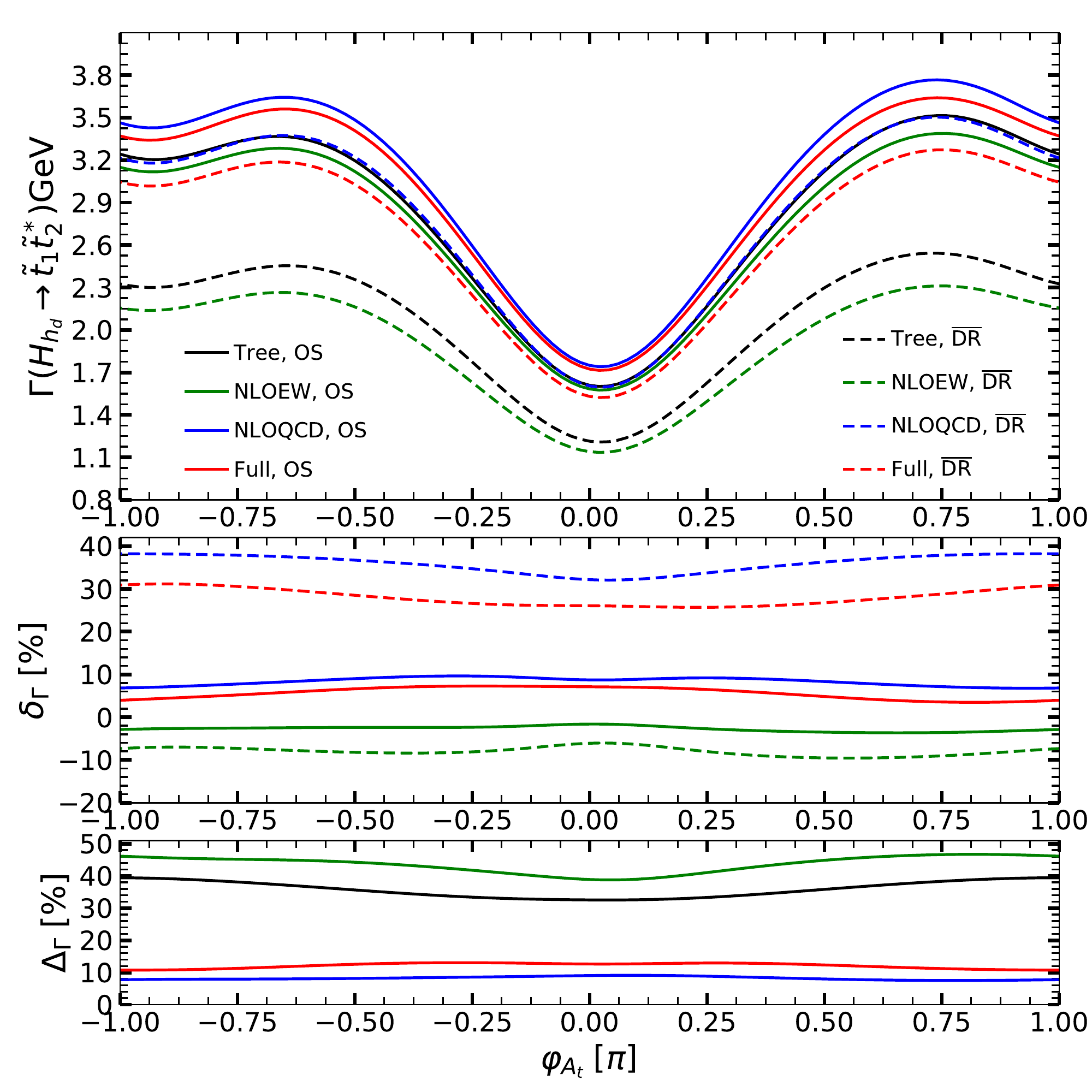}
\includegraphics[width=0.45\textwidth]{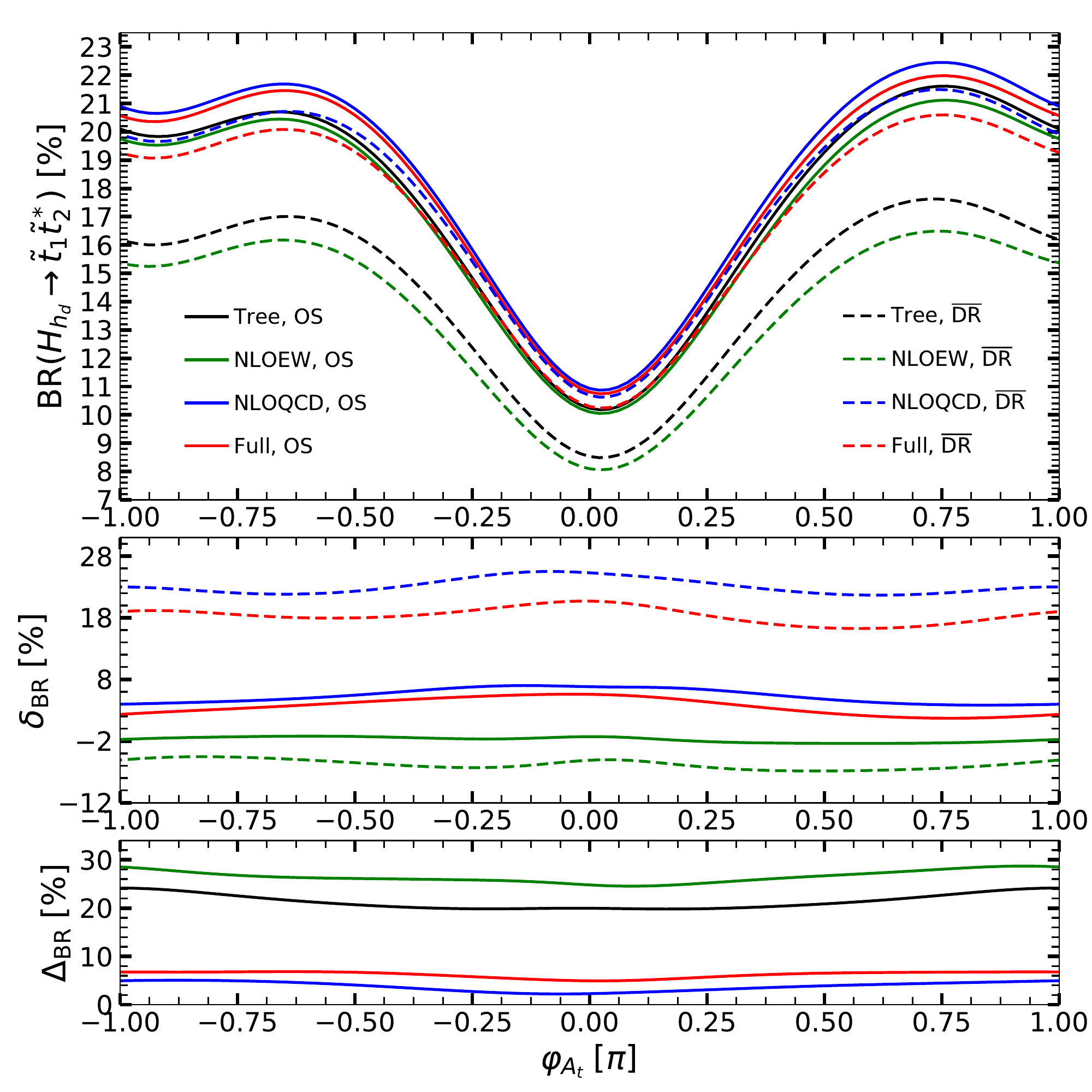}
  \caption{Similar to \figref{fig:plotHast1st2} but for the decay
    $H_{h_d}\to \ti t_1\ti t_2^*$.}
 \label{fig:plotHdst1st2}
\end{figure}

We present in \figref{fig:plotHdst1st2} the same plots for the decay
widths and branching ratios as in \figref{fig:plotHast1st2} but for
the decay of $H_3 \to \ti t_1 \ti t_2^*$ which is the
dominant decay channel of $H_3$. In {\it scenario2}
$H_3$ is $h_d$-like. For both the OS and the $\DRb$
scheme the NLO EW corrections to the decay width are negative and the
relative corrections $\delta_\Gamma$ are around -3\% in the OS and
  -8\% in the $\DRb$ scheme. The NLO QCD corrections on the other
hand are positive and their relative size can reach $8\%$ in the OS
scheme and around $37\%$ in the $\DRb$ scheme. For 
  $\varphi_{A_t}=0$, the difference between the decay widths 
in the OS and the $\DRb$ scheme, $\Delta_\Gamma$, is
about $36\%$ at tree-level and 
reduces to 12\% at full one-loop
level. The corresponding numbers for the branching ratios are similar.  \s

We have observed that for this parameter point the EW and QCD
corrections in the OS scheme are smaller in the $\DRb$
scheme. 
In the OS scheme we have seen that this is due to a cancellation between
the genuine triangle diagram contribution and the counterterm 
contribution while this does not happen in the $\DRb$
scheme. Overall the inclusion of the QCD corrections
in addition to the EW corrections reduces the difference
between the OS and $\DRb$ results. 
\section{Conclusions\label{sec:conclusions}} 
In this paper, we have presented in the framework of the CP-violating NMSSM 
our calculation of the complete (SUSY-)EW to the on-shell Higgs boson decays
into fermion pairs, massive gauge boson final states, gauge and Higgs boson
final state pairs, electroweakino and stop and sbottom pairs. Where
applicable we have included the SUSY-QCD corrections. We
have implemented these new corrections 
into {\tt NMSSMCALC}, a Fortran code for the computation of the Higgs
mass spectrum up to presently two-loop order ${\cal O}(\alpha_t
\alpha_s + \alpha_t^2)$ and the calculation of their branching
ratios. The code already included in the branching ratios the state-of-the-art QCD
corrections and the $\Delta_b$ corrections as well as decays into off-shell
massive gauge boson pairs and decays with off-shell 
heavy quarks in the final 
state. The new code is called {\tt NMSSMCALCEW}. \s

The consistent implementation of our newly computed
corrections provides the presently highest level of precision in the
calculation of the NMSSM Higgs boson decays. In contrast to the
previous {\tt NMSSMCALC} version, we have included the $\ZH$ factor with full momentum 
dependence in order to render the loop-corrected masses on-shell. 
For the decays into the electroweakinos and into squark pairs different
renormalization schemes were implemented. 
The numerical analysis has demonstrated that the relative change in the branching ratios
due to this new treatment and the newly implemented corrections is
significant. The analysis of the decays into chargino/neutralino and
squark pairs for different renormalization schemes has shown that the
one-loop corrections reduce the theoretical uncertainty due to missing
higher-order corrections. The new code {\tt NMSSMCALCEW}
can be obtained at the url: \\
https://www.itp.kit.edu/$\sim$maggie/NMSSMCALCEW/.

\begin{appendix}
\section{Counterterm contribution to the decay of a neutral Higgs boson into a squark pair}
\label{appendix1}
In this appendix, we give explicit  expressions of the counterterm
couplings for the decays of a neutral Higgs boson into a squark
pair. Note that we have used the redefined WFR constants for the
squarks given in Eqs.~(\ref{eq:diagWFRsquark}) and
(\ref{eq:offdiagWFRsquark}). For the EW corrections, the counterterm
entering  \eqref{eq:EWhsqsq} for the Higgs decay into a stop pair is given by
\begin{align}
\calM^{\text{CT ,EW}}_{h_{i}\ti t_{j} \ti t_{k}^*}&= \fr12\sum_{i^\prime=1}^5 g_{h_{i^\prime}
\ti t_{j} \ti t_{k}^*} \delta Z_{h_i h_{i^\prime}}
+\fr12 \sum_{j^\prime=1}^2 \braket{g_{h_{i} \ti t_{j^\prime} \ti t_{k}^*} 
\delta Z_{\tilde{t}_{j^\prime} \tilde{t}_j} (M_{\ti t _j}^2)+
g_{h_{i} \ti t_{j} \ti t_{j^\prime}^*} \delta Z^*_{\tilde{t_{j^\prime}} \tilde{t}_k} (M_{\ti t_k}^2) } \crn
&+g_{h_{i} \ti t_{j} \ti t_{k}^*}\braket{ \de Z_e  + \fr{\de M_Z^2}{ 2 M_Z^2 } - \fr{\de s_{\theta_W}}{s_{\theta_W}}  - \fr{\de c_{\theta_W}}{c_{\theta_W}}  }\crn
& +\fr{eM_Z} {s_{\theta_W}c_{\theta_W}}\bigg[ \fr{m_t^2 \calR_{i2} \braket{U_{j1}^{\ti t*}
U_{k1}^{\ti t} +U_{j2}^{\ti t*} U_{k2}^{\ti t}  } }{\sbeta M_Z^2} \braket{ 2 \fr{\de m_t}{ m_t} -\fr{\de M_Z^2}{M_Z^2} - \fr{\de \sbeta}{ \sbeta}  }  \crn
\nonumber
\end{align}
\begin{align}
\phantom{\calM^{\text{CT ,EW}}_{h_{i}\ti t_{j} \ti t_{k}^*}}
&+\fr{m_t \braket{ U_{j2}^{\ti t*}
U_{k1}^{\ti t} F_1 +U_{j1}^{\ti t*} U_{k2}^{\ti t} F_1^* } }{2 \sbeta M_Z^2} \braket{  \fr{\de m_t}{ m_t} -\fr{\de M_Z^2}{M_Z^2} - \fr{\de \sbeta}{ \sbeta}  } 
\crn
&+
\fr{m_t \braket{ U_{j2}^{\ti t*}
U_{k1}^{\ti t} \de F_1 +U_{j1}^{\ti t*} U_{k2}^{\ti t} \de F_1^* } }{2
  \sbeta M_Z^2} 
\crn
&+ \fr 16 \braket{\de \cbeta \calR_{i1}  - \de \sbeta \calR_{i2}}\braket{ (4c_{\theta_W}^2 - 1 )U_{j1}^{\ti t*} U_{k1}^{\ti t} +
4 s_{\theta_W}^2 U_{j2}^{\ti t*} U_{k2}^{\ti t} }
\crn
&+\fr 43 \braket{ \cbeta \calR_{i1}  - \sbeta \calR_{i2}}\braket{ - U_{j1}^{\ti t*} U_{k1}^{\ti t} +
  U_{j2}^{\ti t*} U_{k2}^{\ti t} }s_{\theta_W}\de s_{\theta_W}
 \bigg],
\end{align}
and for the decay into a sbottom pair 
\begin{align}
\calM^{\text{CT,EW}}_{h_{i}\ti b_{j} \ti b_{k}^*}&=\fr12\sum_{i^\prime=1}^5g_{h_{i^\prime} \ti b_{j} \ti
b_{k}^*} \delta Z_{h_i h_{i^\prime}} +\fr12\sum_{j^\prime=1}^2\braket{g_{h_{i}
\ti b_{j^\prime} \ti b_{k}^*} \delta Z_{\tilde{b}_{j^\prime} \tilde{b}_j} (M_{\ti b_ j}^2)+
g_{h_{i} \ti b_{j} \ti b_{j^\prime}^*} \delta Z^*_{\tilde{b_{j^\prime}} \tilde{b}_k}(M_{\ti b_ k}^2) } \crn
&+g_{h_{i} \ti b_{j} \ti b_{k}^*}\braket{ \de Z_e  + \fr{\de M_Z^2}{ 2 M_Z^2 } - \fr{\de s_{\theta_W}}{s_{\theta_W}}  - \fr{\de c_{\theta_W}}{c_{\theta_W}}  }\crn
&+\fr{eM_Z} {s_{\theta_W}c_{\theta_W}}\bigg[ \fr{m_b^2 \calR_{i1} \braket{U_{j1}^{\ti b*}
U_{k1}^{\ti b} +U_{j2}^{\ti b*} U_{k2}^{\ti b}  } }{\cbeta M_Z^2}  \braket{ 2 \fr{\de m_b}{ m_b} -\fr{\de M_Z^2}{M_Z^2} - \fr{\de \cbeta}{ \cbeta}  }  \crn
& +\fr{m_b \braket{ U_{j2}^{\ti b*}
U_{k1}^{\ti b} F_2 +U_{j1}^{\ti b*} U_{k2}^{\ti b} F_2^* } }{2 \cbeta M_Z^2} \braket{  \fr{\de m_b}{ m_b} -\fr{\de M_Z^2}{M_Z^2} - \fr{\de \cbeta}{ \cbeta}  }  \crn
& +\fr{m_b \braket{ U_{j2}^{\ti b*}
U_{k1}^{\ti b} \de F_2 +U_{j1}^{\ti b*} U_{k2}^{\ti b} \de F_2^* } }{2 \cbeta M_Z^2}\crn
&- \fr 16 \braket{\de \cbeta \calR_{i1}  -\de \sbeta \calR_{i2}}\braket{ (2c_{\theta_W}^2 + 1 )U_{j1}^{\ti b*} U_{k1}^{\ti b} +
2 s_{\theta_W}^2 U_{j2}^{\ti b*} U_{k2}^{\ti b} }\crn
&- \fr 23 \braket{\cbeta \calR_{i1}  -\sbeta \calR_{i2}}\braket{
  - U_{j1}^{\ti b*} U_{k1}^{\ti b} +
 U_{j2}^{\ti b*} U_{k2}^{\ti b} } s_{\theta_W} \de s_{\theta_W}
 \bigg],
\end{align}
where
\begin{align}
\de F_1&=  \de A_t^* e^{-i\varphi_u} \braket{ \calR_{i2}- i\cbeta\calR_{i4} } -\de \mueff \braket{\calR_{i1}  + i\sbeta\calR_{i4}}\crn
& 
-\fr{v \lambda\cbeta \braket{\calR_{i3} + i\calR_{i5} } e^{i\varphi_s} }{\sqrt{2} }\braket{\fr{\de \lambda}{\lambda}+
\fr{\de v}{v} +\fr{\de \cbeta}{\cbeta} },\\ 
\de F_2&= \de A_b^*  \braket{ \calR_{i1}- i\sbeta\calR_{i4} } -\de\mueff e^{i\varphi_u} \braket{\calR_{i2}  + i\cbeta\calR_{i4}} \crn
&
-\fr{v \lambda  \sbeta \braket{\calR_{i3} + i\calR_{i5} } e^{i(\varphi_s +\varphi_u )}} {\sqrt{2} } \braket{\fr{\de \lambda}{\lambda}+
\fr{\de v}{v} +\fr{\de \sbeta}{\sbeta} }.
\end{align}
For the QCD corrections, the counterterm couplings are obtained from
$\calM^{\text{CT,EW}}_{h_{i}\ti q_{j} \ti q_{k}^*}$ by setting the set of EW counterterms, 
$\de Z_e, \de M_Z^2, \de v, \de v_s, \de \lambda, \de s_{\theta_W}, \de
c_{\theta_W}, \de \sbeta, \de \cbeta$ and $ \delta Z_{h_i h_{i'}}$ to zero.
Note that the counterterm $\delta
  \mu_{\text{eff}}$ is given in terms of $\de \lambda$ and $\de v_s$ as
\be \delta
  \mu_{\text{eff}}= e^{i \varphi_s}\fr{( v_s\de \lambda + \lambda \de
    v_s)}{\sqrt{2}}. \ee

\section{The Code {\tt NMSSMCALCEW} \label{app:nmssmcalcew}}
We here give a brief introduction into our new code {\tt NMSSMCALCEW}
that includes the newly calculated and here presented corrections to
the decay widths of the neutral NMSSM Higgs bosons in the CP-violating
NMSSM, as well as the newly calculated and here presented
one-loop corrections to the chargino, neutralino, stop, and sbottom
masses. It is derived from the code {\tt NMSSMCALC}, which is described
in detail in Ref.~\cite{Baglio:2013iia}. We here concentrate on the new
features in {\tt NMSSMCALCEW} with respect to {\tt NMSSMCALC}. \s

{\tt NMSSMCALCEW} requires 'LoopTools' version 2.14 (or
higher) \cite{vanOldenborgh:1989wn,Hahn:1998yk} to
work with the EW corrections in the decay part. If 'LoopTools' is not
installed yet, it can be obtained from the url: 
http://www.feynarts.de/looptools/ \s

In order to generate the executable, download and unpack the tar file
'nmssmcalcew.tar.gz'. It contains two subdirectories called
'nmssmcalc\_rew\_alphat2-master' and 'nmssmcalc\_cew\_alphat2-master' for
the real and complex NMSSM, respectively. Go to the subdirectory of
the version in which you want to work in. Open in a text editor the
file 'makefile' and in line 
31 provide the absolute path to the
'LoopTools' binary directory located in the main directory of LoopTools.
Modify also the line 66 (to make sure it refers to the correct 'lib'
sub-directory within the 'LoopTools' binary directory). 
In case the package is compiled without the EW corrections in the
decay widths, the flag 'yesEW' can be switched to 'FALSE' on line 19 
and 'LoopTools' is not needed anymore. 
Subsequently, all files are compiled by typing 'make'. An executable
'run' is created. By typing'./run' the executable is run. \s

For the code to be run, the user has to provide the input files for
'CalcMasses.F' (default name 'inp.dat') and for bhdecay(\_c).f (to be
named 'bhdecay.in'). The user also has the choice to provide in the
command line the names of the input and output files for
'CalcMasses.F' (first and second argument) and the name of the output
file provided by the decay routine (third argument). Hence the command
will be 'run name\_file1 name\_file2 name\_file3' in this case. Sample
input files 'inp.dat' and 'bhdecay.in' are included in the .tar files. 
By typing 'make clean' the executable as well as the object
files generated in the 'obj' directory are removed. \s

In 'bhdecay.in' that is used by {\tt NMSSMCALCEW}, new options
have been included. They are 

\begin{itemize}
\item 'ischhXX' to choose the renormalization scheme for the loop
  corrected electroweakino masses. The options are 1 (OS1), 2 (OS2)
  and 3 (DRbar). The two OS schemes are specified in Section
  \ref{sssect:NCrenor}.
\item 'ischhst' to choose the renormalization scheme for the stop
  sector. The options are 1 (OS NLO-EW, OS NLO-QCD), 2 (OS NLO-EW,
  DRbar NLO-QCD), 3 (DRbar NLO-EW, DRbar NLO-QCD). 
\item 'ischhsb' to choose the renormalization scheme for the sbottom
  sector. The options are 1 (OS NLO-EW, OS NLO-QCD), 2 (OS NLO-EW,
  DRbar NLO-QCD), 3 (DRbar NLO-EW, DRbar NLO-QCD). 
\item 'iewh' to choose the levels of NLO SUSY-EW (SUSY-QCD)
  corrections that are included. The options are 0 (as in {\tt NMSSMCALC}
  3.00), 1 (decays as implemented in {\tt NMSSMCALC}, but with the $\ZH$
  factor), 2 (full NLO corrections as described in this paper). Both
  for option 1 and 2, the loop-corrected electroweakino and
  stop/sbottom masses are used in the phase space factor. 
\end{itemize}

Further information on the organization of the files for the code and
their functionalities as well as modifications on the code (which are
constantly updated) can be found at the webpage of {\tt NMSSMCALCEW}.
The code has been tested on a Linux machine.

\end{appendix}


\paragraph{Acknowledgments\newline}

JB acknowledges the support from the Carl-Zeiss foundation. 
TND and MM acknowledge financial support from the DFG project
Precision Calculations in the Higgs Sector - Paving the Way to the New
Physics Landscape (ID: MU 3138/1-1).  TND's work is funded by the Vietnam 
National Foundation for Science and Technology Development (NAFOSTED)
under grant number 103.01-2017.78. \s

{\bf Note added} While this paper was being completed,
Ref.~\cite{Domingo:2019vit} appeared, where the impact of the EW
corrections to heavy Higgs boson decays into fermion pairs was
discussed. It was shown that these contributions are dominated by
Sudakov double logarithms. The authors furthermore stressed the
relevance of three-body decays for a consistent evaluation of the
total widths and branching ratios at complete one-loop order.


\bibliographystyle{h-physrev}


\end{document}